\documentclass[11pt,a4paper]{article}

\usepackage{ascmac}
\usepackage{amsmath}
\usepackage{amssymb}
\usepackage{bm}
\usepackage[footnotesize,bf]{caption}
\usepackage{fullpage}
\usepackage{color}
\usepackage{float}
\usepackage{graphicx}
\usepackage[]{natbib}
\usepackage{subfigure}
\usepackage{txfonts}
\usepackage{threeparttable}
\usepackage{wrapfig}
\usepackage[]{multicol}
\usepackage{wrapfig}
\usepackage{authblk}

\usepackage{floatflt}

\title{\large Complete list of the ASTRO-H Science Working Group}
\date{\vspace{-0.5cm}}
\newcommand{\MakeWhitePaperTitle}{
	\begin{center}
		\begin{figure}
			\vspace{1cm}
			\begin{center}
				\includegraphics[width=0.2\hsize]{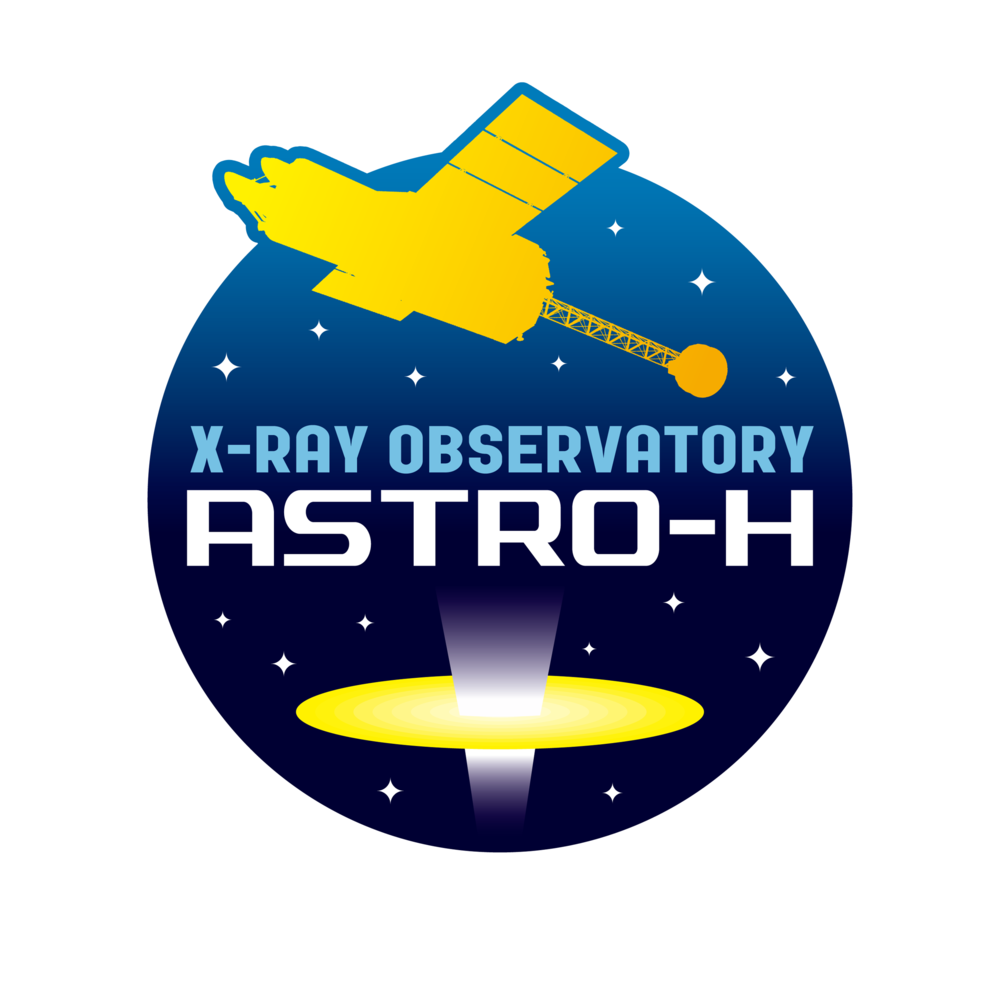}
			\end{center}
		\end{figure}
		\vspace{1cm}
		{\LARGE
		ASTRO-H Space X-ray Observatory\\
		White Paper\\
		}
		\vspace{5mm}
		{\large
		\WhitePaperTitle\\
		}
		\vspace{1cm}
		{
		\WhitePaperAuthors\\
		on behalf of the ASTRO-H Science Working Group
		}
	\end{center}
}

\usepackage{authblk}
\author[a]{Tadayuki~Takahashi}
\author[a]{Kazuhisa~Mitsuda}
\author[b]{Richard~Kelley}
\author[c]{Felix~Aharonian}
\author[d]{Hiroki~Akamatsu}
\author[e]{Fumie~Akimoto}
\author[f]{Steve~Allen}
\author[g]{Naohisa~Anabuki}
\author[b]{Lorella~Angelini}
\author[h]{Keith~Arnaud}
% ...
\affil[a]{Institute of Space and Astronautical Science (ISAS), Japan Aerospace Exploration Agency (JAXA), Kanagawa 252-5210, Japan}
\affil[b]{NASA/Goddard Space Flight Center, MD 20771, USA}
\affil[c]{Astronomy and Astrophysics Section, Dublin Institute for Advanced Studies, Dublin 2, Ireland}
\affil[d]{SRON Netherlands Institute for Space Research, Utrecht, The Netherlands}
\affil[e]{Department of Physics, Nagoya University, Aichi 338-8570, Japan}
\affil[f]{Kavli Institute for Particle Astrophysics and Cosmology, Stanford University, CA 94305, USA}
\affil[g]{Department of Earth and Space Science, Osaka University, Osaka 560-0043, Japan}
\affil[h]{Department of Astronomy, University of Maryland, MD 20742, USA}
% ...

\usepackage{subfigure}
\usepackage{txfonts}
\usepackage{threeparttable}
\usepackage{wrapfig}
\usepackage{comment}
\usepackage{tabularx}

\newcommand{\apj}{Astrophysical Journal}
\newcommand{\apjl}{Astrophysical Journal Letter}
\newcommand{\physrep}{Physics Reports}
\newcommand{\pasj}{Publication of the Astronomical Society of Japan}

\newcommand{\mnras}{Monthly Notices of the Royal Astronomical Society}
\newcommand{\aapr}{The Astronomy and Astrophysics Review}
\newcommand{\apss}{Astrophysics and Space Science}
\newcommand{\nat}{Nature}
\def\jnl@style{\it}
\def\aaref@jnl#1{{\jnl@style#1}}

\def\aaref@jnl#1{{\jnl@style#1}}

\def\aj{\aaref@jnl{AJ}}                   
\def\araa{\aaref@jnl{ARA\&A}}             
\def\apj{\aaref@jnl{ApJ}}                 
\def\apjl{\aaref@jnl{ApJ}}                
\def\apjs{\aaref@jnl{ApJS}}               
\def\ao{\aaref@jnl{Appl.~Opt.}}           
\def\apss{\aaref@jnl{Ap\&SS}}             
\def\aap{\aaref@jnl{A\&A}}                
\def\aapr{\aaref@jnl{A\&A~Rev.}}          
\def\aaps{\aaref@jnl{A\&AS}}              
\def\azh{\aaref@jnl{AZh}}                 
\def\baas{\aaref@jnl{BAAS}}               
\def\jrasc{\aaref@jnl{JRASC}}             
\def\memras{\aaref@jnl{MmRAS}}            
\def\mnras{\aaref@jnl{MNRAS}}             
\def\pra{\aaref@jnl{Phys.~Rev.~A}}        
\def\prb{\aaref@jnl{Phys.~Rev.~B}}        
\def\prc{\aaref@jnl{Phys.~Rev.~C}}        
\def\prd{\aaref@jnl{Phys.~Rev.~D}}        
\def\pre{\aaref@jnl{Phys.~Rev.~E}}        
\def\prl{\aaref@jnl{Phys.~Rev.~Lett.}}    
\def\pasp{\aaref@jnl{PASP}}               
\def\pasj{\aaref@jnl{PASJ}}               
\def\qjras{\aaref@jnl{QJRAS}}             
\def\skytel{\aaref@jnl{S\&T}}             
\def\solphys{\aaref@jnl{Sol.~Phys.}}      
\def\sovast{\aaref@jnl{Soviet~Ast.}}      
\def\ssr{\aaref@jnl{Space~Sci.~Rev.}}     
\def\zap{\aaref@jnl{ZAp}}                 
\def\nat{\aaref@jnl{Nature}}              
\def\iaucirc{\aaref@jnl{IAU~Circ.}}       
\def\aplett{\aaref@jnl{Astrophys.~Lett.}} 
\def\apspr{\aaref@jnl{Astrophys.~Space~Phys.~Res.}}
\def\bain{\aaref@jnl{Bull.~Astron.~Inst.~Netherlands}} 
\def\fcp{\aaref@jnl{Fund.~Cosmic~Phys.}}  
\def\gca{\aaref@jnl{Geochim.~Cosmochim.~Acta}}   
\def\grl{\aaref@jnl{Geophys.~Res.~Lett.}} 
\def\jcp{\aaref@jnl{J.~Chem.~Phys.}}      
\def\jgr{\aaref@jnl{J.~Geophys.~Res.}}    
\def\jqsrt{\aaref@jnl{J.~Quant.~Spec.~Radiat.~Transf.}}
\def\memsai{\aaref@jnl{Mem.~Soc.~Astron.~Italiana}}
\def\nphysa{\aaref@jnl{Nucl.~Phys.~A}}   
\def\physrep{\aaref@jnl{Phys.~Rep.}}   
\def\physscr{\aaref@jnl{Phys.~Scr}}   
\def\planss{\aaref@jnl{Planet.~Space~Sci.}}   
\def\procspie{\aaref@jnl{Proc.~SPIE}}   

\begin{document}

\newcommand{\WhitePaperTitle}{Accreting Pulsars, Magnetars, \& Related Sources}
\newcommand{\WhitePaperAuthors}{
S.~Kitamoto~(Rikkyo~University),
T.~Enoto~(RIKEN~\&~NASA/GSFC), \\
S.~Safi-Harb~(University~of~Manitoba),
K.~Pottschmidt~(UMBC~\&~NASA/GSFC),
C.~Ferrigno~(University~de~Geneve),
M.~Chernyakova~(Dublin~Institute~for~Advanced~Studies),
T.~Hayashi~(JAXA),
N.~Hell~(LLNL\footnote{Also at Remeis~Observatory,~FAU.}),
K.~Kaneko~(Tokyo~University~of~Science),
D.~Khangulyan~(JAXA),
T.~Kohmura~(Tokyo~University~of~Science),
H.~Krimm~(NASA/GSFC),
K.~Makishima~(University~of~Tokyo),
T.~Nakano~(University~of~Tokyo),
H.~Odaka~(JAXA),
M.~Ohno~(Hiroshima~University),
M.~Sasano~(University~of~Tokyo),
S.~Sugita~(Ehime~University),
Y.~Terada~(Saitama~University), 
T.~Yasuda~(Saitama~University), and
T.~Yuasa~(JAXA)
}
\MakeWhitePaperTitle

\begin{abstract}
  As the endpoints of massive star evolution, neutron stars are
  enigmatic celestial objects characterized by extremely dense and exotic
  nuclear matter, magnetospheres with positrons
  (antimatter), rapid rotation and ultra-strong magnetic fields.  Such an extreme
  environment has provided an accessible astrophysical laboratory to study
  physics under conditions unattainable on Earth and
  to tackle a range of fundamental questions related to: the aftermath of stellar
  evolution and the powerful explosions of massive stars, the equation
  of state and physics of some of the most exotic and magnetic stars
  in the Universe, the workings of the most powerful particle
  accelerators in our Galaxy and beyond, and the sources of
  gravitational waves that are yet to be detected.  Recent
  observations revealed a great diversity of neutron stars, including
  ultra-strongly magnetized pulsars, referred to as ``magnetars", and
  unusual types of accreting X-ray pulsars.  In this white paper, we
  highlight the prospects of the upcoming X-ray mission, \textsl{ASTRO-H}, in
  studying these highly magnetized neutron stars.
\end{abstract}
\clearpage

\maketitle
\clearpage

\tableofcontents
\clearpage

\section{Overview: Strongly Magnetized Neutron Stars}
\subsection{Introduction}

Since their discovery, neutron stars (NSs) have excited a broad range
of interests not only in the astrophysical context, but also in terms
of fundamental physics: for example, confirmation of the existence of
gravitational waves \citep{1982ApJ...253..908T}, high-density nuclear matter
inside NSs \citep{2007PhR...442..109L}, and high-magnetic field
effects around these enigmatic objects \citep{2006RPPh...69.2631H}.
Since NSs are characterized by extreme conditions, such as dense
matter, rapid rotation, and high magnetic field, they have proved to
be ideal laboratories to test fundamental physics, which cannot be
achieved by ground-based experiments. Even 45 years into their
discovery, space-based observations are becoming more important
in understanding the growing diversity of these enigmatic
objects.

\begin{figure}[htb]
\begin{center}
  \includegraphics[width=0.5\hsize]{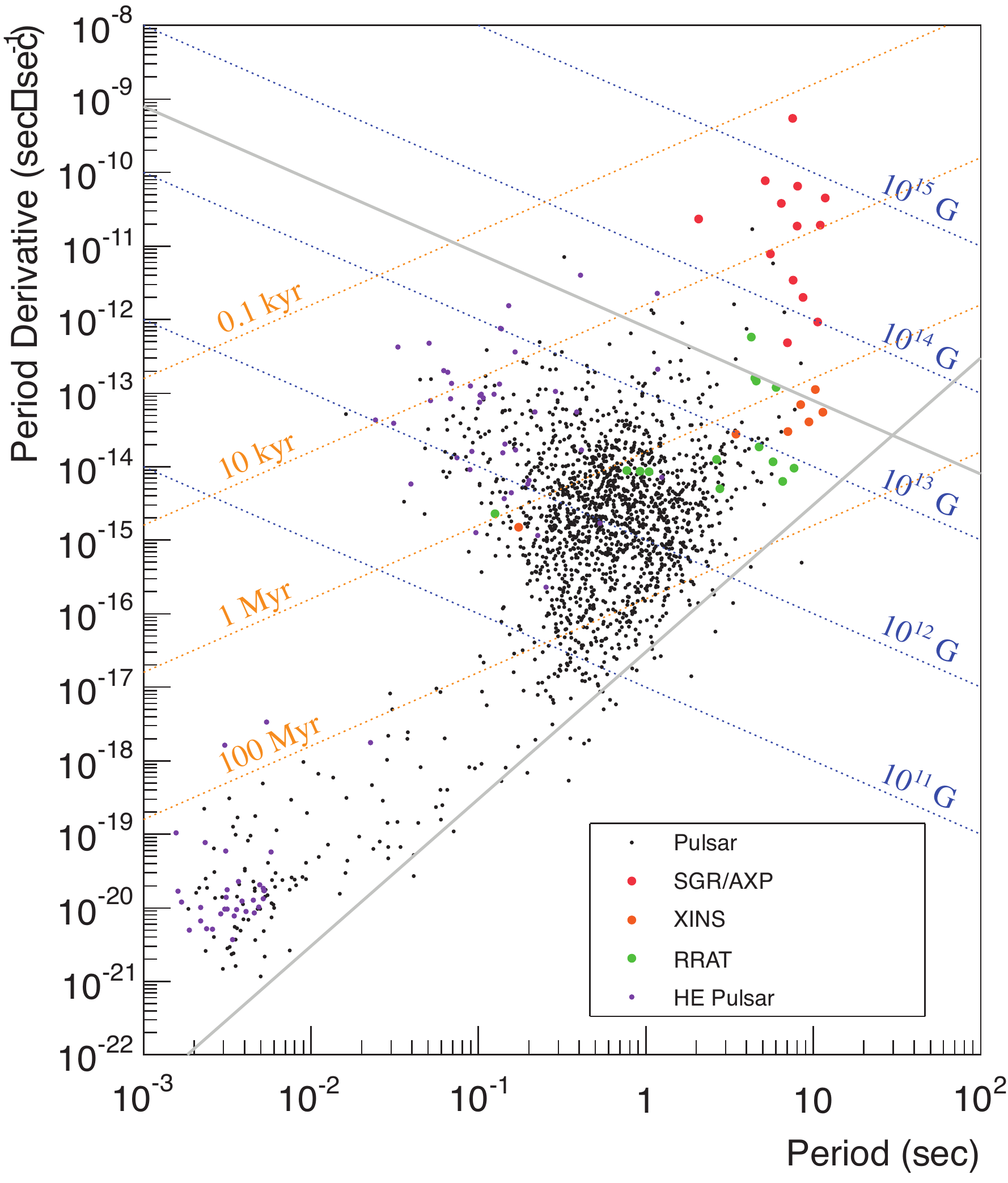}
\caption{ The $P$-$\dot{P}$ diagram of pulsars shown with lines of
  constant dipole magnetic field, $B$, and spin-down age
  ($\tau_c$=$\frac{P}{2~\dot{P}}$). The black dots show the majority
  of radio-discovered pulsars believed to be rotation-powered, and the
  red circles show the X-ray or gamma-ray discovered magnetars
  addressed in this white paper (see \S1.1). Source: ATNF pulsar
  catalog \citep{Manchester2005AJ....129.1993M}.}
\label{fig1:ppdot}
\end{center}
\end{figure}
	
Recent multi-wavelength observations from radio to the highest energy
gamma-rays have revealed a remarkable diversity of NSs
\citep{2010arXiv1005.0876K}. Figure~\ref{fig1:ppdot} shows the
distribution of known pulsars based on their measured spin ($P$) and
spin-down ($\dot{P}$) properties.  So far over $\sim$1700 pulsars have
been discovered in the radio and most of them are thought to be
powered by their rotational energy (i.e. Rotation Powered Pulsars;
RPPs). Their magnetic field is usually inferred from their spin
properties\footnote{The pulsar's dipole surface magnetic field is
  estimated as $B_{\rm s}$ (Gauss)
  $\approx$3.2$\times$10$^{19}$($P/\dot{P}$)$^{1/2}$ where $P$ is in
  sec.}, in the 10$^{11}$--10$^{13}$~Gauss
(1~Tesla~=~10$^4$~Gauss). In addition some accretion-powered pulsars
show X-ray ``Cyclotron Resonance Scattering Features'' allowing us to
estimate the magnetic field of the neutron star, see
\S\ref{sec:binaries_intro} and \S\ref{sec:crsf}. On the other hand,
Soft Gamma-ray Repeaters (SGRs) and Anomalous X-ray Pulsars (AXPs)
have become a rapidly growing new subclass with much higher magnetic
fields, $B\sim 10^{14}-10^{15}$ Gauss and dubbed as ``magnetars".
Unlike for rotation-powered or accretion-powered pulsars, the bulk of
their X-ray emission appears to be powered by their super-strong
magnetic fields. The growing diversity of NSs includes the Rotating
Radio Transients (RRATs), X-ray Dim Isolated NSs (XDINSs), and Central
Compact Object (CCOs). Understanding the connection between these
classes remains one of the most important questions in this field;
multi-wavelength observations continue to provide clues on their
nature, emission mechanisms and physical properties. A unified
understanding requires a better understanding of their birth
environment, evolutionary path, and interaction with their (binary, if
applicable) environment.
	
In this White Paper, we focus on highly magnetized NSs with magnetic
fields $B>10^{12}$\,Gauss. These include magnetars
(\S\ref{sec:magnetars_intro}, \S\ref{sec:magnetars}) as well as
accreting pulsars (\S\ref{sec:binaries_intro},
\S\ref{sec:binaries}). The latter are mostly found in High Mass X-ray
Binaries (HMXBs). NS with relatively weaker fields
($B\leq10^{11}$\,Gauss) are usually observed in Low Mass X-ray
Binaries (LMXBs), these sources will be reviewed in another White
Paper (WP\#3, Done, Tsujimoto et al.). There are exceptions, though,
like the Low and Intermediate Mass X-ray Binary pulsars GX\,1$+$4
(\S\ref{sec:gx1p4}) and Her~X-1 (\S\ref{sec:crsf}). We also include
new and not yet fully characterized accreting binary classes like
Super-giant Fast X-ray Transients (SFXTs, \S\ref{sec:sfxt}) and
Gamma-ray Loud Binaries (\S\ref{sec:gamma}) in the current White
Paper. Last but not least we also consider peculiar sources that are
not necessarily known to harbor a pulsar and can therefore serve as
comparison sources (e.g., regarding wind accretion) as well as being
unique study objects in their own right, e.g., Cyg~X-3
(\S\ref{sec:wind}) or SS\,433.

\subsection{Accreting Pulsars}\label{sec:binaries_intro}

Many XRBPs comprise a young neutron star (NS), endowed with a strong
$B$-field ($\sim$$10^{12}$ G) and a supergiant or main sequence,
``donor" star. The donor star can lose a conspicuous amount of matter
through a strong stellar wind and/or via the Roche lobe overflow
\citep[see, e.g.,][]{frank2002}. A large part of this material is
focused toward the NS as a consequence of the strong gravitational
field of this object, and then threaded by its intense magnetic field
at several thousand kilometers from the stellar surface. The region in
which the ram pressure of the plasma equals the magnetic field
pressure is known as the Alfven surface and in its vicinity all the
exchange of angular momentum happens. Once funneled down to the
magnetic poles of the NS, accretion columns may form. Here, the
gravitational potential energy of the accreting matter is first
converted into kinetic energy and then dissipated in the form of
X-rays \citep[see, e.g.,][]{Pringle72,Davids73}.


To a first approximation, the $B$-field lines of the NS rotate rigidly
with the surface of the star, and thus the X-ray emission emerging
from the accretion column is received by the observer modulated at the
spin period of the compact object, if the magnetic and rotational axes
are misaligned. Important effects are expected from the interaction
of radiation with the strongly magnetized plasma (see below).  These
effects can be proved on observational grounds studying pulse profiles
and time resolved spectra.

As a more important progress expected with the \textsl{ASTRO-H}, SXS will probe
with high significance the characteristics of the plasma in the
stellar wind and its coupling with the Alfven surface. This is an open
field of study, which is receiving further thrust by novel simulations
\citep[e.g.,][]{2012A&A...547A..20M} and will receive an observational
breakthrough from the high spectral resolution and high effective area
of the SXS, which can measure precisely, e.g., the ionization
parameter \citep{Turner1968}. The strong wind, with the velocity of
more than $10^8$ cm s$^{-1}$, has a strong impact on the emission line
profiles. Doppler shift and broadening, coupled with self absorption
by the wind itself \citep{2001ApJ...559.1108O}, and P-Cygni profile
will be observed \citep{2000ApJ...544L.123B}. The existence of
accretion wakes and ionization front of a photo-ionized plasma can be
also studied with a modulation of the line profiles induced by the
orbital motion. A subject which looks particularly interested is the
study of the time variation of iron fluorescence line at the spin
period timescale, which might help tracing the ultimate fate of the
plasma before being captured by the $B$-field and unveil the location and
extension of the transition zone between the wind and the NS
magnetosphere.

The understanding of the broad-band energy spectrum and its
time variations can be used to track the X-ray emission pattern close
to the NS surface and ultimately constrain the fundamental parameters
of a neutron stars, such as its mass, magnetic moment and radius.
Among the different radiation processes, the interaction with the
magnetic field is probably the most crucial for the XRBPs, as it gives
rise to the cyclotron resonant scattering features \citep[CRSFs; see,
e.g.,][ for recent models]{isenberg1998,araya2000,gabi2007}. The CRSFs
provide a unique tool to directly estimate the magnetic field strength
in the X-ray emitting region close to the NS surface, because the
centroid energy of the fundamental appears at $E_{cyc} = 11.6 B_{12}
\times (1 + z)^{-1}$ keV. \citep[Here $B_{12}$ is the magnetic field
strength in units of $10^{12}$\,G and $z$ is the gravitational
redshift in the line-forming region,][]{wasserman1983}. If a cyclotron
line is detected with high significance in the energy range of the SGD
(many are known, see \citealt{2012MmSAI..83..230C} for a compilation),
it can be possible to measure the degree of polarisation, which is
induced by the scattering with electrons trapped in a strong magnetic
field. This would be a major breakthrough for our understanding of the
source emission mechanism, as it would be a very strong constraint on
the geometry of the X-ray emitting region, which is currently unknown.

\subsection{Magnetars}\label{sec:magnetars_intro}

We expect NSs to acquire their strong magnetic fields up to
$\sim$10$^{12}$~Gauss or more when their massive progenitors collapse,
but the origin of strong magnetism in NSs is one of the biggest
unsolved problems in astrophysics.  Some of the major challenges for
understanding their fundamental properties lie in the difficulties
associated with probing observationally their internal structure, the
state of nuclear matter, and their magnetic field configuration.
Similarly, we know little about how their magnetic fields evolve and
whether they decay with time.  To tackle these questions, Soft
Gamma-ray Repeaters (SGRs) and Anomalous X-ray Pulsars (AXPs),
collectively called ``magnetars" \citep{1992ApJ...392L...9D,
  1995MNRAS.275..255T}, have in the past decade provided an ideal
laboratory, particularly in the light of recent multi-wavelength
observations and monitoring programs.

Magnetars are a fascinating subclass of NSs
\citep{2006csxs.book..547W, 2008A&ARv..15..225M} characterized by (i)
spin periods in a narrow range of $P$$\sim$2--12 s; (ii) high
spin-down rate, $\dot{P}$, indicative of a young characteristic age
$\tau_c\equiv P/2\dot{P} \lesssim$100 kyr; (iii) dominant emission in
X-rays, with luminosity $L_X \sim 10^{34-36}$ erg s$^{-1}$ which
largely exceeds their spin-down luminosity, $\dot{E}_{\rm
  sd}$~=~$\frac{-2\pi~I~\dot{P}}{P^3}$ (where $I$ is the pulsar's
moment of inertia); (iv) sporadic X-ray activities on time-scales of
msec to years; and (v) no evidence for accretion indicating
\textit{isolated} compact objects, unlike the HMXBs. 
The surface dipole magnetic field of magnetars
estimated from $P$ and $\dot{P}$ (see footnote in the Introduction and
Figure~1) turns out to extremely high, $B_{\rm s} > B_{QED}$, the
so-called quantum electrodynamic critical field of
4.4~$\times$~10$^{13}$~G at which the Landau level separation exceeds
the rest mass energy of an electron, $m_ec^2=511$ keV. We note however
the recent discovery of lower magnetic field magnetars, discussed
later.

       

Despite the wealth of observational studies dedicated for the study of
magnetars, their true nature and their evolutionary link to the other
classes of NSs is still not clear. Moreover, the magnetar hypothesis
itself remains an open issue.  While the magnetar model
\citep{1992ApJ...392L...9D, 1995MNRAS.275..255T} remains the most
popular theory for explaining the overall properties of SGRs and AXPs,
other models have been proposed and are not yet completely ruled out.
These include the fallback disk model \citep{2001ApJ...557L..61A,
  2009ApJ...702.1309E}, quark stars model
\citep{2006ApJ...653..558O}, and white dwarfs model
\citep{2012PASJ...64...56M}. Therefore, a truly fundamental question
related to their intrinsic magnetic fields and the powering mechanism
for their observed emission remains to be answered.  

One straightforward way of confirming their ``magnetar" nature is
through a \textit{direct} measurement of their magnetic fields.  This
can be performed by detecting cyclotron resonance lines as already
established in accreting-powered pulsars using the ``electron"
cyclotron resonance appearing at $E_{\rm ec}=11.6\times (B/10^{12}
{\rm \ G})$ keV; see e.g. \citep{1978ApJ...219L.105T}.
Correspondingly, the ``proton" cyclotron resonance is expected to
appear at an energy
\begin{equation}
	E_{\rm pc} = 0.63\times(B/10^{14} {\rm \ G}) {\rm \ keV}.
	\label{eq:proton_cyclotron}
\end{equation}
Although signatures of the proton cyclotron resonance have been
reported from only a few magnetars \citep{Ibrahim2002ApJ...574L..51I,
  Ibrahim2003ApJ...584L..17I, 2003ApJ...586L..65R, 2008AIPC..983..234G}, none
of them are considered to be convincing due to the limited energy
bands, their transient nature, or insufficient statistics. We are thus
urged to search for a much firmer evidence of proton cyclotron
resonances using the SXS of \textsl{ASTRO-H}. The detailed discussion will be
performed in \S3.3. 


If this subclass indeed exhibits such extreme fields, how do these
stars sustain their strong magnetism, and how
are the magnetars linked to the canonical NSs with typical $B_{\rm s}
\sim 10^{12}$ G?  Since we do not know whether they are intrinsically
different from other ordinary pulsars, an effective approach is to
search for transition objects between magnetars and canonical pulsars.
Such an object has in fact been discovered and caught in the act of
transitioning from a rotation-powered pulsar to a magnetar: the
high-magnetic field pulsar J1846--0258 in the supernova remnant Kes~75
\citep{2008ApJ...678L..43K, 2008Sci...319.1802G}.  Furthermore, more
recently, the {\it Swift} satellite is allowing the detection of
sporadic X-ray outbursts, thus increasing the number of magnetars by a
few new sources per year. These recent on-going discoveries of the
magnetar class implies that this population would be much larger in
our Galaxy than we expected before, requiring us to revise our
understanding of the formation and evolution both of magnetars and
NSs.

\begin{figure}[htb]
\begin{center}
\includegraphics[height=0.32\hsize]{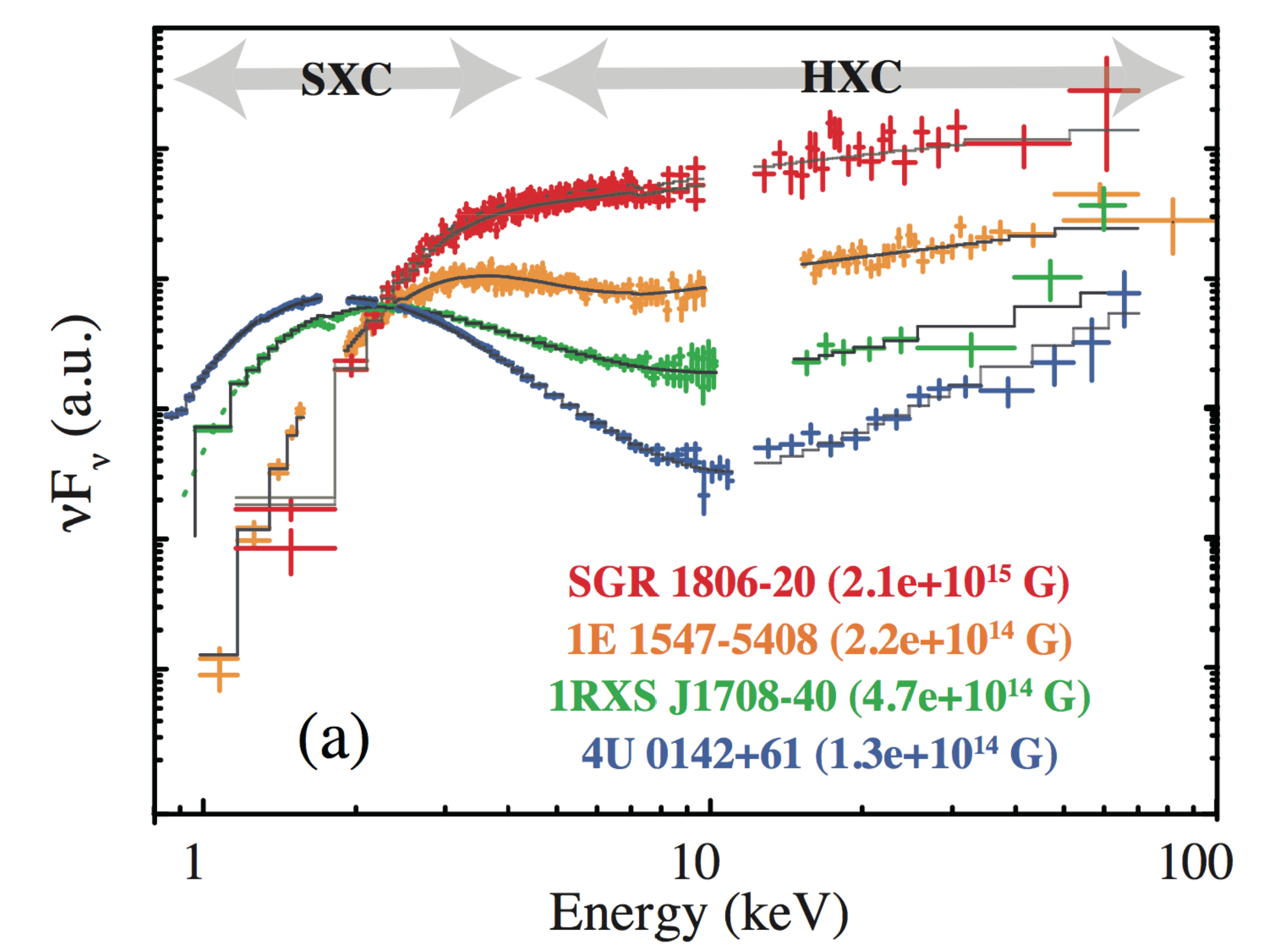}
\includegraphics[height=0.32\hsize]{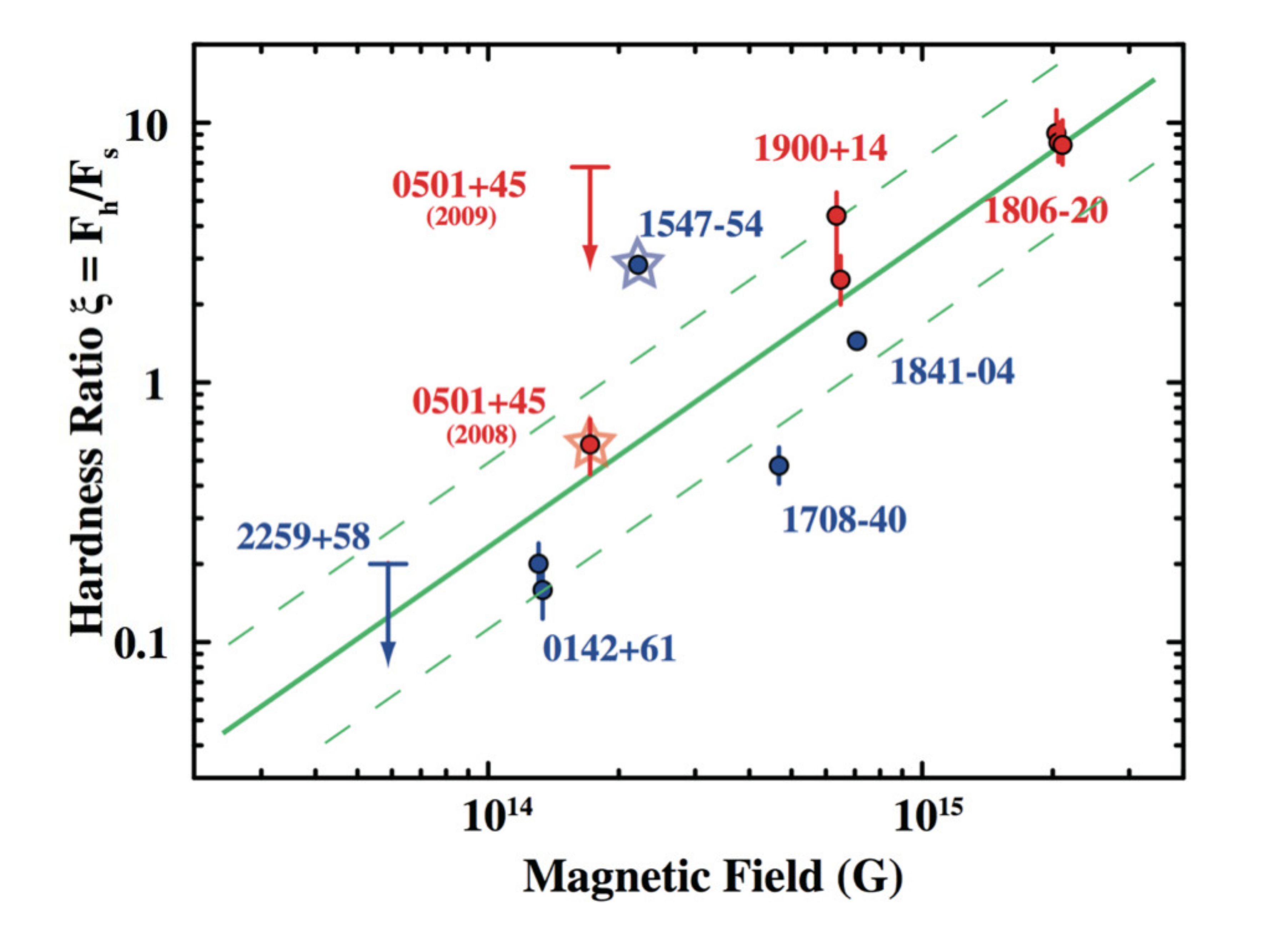}
\caption{(left) Broadband $\nu F_{\nu}$ spectra of 4 magnetars
  observed with \textsl{Suzaku} (modified from \citep{Enoto2010ApJ...722L.162E}).
  To clearly illustrate the difference, each spectrum is normalized at
  2 keV. Individual sources are shown in different colors with their
  magnetic field. (right) Correlation of the broadband HR
  (=HXC/SXC) to the surface magnetic field. 
  }
\label{fig1:suzaku_magnetar_nuFnu}
\end{center}
\end{figure}

The discovery of transient magnetars \citep{2004ApJ...609L..21I} and,
more recently, the discovery of a low-field (i.e.\ $B<B_{QED}$)
``magnetar" SGR 0418+5729 \citep{2010Sci...330..944R} (and a few
others, see later) suggest that there is a larger population of
magnetar-like objects or ``fossil magnetars" in our Galaxy and that
the spin-down dipole magnetic field is \textit{not} the only factor
determining the magnetar properties.  These ``fossil" magnetars will be
interesting potential targets for \textsl{ASTRO-H}. 
As suggested by such weaker field magnetar activities, the NSs are
suggested to store a huge hidden poloidal field component as energy
reservoir in addition to the canonical toroidal field component.  This
is an interesting topic to be related with the supernova explosion
mechanism and the formation of magnetars or even the more canonical
NSs.
In order to observationally address the supernova explosion mechanism,
the birth environment and progenitors of magnetars, one promising
approach is through X-ray spectroscopy of supernova remnants (SNRs)
associated with magnetars. Recent \textit{Chandra} and
\textit{XMM-Newton} observations (e.g. \citealt{2012ApJ...754...96K})
suggested a high-mass progenitor ($\gtrsim$30 $M_\odot$) for an
SNR associated with a high-B pulsar; a result that is also supported
by multi-wavelength studies of a few magnetars (see
e.g. \citealt{2005ApJ...620L..95G, 2007ApJ...667..219M}).  Related
\textsl{Suzaku} studies have been also performed for the SNRs Kes~73 and
CTB~109 hosting AXP (\S3.1).


So far $\sim$20 SGRs and AXPs have been discovered in our
Galaxy and the Magellanic Clouds. For a long time, they were
thought to emit X-rays only below $\sim$10~keV.  This
bright soft X-ray component (hereafter SXC) is represented by a
quasi-thermal spectrum with $kT \sim 0.5$ keV hotter than other
isolated NSs. A new breakthrough came with \textsl{INTEGRAL}'s discovery of
an unexpected hard X-ray component (HXC) above $\sim$10~keV
\citep{2006ApJ...645..556K}. This HXC was confirmed by follow-up
\textsl{Suzaku} observations, covering the SXC and HXC simultaneously thanks
to combining the X-ray Imaging Spectrometer (XIS) and the Hard X-ray
Detector (HXD) \citep{2011PASJ...63..387E}. Figure
\ref{fig1:suzaku_magnetar_nuFnu} shows $\nu F_{\nu}$  spectra
of four magnetars. The HXC has an extremely hard photon index,
$\Gamma_{\rm h} \sim 1$, almost the hardest among known X-ray sources,
and extends up to $\sim$100 keV. Although theoretical
accounts have not yet been reached (e.g.,
\citealt{2005ApJ...634..565T, 2007Ap&SS.308..109B}), its near-absence in
other types of X-ray sources suggests its connection to the strong
magnetic fields of magnetars.



It is hence imperative  to examine their broad-band
spectral properties including both the SXC and the HXC to understand
this enigmatic object.  \textsl{Suzaku} has
performed broad-band (0.8--70 keV) observations of $\sim$9 magnetar
sources, providing some evidence that two components in the magnetars
spectra (i.e.\ the SXC and HXC described above) systematically change
depending on their magnetic field and characteristic age
\citep{Enoto2010ApJ...722L.162E}. In particular, the HXC becomes weaker
relative to the SXC, but harder for older objects. Although SGRs and
AXPs first came into astrophysics as unrelated objects, where the
former class was discovered through the burst activity while the
latter class was recognized as unusually bright X-ray pulsars, the
present correlation implies a unification of SGRs and AXPs into one
evolutionarly path.

Recent exciting discoveries of X-ray outbursts from magnetars also
provide us rare opportunities to search for the HXC during active
states. Following the {\it Swift} discoveries of magnetar outbursts,
\textsl{Suzaku} and \textsl{INTEGRAL} follow-up observations successfully
detected the enhanced HXC and their gradual decays coinciding with the
decay of the SXC: e.g., SGR~0501+4516 \citep{2009MNRAS.396.2419R,
  2010PASJ...62..475E} and 1E~1547.0-5408 \citep{2010PASJ...62..475E,
  2012ApJ...748..133K, 2012PASJ_Iwahashi}. Although the number of the
HXC-confirmed magnetars is gradually increasing, we still don't know
whether all magnetars actually exhibit the HXC in all the states;
e.g., the HXC has not yet been confirmed from some persistent emitting
sources, such as 1E~2259+586 and 1E~1048.1--5937. In the \textsl{ASTRO-H}
era, our next step is to investigate the broadband spectra of
transient magnetars in their outbursts. In particular, to bring our
discoveries during the quiescent states into the next stage, our main
strategy will be to perform prompt observations of the HXC just after
the onset of the outburst. These questions on a unified understanding
of the magnetar evolution will be discussed in detail in \S3.2.
	
Another distinctive feature  is sporadic burst activities,
short bursts, intermediate flares, and giant flares, usually
coinciding with X-ray outbursts. The spectra of SGR bursts detected
with {\it HETE}-2 and {\it Suzaku} were described well by two
blackbody components \citep{2004ApJ...616.1148O, Nakagawa2007PASJ...59..653N}.
The blackbody temperature can much exceed the normal Eddington limit,
probably due to suppression of the Thomson scattering. Thus, the
strong magnetic field significantly affects the familiar blackbody
radiation. Also, the blackbody radius sometime increases up to 30
km during bursts, suggesting the photosphere to expand to fill the
magnetosphere. At the moment, we don't know why two blackbody
components appear. Do they represent the two photo polarizations (O-
and X-modes)\footnote{Two photon polarizations of which the electric
  vector is parallel and perpendicular to the magnetic field.}  which
acquire different ``photospheres" when $B>B_{\rm c}$?  More
fundamentally, is the two-blackbody fit physically meaningful or
mimicking a more complex process, e.g., resonant electron cyclotron
scattering?  In fact, the persistent emission from some magnetars
prefer blackbody $+$ power-law modeling to the two-blackbody fit,
requiring investigation (\S3.5).
	



\section{Probes into Accreting Pulsars and their Environment}\label{sec:binaries}
\subsection{X-raying the Environment}



\subsubsection{Mapping the Stellar Wind}\label{sec:wind}

\begin{figure}[htb]
  \begin{center}
    \includegraphics[width=0.7\hsize]{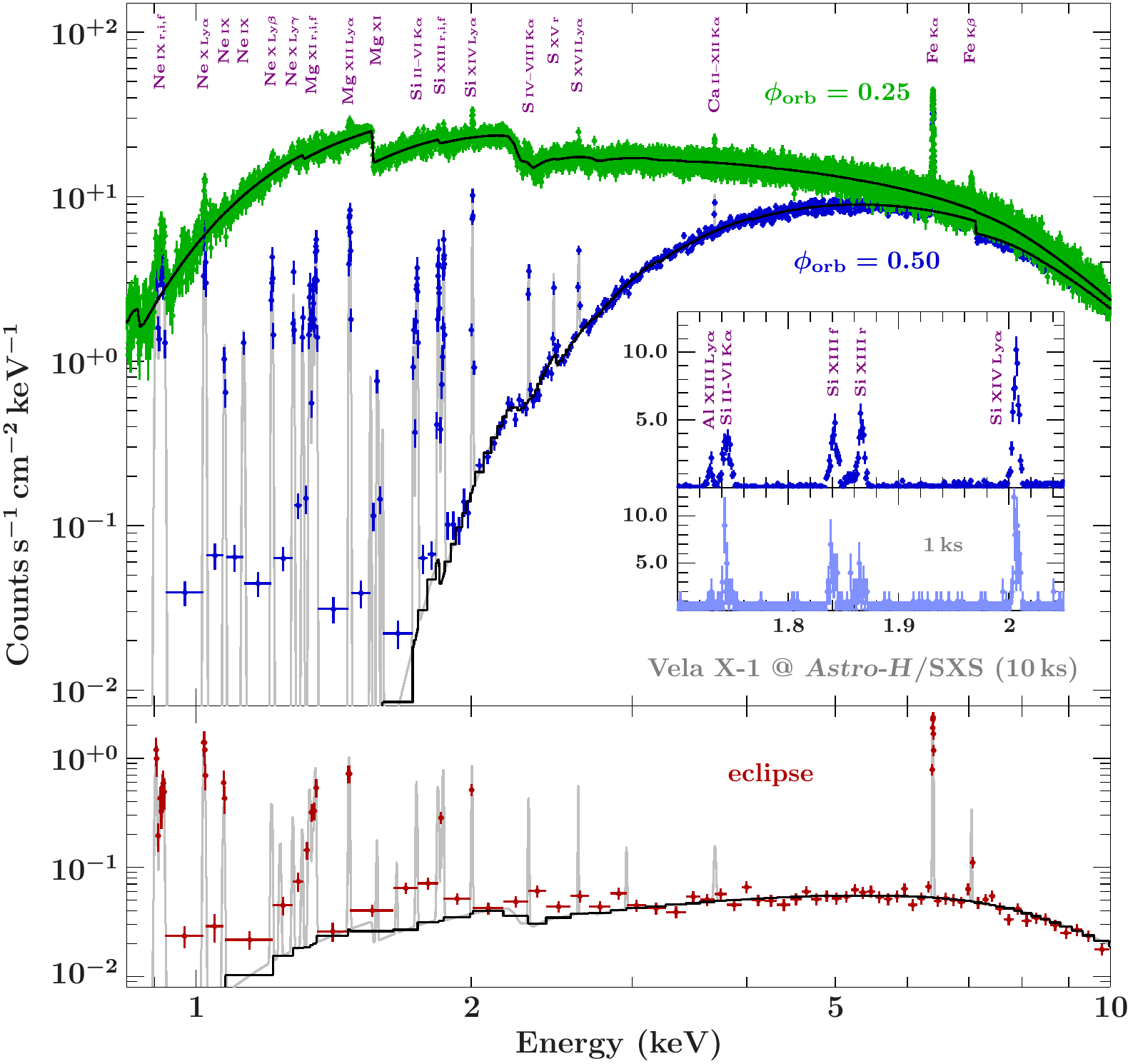}
  \end{center}
  \caption{SXS simulations of three 10\,ks observations of the
    accreting pulsar and HMXB Vela~X-1 at different orbital phases,
    based on longer \textsl{Chandra} observations
    \citep{watanabe2006}. The spectra have been rebinned for clarity,
    excesses at low energies indicate additional lines. Courtesy
    M. K\"uhnel (FAU) and N. Hell (FAU \& LLNL).}
  \label{fig:velax1_sxs}
\end{figure}

Many accreting pulsars are part of a High Mass X-ray Binary
(HMXB). HMXBs posses a strong stellar wind that can be photoionised by
the X-ray emission from the compact object. Hundreds of X-ray emission
and absorption lines have been observed with the \textsl{Chandra} and
\textsl{XMM-Newton} gratings for bright wind NS and black hole
accreters like \textbf{Vela~X-1}, \textbf{Cen~X-3}, \textbf{Cyg~X-3},
\textbf{Cyg~X-1}, \textbf{4U\,1700$-$37}, or \textbf{GX\,301$-$2}
\citep{sako2002,schulz2002,hanke2009a,fuerst2011}.

From modelling those lines as well from studying the sometimes extreme
flux variability -- as, e.g., shown by the supergiant fast X-ray
transients (\S\ref{sec:sfxt}) or sources like Vela X-1
\citep{kreykenbohm2008,fuerst2010} -- it is known that these winds are
often structured. They can show focused material streams towards the
compact object \citep{hanke2009b}, can have material trailing the
compact object in an accretion wake \citep{nagase1992,manousakis2012},
and can exhibit instabilities leading to clumping
\citep{feldmeier2008,hanke2009b}. In addition several HMXBs are
eclipsing sources (e.g., Vela~X-1, Cen~X-3, 4U\,1700$-$37).  Observing
the ingress or egress can in principle allow for a detailed look at
the atmosphere of the companion star, e.g., its density and ionisation
structure resulting from the interaction with the compact object
\citep{wojdowski2003,naik2011}.

Modelling lines and absorption edges from photoionised plasmas,
ideally fitting series from the same ion together, has already
provided important constraints, e.g., on predominant ionisation
stages, ion abundances, and equivalent widths for different sources
and orbital phases. However, this is just the beginning of what will
be possible using the \textsl{ASTRO-H} SXS instrument. Pre-SXS
limitations include that wind properties have not yet been spatially
resolved. This requires high time resolution line mapping, e.g.,
through eclipse egress/ingress or of the photoionised accretion
wake. The comparatively small effective area of grating spectrometers
cannot provide sufficient statistics for such a study, contrary to the
SXS, (see below). Such intriguing studies as resolving individual
clumps will also become feasible for the first time.

\begin{figure}[htb]
  \begin{center}
    \includegraphics[width=0.6\hsize]{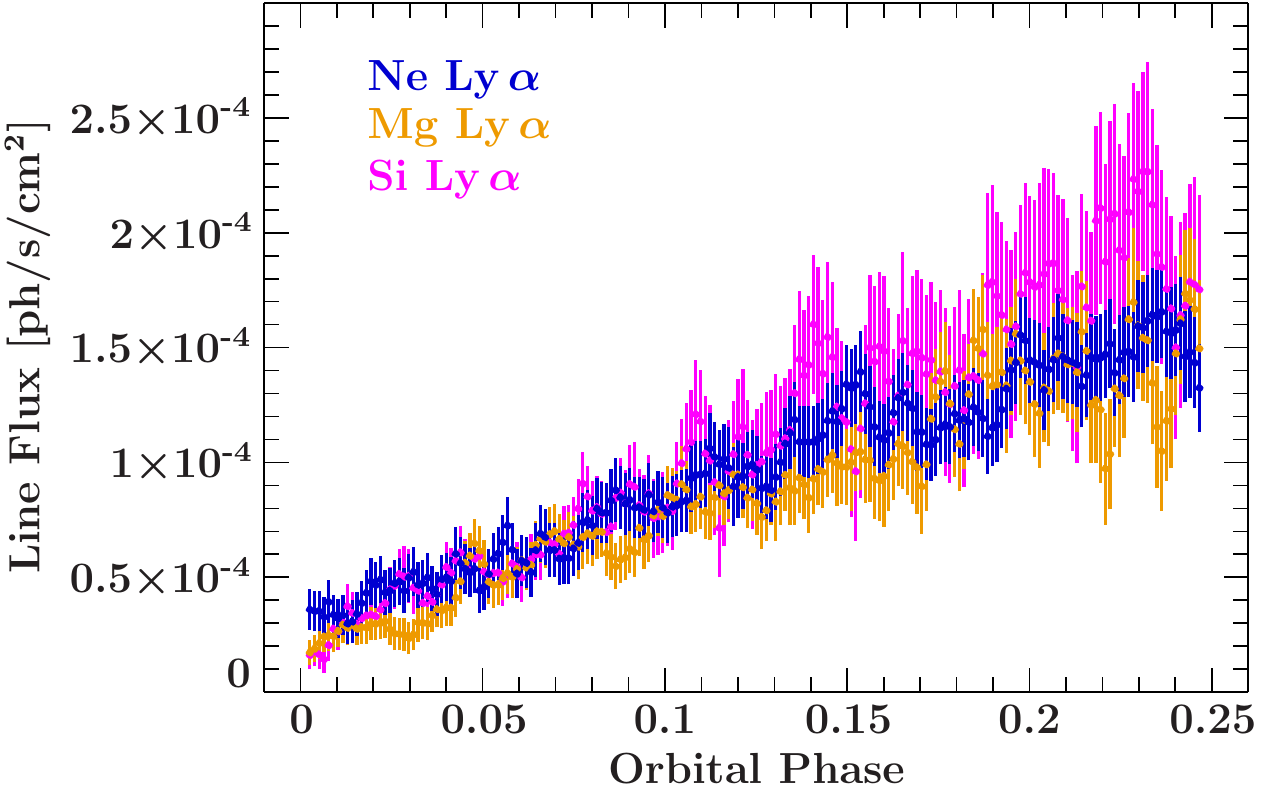}
  \end{center}
  \caption{Simulated evolution of the line flux with orbital phase
    during egress for three prominent lines of Vela~X-1. A 5\,ks
    gliding window was applied to 1\,ks simulations.  The simulated
    line flux was linearly interpolated between \textsl{Chandra}
    observations at phase 0 and 0.25. The simulated absorbed continuum
    has been subtracted ($N_\textrm{H}$ was assumed to be modulated
    with phase according to spherical wind model). The wavy structures
    are a numerical artifact. Courtesy N. Hell (FAU \& LLNL).}
  \label{fig:velax1_egress}
\end{figure}

\noindent\textbf{Line spectra of exemplary accreting pulsars} ---
\textbf{Vela~X-1:} For a description of system properties see
\S\ref{sec:crsf}. Figure~\ref{fig:velax1_sxs} shows the SXS view of
lines from many highly ionised ions as well as their strong
variability over the $\sim$9\,d orbit. The X-ray continuum spectrum is
highly absorbed, especially near phase 0.5 (accretion wake) but the
lines still have high equivalent widths, i.e., they are not fully
absorbed by this material. SXS observations during the line-rich
(near-)eclipse as well as around phase 0.5 can be expected to maximize
the scientific return.  Figure~\ref{fig:velax1_egress} (egress) and
the inset in Figure~\ref{fig:velax1_sxs} (wake) show that SXS
observations will allow us to determine the evolution of line
parameters with unprecedented time resolution of 1-5\,ks over a broad
energy range. --- \textbf{Cen X-3}: This system consists of a NS
spinning at a period of $\sim$4.8\,s in an $\sim$2.1\,d orbit with a
O-type supergiant star\citep[see, e.g., references
in][]{suchy2008}. All the phenomena described above -- from eclipse to
eclipse -- could thus potentially be covered by one SXS
observation. Near-neutral, He-like, and H-like iron lines have been
observed by \textsl{Chandra} and \textsl{Suzaku}
\citep{iaria2005,naik2011}. Outside of eclipse the X-ray spectrum is
dominated by sometimes highly variable continuum emission, exhibiting
very weak absorption lines mostly from H-like ions
\citep{sako2002}. During eclipse, however, the spectrum is dominated
by line emission \citep{wojdowski2003}. --- \textbf{GX\,301$-$2}: For
a description of system properties see \S\ref{sec:crsf}. That section
also presents an SXS simulation of the continuum and line spectrum
expected during the bright pre-periastron flare that is a regular
feature of the 41.5\,d orbital flux variation, see
Figure~\ref{fig:gx301}. GX~301$-$2 is characterised by several
(near-neutral) fluorescent emission lines, a Compton shoulder of the
iron line, and a highly absorbed ionising continuum
\citep{fuerst2011,watanabe2003}. --- \textbf{CRSF:} Note that all
three of these accreting pulsars are also known cyclotron line (CRSF)
sources. Observing their broad band spectra with \textsl{ASTRO-H}
would thus also allow us to address the scientific objectives
described in \S\ref{sec:crsf}.

\begin{figure}[htb]
\begin{center}
\includegraphics[width=0.6\hsize]{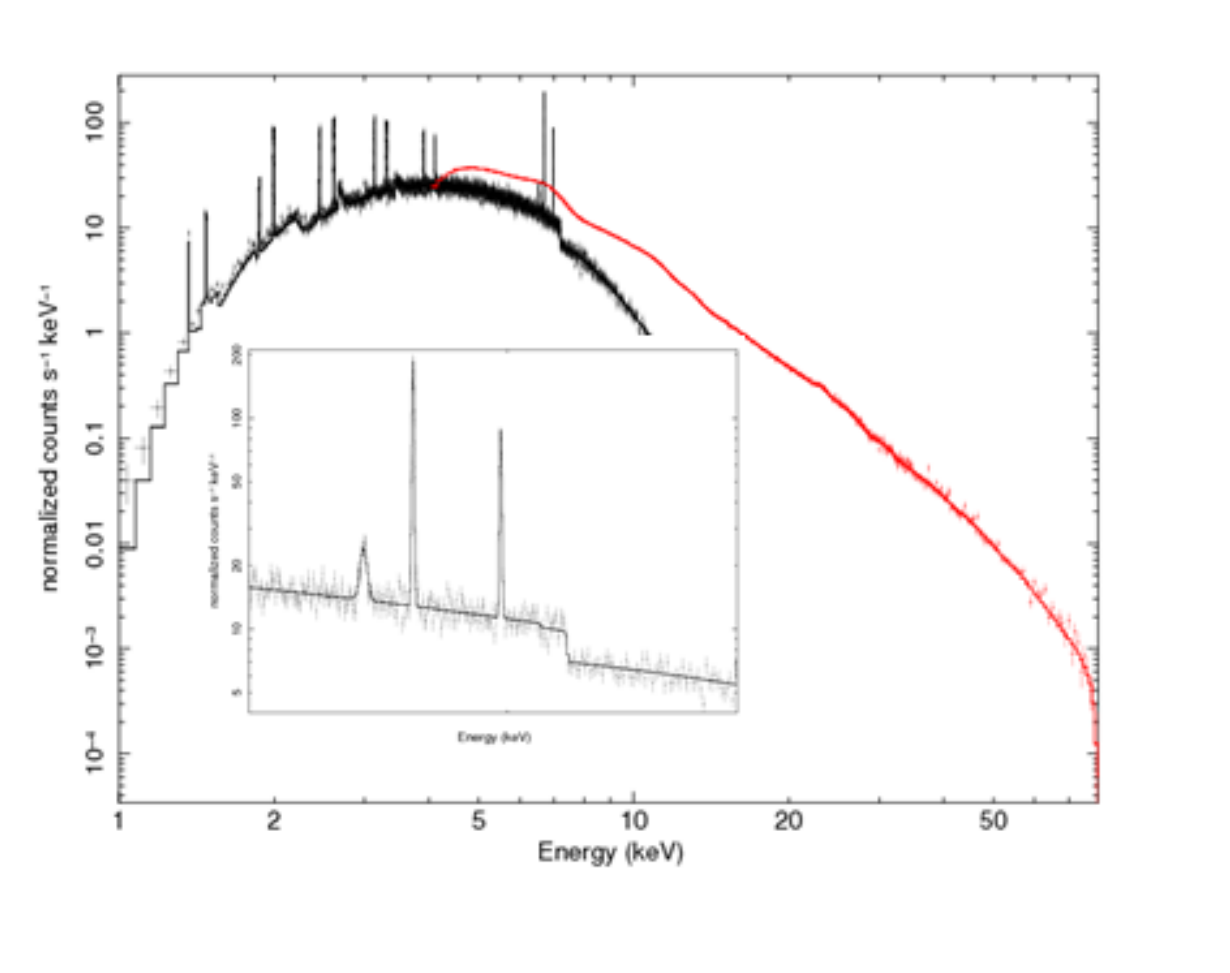}
\caption{SXS and HXI simulation of a 2\,ks observation of Cyg~X-3.}
\label{fig:cygx3_sxs}
\end{center}
\end{figure}

\noindent\textbf{A unique wind accreter} --- \textbf{Cyg~X-3:} Located at a
distance of 8--10\,kpc in the plane of the Galaxy, Cyg~X-3 is composed
of a compact object in a 4.8\,hours orbit with a Wolf-Rayet star. The
emission from the system is detected from radio up to GeV energy
band. The soft X-rays are thought to arise from the accretion disc,
the hard X-rays from the corona, and the radio from the jet
\citep{szostek2008}. This makes interpretation of the data complex and
one needs to use timing information along with the spectral one to be
able to reconstruct the data in a unique way
\citep{koljonen2013}. Cyg~X-3 harbours a large number of emission
lines that are especially prominent in the high resolution energy
spectra \citep{paerels2000}. However, due to the low sensitivities of
the \textsl{RXTE} PCA below 3.5\,keV, \cite{koljonen2013} were able to
use only the iron line complex in their analysis. Despite this fact
they still have seen dips in the variance spectrum in the energy bins
1.8--1.9 and 2.3-2.4\,keV. The dips correspond roughly to the location
of some of the strongest emission lines in the X-ray spectra (H-like
Si at $\sim$2.0\,keV and H-like S at $\sim$2.5\,keV) and could be
interpreted as a reduction in variability indicating line production
further out from the compact object in the photoionised stellar
wind. However, the dips could also be due to the low flux and wide
energy bins. \cite{paerels2000} showed that the iron line complex
consists of He-like and H--like iron ions (XXV/XXVI) at 6.7\,keV and
6.9\,keV, respectively, and cold iron K$_\alpha$ at 6.4\,keV. However,
these lines blend into a single broad feature in the PCA, which makes
it difficult to disentangle the two possible emission regions.

Thus a similar exercise done with \textsl{ASTRO-H} data will provide a
unique chance to study the spectral-timing variability of the source
along the orbit using the whole wealth of information coming from
spectral lines and using the simultaneous time series for a very broad
energy range, which will make it possible to disentangle these
different components and see all of them in play simultaneously. The
sensitivity of the SXS instrument allows us to get detailed spectra
already with a 2\,ks exposure (see Figure \ref{fig:cygx3_sxs}), and
thus with a 20\,ks observation \textsl{ASTRO-H} would be able to trace
the spectral variability of the source along the whole orbit.

\subsubsection{Searching for Signatures of the Alfven Shell}\label{sec:gx1p4}

Fluorescent lines of low ionized iron ions have been observed from
some of accretion powered X-ray pulsars. Although its emission region
is not well understood, its candidate is an Alfven shell or an inner
part of their accretion disk. In these regions, magnetic field and the
matter are fiercely competing with each other and information of their
dynamical motion of the gas provide us new insights of the environment
of accreting neutron stars. The information of the dynamics around the
Alfven shell is relating to the strength of the magnetic moment and
mass of the neutron stars.

GX\,1+4 is not a High Mass X-ray Binary, but is considered to contain
a highly magnetized neutron star with $\sim$10$^{13}$ G
\citep{1988Natur.333..746M}. Its prominent $K \alpha$ line has an
equivalent width of $\sim$200~eV. Its central energy and the 7.1~keV
absorption edge indicate the iron ions are almost neutral and suggest
its fluorescent origin.


\begin{figure}[htb]
\begin{center}
\includegraphics[width=0.55\hsize]{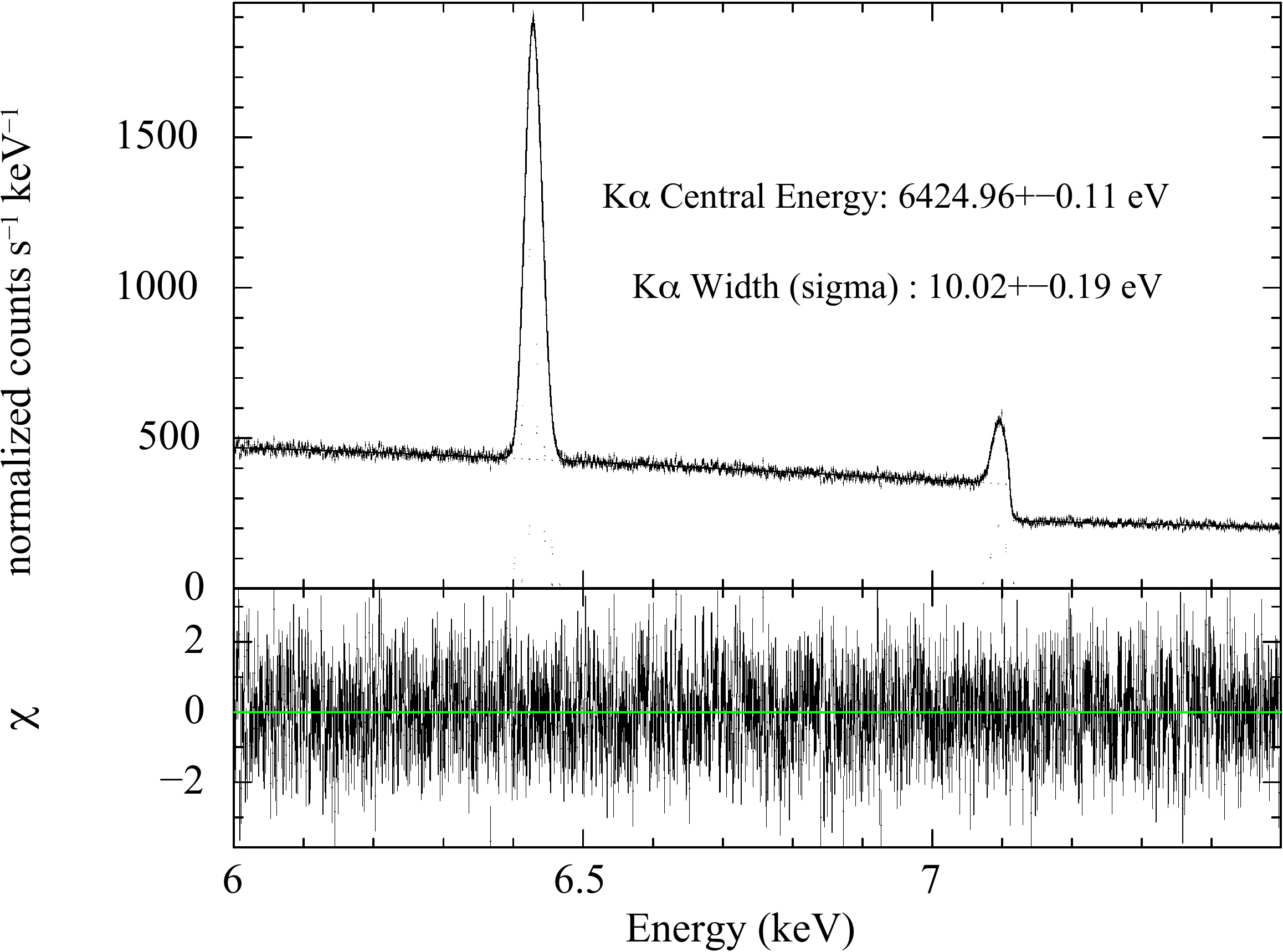}
\caption{Simulation of the SXS spectrum of GX\,1+4. The exposure time is assumed
  to be 4~ks and the model is an absorbed power law plus three
  Gaussian lines ($K _{\alpha 1}$, $K_{\alpha 2}$ and $K_{\beta}$).
  The line widths are all assumed to be 10~eV (standard deviation of a
  Gaussian function). }
\label{fig:GX14_sxs}
\end{center}
\end{figure}

Although \textsl{Suzaku} XIS data did not show the energy modulation of the
line central energy ($<$10~eV), it shows a line broadening with
roughly 40$\sim$50~eV in standard deviation. These results, the
broadening and the small energy modulation, indicate that the line
emission region is in a widely extended region of the accretion flow
or of the Alfven shell. However, current CCDs, including XIS, have
ambiguity of the energy response function for bright point sources due
to the SCF-effect \citep{2012PASJ...64..101T}. Therefore, SXS, for the first
time, makes it possible to study detailed spectroscopy of spin-phase
sliced spectra.

The expected Kepler velocity around the Alfven shell is $\sim$1000 km
s$^{-1}$. Thus the energy shift by this velocity is $\sim$200~eV.  If
the emission lines uniformly come from the Alfven shell, the line
width is expected to be $\sim \frac{200}{\sqrt[]{12}} \sim 60$~eV in
standard deviation. By the observation with the SXS, we can expect the
detection of a significant broadening and probably a modulation of the
width as well as the central energy.

Assuming a 40~ksec observation and analysis of 10 phase sliced
spectra, we simulated a 4~ksec observation using best fit parameters
obtained by \textsl{Suzaku}. An expected spectrum is shown in
Figure~\ref{fig:GX14_sxs}, We assumed a power law continuum with a
photon index of 2.9 suffered photo-electric absorption of
4$\times$10$^{23}$ H atoms cm$^{-2}$, and neutral iron fluorescent
lines: $K_{\alpha 1}$, $K_{\alpha 2}$, and $K_{\beta}$ (this line
contains several lines but here we represented them by one Gaussian).
The intensity-ratio of the three lines is fixed to be 100:50:17 and
the line width is fixed to be 10~eV (standard deviation).  One sigma
error of the line central energy is estimated to be $\sim$ 0.1~eV,
which corresponds to 4.7 km sec$^{-1}$.  The line width can be also
determined with an error of $\sim$ 0.2~eV, which is again much less
than the expected width. Therefore we can precisely determine an
average velocity and velocity dispersion of gas which emits
fluorescent lines with only a 1 day observation.

If we can estimate the Kepler velocity at the Alfven radius from the
observed modulation of the central energy, by assuming some accretion
flow geometry, we can discuss the relation among mass, magnetic moment
of the neutron star and accretion rate.  Since the accretion rate can
be estimated from the bolometric luminosity, the relation between the
mass and the magnetic moment of the neutron star can be deduced.

\subsection{Cyclotron Line Sources}\label{sec:crsf}

\subsubsection{Background and Previous Studies}

About 25 X-ray binaries show broad absorption-line-like features in
their hard X-ray spectrum due to the interaction of the emerging
photons with the electrons trapped along the $10^{12}$\,G magnetic
field. The energy of these features is the only direct measurement of
a magnetic field in the X-ray emitting region ($E \simeq 11.6 \times n
\times B_{12}$\,keV, where $n$ is the harmonic number and $B_{12}$ is
the magnetic field in units of $10^{12}$\,G). The underlying continuum
emission is produced by Comptonization of the radiation produced in
the optically thick part of the accretion flow or at its footprint on
the NS surface and has a smooth power-law shape exponentially
attenuated at high energy. Although its phenomenological description
is fairly simple, the detailed physics are still a matter of debate
\citep[][ and references therein]{becker2012}, as well as the strong
variability at all time scales, which is thought to be due to the
clumpy nature of the plasma, when it flows within the
magnetosphere. Reproducing on theoretical grounds the observables is
still a challenge which would yield informations on basic properties
of the neutron star such as the magnetic field configuration and its
interaction with the accreted matter.

Timing signatures are of paramount importance to understanding these
objects. In particular the ``pulse profile'', which is the X-ray
emission of the XRBPs folded at the spin period of the NS. Pulse
profiles are remarkably stable when averaged over several rotational
cycles, reflecting the trapping of plasma along the magnetic field
lines. Significant variations in the shape of profiles are observed
when the XRBPs undergo transitions between different X-ray luminosity
states, such as reported in the cases of EXO\,2030$+$275 and
V\,0332$+$65, \citep[][ and references
therein]{dima2008,tsygankov2010}. At high luminosity, a radiative
shock forms in a relatively extended (a few km) accretion column, and
the radiation is emitted mainly from its lateral walls in the form of
a ``fan beam''.  At lower luminosities, the radiative shock is
suppressed, matter reaches the base of the column with the free-fall
velocity and X-rays are emitted nearly vertically in a ``pencil
beam''. In this state, the height of the column can be significantly
reduced, and for very weak sources, it reduces to a hot-spot on the NS
surface. Correspondingly, variations of the centroid energy of the
cyclotron scattering features with the luminosity have been observed
and tentatively been interpreted as variations of the accretion column
height \citep{staubert2009,klochkov2011,klochkov2012}. This picture
still needs to be confirmed by further observations.


The majority of XRBPs show significant changes in the spectral energy
distribution of the X-ray emission at different pulse phases. This is
due to a variety of factors that change during the rotation of the
star. In particular, the cross section of the scattering between
photons and electrons trapped in a magnetic field strongly depends on
the angle between the photon trajectory and the magnetic field lines.

The interplay between the extraordinary and ordinary polarisation
modes of photons propagating in a strongly magnetised plasma is
expected to produce phase variable linear polarisation at a level as
high as 80\% in correspondence to the cyclotron scattering features
and for particular pulse phases \citep[the polarisation fraction is
lower for energies out of resonance,][]{meszaros1988p}. Different
imprints are expected to appear in the spin-phase resolved emission
for different emission patterns: an anti-correlation between
polarisation fraction and X-ray intensity in the case of pencil beam,
and anti-correlation in case of a fan beam. The linear polarisation
fraction can be lowered of $\sim$30\% by relativistic light bending,
which causes emission from both magnetic poles to be simultaneously
visible. The theoretical predictions are highly uncertain as they
depend on the unknown system geometry and emission beaming.

Advances in our understanding of these objects have come from
\textsl{BeppoSAX} \citep{sax} and \textsl{RXTE} \citep{rxte}
observations, which have however a limited spectral resolution, while
the superb performance of \textsl{Suzaku} \citep{suzaku} is hampered
by the spectral gap between the soft and the hard
X-rays. \textsl{NuSTAR} \citep{nustar} has provided a quantum leap in
the 5--60\,keV energy domain spectral analysis, which is comparable to
the possible achievements of the HXI.

\subsubsection{Prospects \& Strategy}

\textsl{ASTRO-H} provides a wide spectral coverage, coupled with
excellent timing capabilities (the latter for SXS, HXI, and SGD) and
unprecedented spectral resolution in the broad energy range relevant
for the physics of HMXB (0.1--100\,keV). The significant overlap in
the bands of the instruments provides a very robust inter calibration
tool, which helps in reducing the systematic uncertainties. It is,
therefore, an ideal laboratory for studies of objects emitting over a
wide energy range combining the excellence of previous missions. The
limited effective area with respect to past (e.g., \textsl{RXTE}) and
current (e.g, \textsl{XMM-Newton}) instruments is not an issue for
bright X-ray binaries.

\begin{figure}[htb]
  \begin{center}
   \includegraphics[width=0.8\hsize]{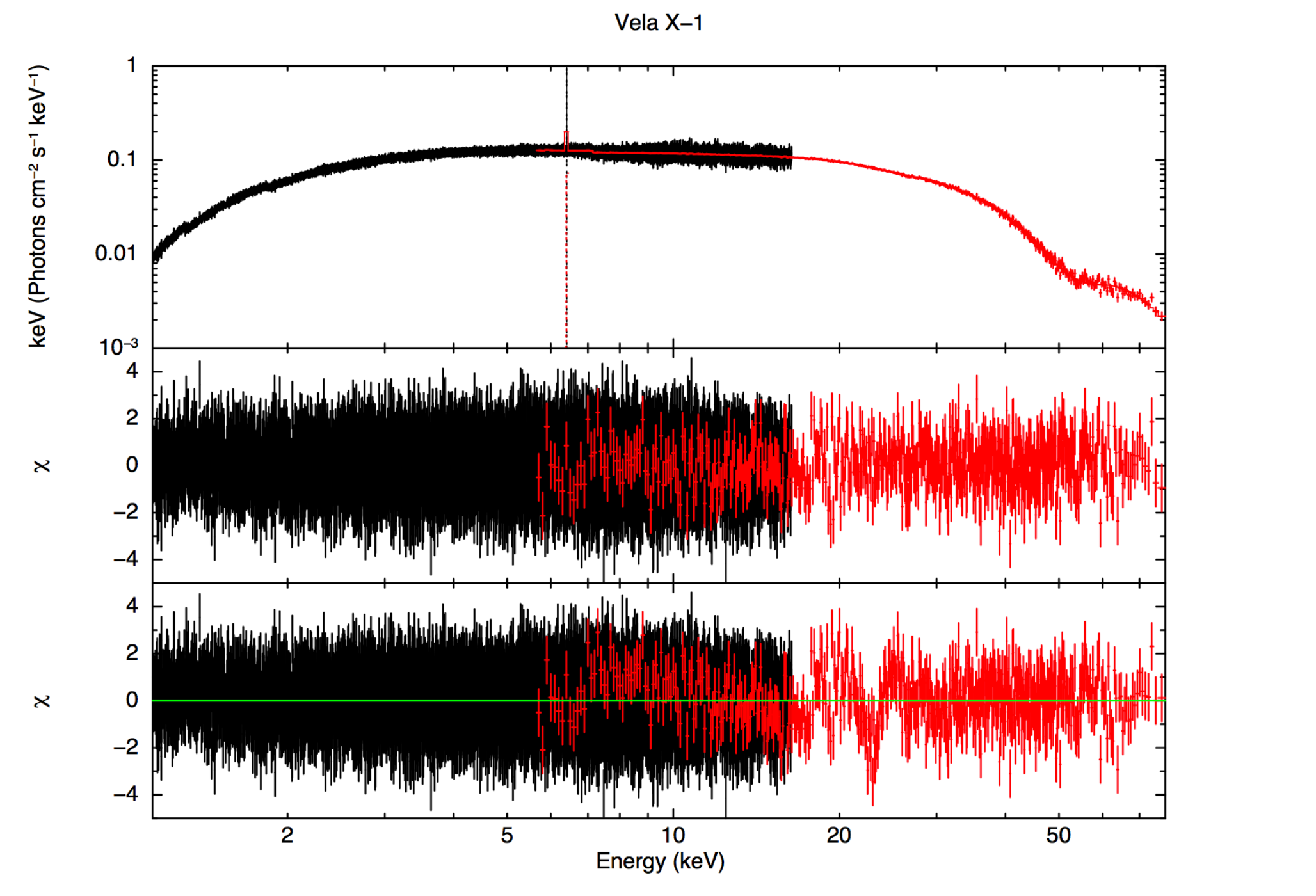}
  \end{center}
  \caption{Simulated SXS and HXI spectrum for a 100\,ks exposure of
    Vela~X-1 using the continuum model and two CRSFs at $\sim$25\,keV
    and $\sim$55\,keV from \cite{indiani2013}. Middle panel: Residuals
    using Lorentzian cyclotron line profiles, as for the
    simulation. Bottom panel: Residuals from a model with Gaussian
    optical depths lines, structure is present around the CRSF
    fundamental at $\sim$25\,keV.}
    \label{fig:velax1}
\end{figure}

Polarisation has not been measurable in X-rays below a few hundreds
keV so far. The Soft Gamma-ray Detector (SGD) on board
\textsl{ASTRO-H} can perform this measurement using the Compton
kinematics above $\sim$50\,keV. This will open an
unprecedented observational window, which will provide stringent
constraints on the theoretical models when combined with the
pulse-phase and luminosity dependent spectral variations of the
cyclotron line energy and underlying continuum. This will potentialy
trim down the enormous parameter space which is currently left almost
unconstrained in the theoretical interpretation.  In particular, the
geometrical configuration of the magnetic field and thus of the
emitting regions on the NS is largely unknown as well as the shape of
the accretion stream close to the NS surface: matter could fall in a
filled or hollow column, or in portions of it
\citep{basko1976}. Theoretical models are already challenged by
observation, which in turn are still not sufficient to find a solution
due to limited spectral resolution or band pass limitations.

The broad band data with high spectral and timing resolution provided
by \textsl{ASTRO-H} will provide an unprecedented robust benchmark for
theory. Many of the sources that display cyclotron lines or are strong
candidates are transient, e.g., \textbf{4U\,0115$+$63},
\textbf{V\,0332$+$53}, \textbf{GRO\,J1008$-$57}, \textbf{GX\,304$-$1},
\textbf{EXO\,2030$+$375}, or \textbf{A\,0535$+$26}. In the case
that one of them shows a giant outburst it will be an excellent target
for \textsl{ASTRO-H}, providing the best chance of observing
polarisation, see \S\ref{sec:cyc_3}. Bright persistent sources that
will allow for detailed time-, luminosity-, and pulse-phase-resolved
studies of the cyclotron lines and the X-ray continuum, are, e.g.,
\textbf{Vela~X-1}, \textbf{GX\,301$-$2}, \textbf{Cen~X-3},
\textbf{Her~X-1}, or \textbf{4U 1626$-$67}. In the following we
present simulations of \textsl{ASTRO-H} observations for the
persistent cyclotron line sources Vela~X-1 and GX~301$-$2. Vela X-1
is characterised by significant orbital-phase-dependent variability
(\S\ref{sec:wind}) as well the irregular occurrence of strong flares
and off-states \citep{fuerst2010}. GX~301$-$2 presents a marked
pre-periastron flare and heavy intrinsic absorption in the stellar
wind, thought to be due to an accretion stream preceding the NS
\citep[][and references therein]{fuerst2011}. Both sources also show
pronounced emission lines due to neutral and ionised material
(Figure~\ref{fig:velax1_sxs} and Figure~\ref{fig:gx301}).

\subsubsection{Targets \& Feasibility}

\begin{figure}[htb]
  \begin{center}
    \includegraphics[width=80mm]{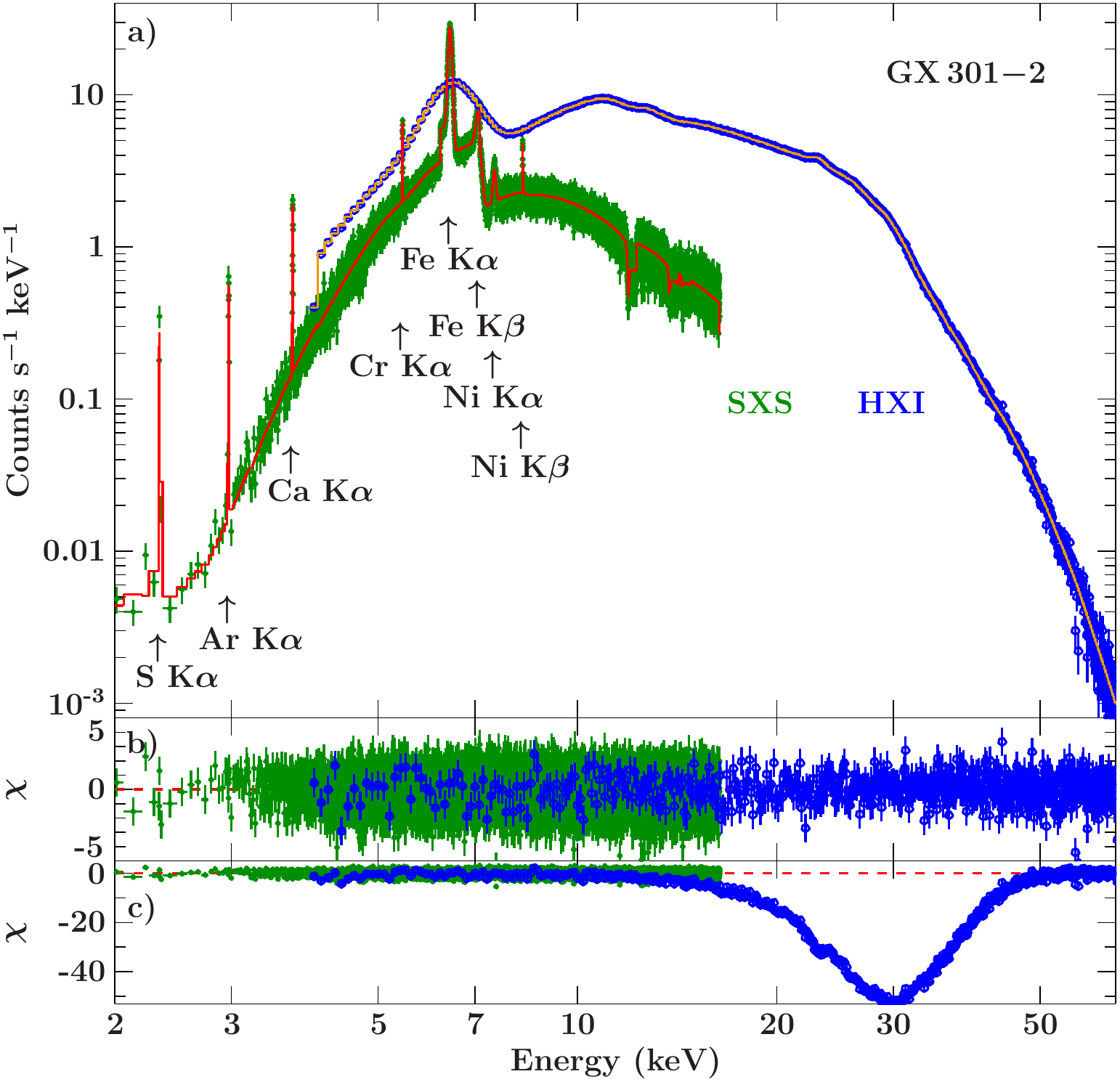}
    \caption{a) Simulated SXS and HXI spectra for a 50\,ks exposure on
      GX 301$-$2 based on the analyses by \cite{fuerst2011} and
      \cite{suchy2012}. Emission lines are indicated.  b) Residuals
      from a best fit. c) The residuals from a model with the
      exclusion of the cyclotron absorption feature at
      $\sim$30\,keV. Courtesy M. K\"uhnel (FAU) and N. Hell (FAU \&
      LLNL).}
    \label{fig:gx301}
  \end{center}
\end{figure}

\textbf{Vela X-1} (4U~0900$-$40) is a an eclipsing and persistently
active system consisting of a massive (23$M\odot$; 34$R_\odot$) B0.5
1b supergiant and a neutron star. It has an orbital period of 8.964
\,days and shows only slight eccentricity ($e\sim$0.1). The neutron
star is deeply embedded in the strong stellar wind of the companion;
the typical X-ray luminosity is $\sim4\times10^{36}$erg/s. It exhibits
pulsations with a period 283\,s.  Cyclotron lines appear at $\sim$25
and 55\,keV \citep{labarbera2003,kreykenbohm2002}.  \cite{indiani2013}
analysed a 100\,ks \textsl{Suzaku} observation and found an
interesting dip-feature linked to enhanced absorption form a partial
covering component at a particular pulse phase. This will be better
constrained by the SXS. A 100\,ks simulation with SXS and HXI shows
that \textsl{ASTRO-H} will provide an improved determination of the
energy and width of the cyclotron scattering features in comparison to
existing datasets (Figure~\ref{fig:velax1}; see \S\ref{sec:wind} for a
simulation including the emission lines from the stellar wind). With
such an observation it would be possible to discriminate between
different line profile models for the first time.

Relatively strong polarisation could be present in the SGD band due to
the harmonic cyclotron scattering feature at $\sim50$\,keV. In Vela
X-1 the first harmonic is unusually strong. We have verified that in
50\,ks, it would be possible to detect a polarisation fraction of
35\%.

\textbf{GX 301$-$2} consists of an accreting NS with a period of
$\sim$685\,s fed by the surrounding stellar wind of the B type
emission line companion Wray 977. \cite{doroshenko2010a} determined
an orbital period of 41.482$\pm$0.001\,days. In the SXS band,
GX~301$-$2 is characterised by several fluorescent emission lines,
rising above the highly absorbed ionising continuum
\citep{fuerst2011}. A Compton shoulder of the iron line is visible
both in the \textsl{XMM-Newton} CCD spectrum and in the
high-resolution grating spectrum of \textsl{Chandra}
\citep{watanabe2003} and could be studied in detail by the SXS. The
cyclotron absorption feature is located at about 30\,keV.  We have
simulated a 50\,ks observation based on existing analyses
\citep[Figure~\ref{fig:gx301}][]{fuerst2011,suchy2012} and verified
that the emission lines are detected by the SXS with high
significance, allowing the observer to fully characterise the wind
environment, while the HXI is able to constrain the shape of the
cyclotron absorption line with high accuracy.

\subsubsection{Measuring Polarisation during a Giant Outburst}\label{sec:cyc_3}

\begin{figure}[htb]
  \begin{center}
    \includegraphics[width=80mm]{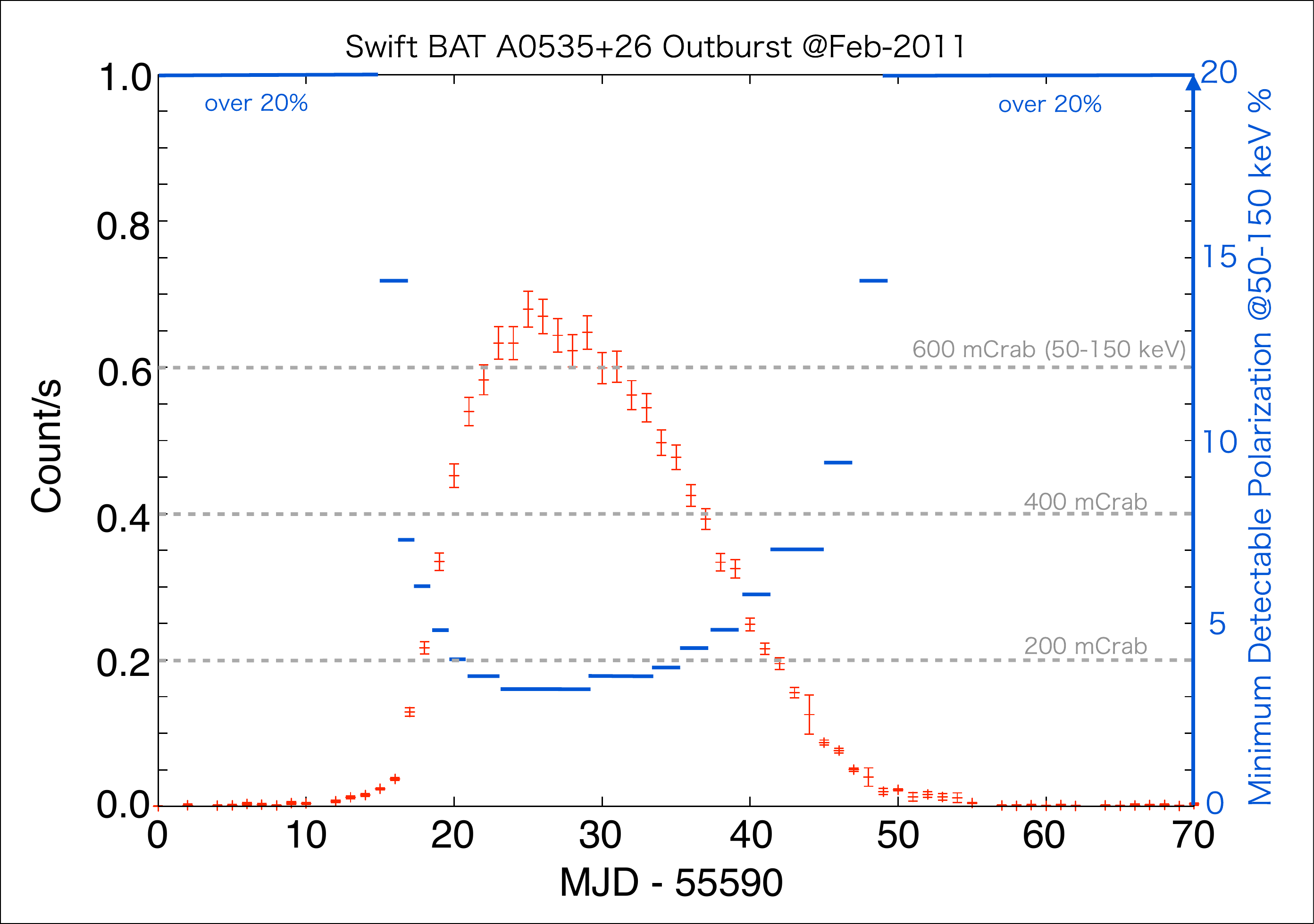}
    \caption{Measurable polarisation fraction of the X-ray signal in
      the 50--100\,keV energy band by the SGD in a 50\,ks exposure
      time (courtesy Sasano).}
    \label{fig:A0535}
  \end{center}
\end{figure}

Giant outbursts of Be/X-ray binaries are thought to be due to an
exceptional extension of the equatorial disc of the donor star
providing material to be accreted on the NS through a viscous
accretion disc \cite[see][for a review]{reig2011}. Outbursts are not
predictable although a correlation is observed with the strength of
the H-$\alpha$ emission of the system. In 4U\,0115$+$63, these
episodes happen every few years and appear to be linked to a
semi-period perturbation of the equatorial disc of the Be donor
\citep{negueruela2001b}. Longer periods of quiescence followed by
repeated outbursts have been observed for A\,0535$+$26, while
EXO\,2030$+$375 shows periodic outbursts at each periastron passage
during which the source does not exceed $\sim$250\,mCrab. The
spectacular source V\,0332$+$53 has been observed in outburst only
once with modern facilities. GRO\,J1008$-$57 has recently exhibited a
giant outburst \citep{mathias2012}.  All these objects are potential
targets for \textsl{ASTRO-H} in case they go in outburst. Triggers can
be provided by \textsl{Swift}/BAT, \textsl{INTEGRAL}, or \textsl{MAXI}
and the X-ray outbursts usually last for a few weeks.

\begin{wraptable}{}{80mm}
  \vspace{-5mm}
  \caption{ Estimated minimum detectable polarisation fraction in the
    50--100\,keV energy range for bright Be/X-ray binaries.}
  \begin{center}
    \vspace{-2mm}
    \begin{tabular}{cccc}
      \hline
      \hline
      \scriptsize Exposure (ks) & \scriptsize A\,0535$+$26 & \scriptsize GX\,304$-$1 & \scriptsize EXO\,2030$+$375 \\
      \hline
      25 & 7\% & 23\% & 8\% \\
      50 & 5\% & 16\% & 5\% \\
      100 & 3\% & 15\% & 4\% \\
      \hline
      \vspace{-13mm}
    \end{tabular}
  \end{center}
  \label{mdp_table}
\end{wraptable}%

Observations during these episodes would provide unprecedented data
sets to study X-ray binaries. We emphasise that the SGD can measure
X-ray polarisation above $\sim$50\,keV where cyclotron lines appear in
a handful of very bright transient sources, for which we have
simulated the feasibility of detection in different observing times
(see Table~\ref{mdp_table}). For the X-ray binary A\,0535$+$26
($E_\mathrm{cyc}\simeq45$\,keV), at the maximum of its powerful
outbursts (0.9\,Crab above 50\,keV) and in 25\,ks of observation, it
is possible to measure a polarisation fraction as low as 7\%. In
Figure~\ref{fig:A0535}, we show how this measurement would be possible
throughout the outburst, helping us to contain the details of the
emission mechanism at different luminosity levels.
In systems as EXO\,2030$+$375, in which the detection of a scattering
feature is controversial, \citep[][and references
therein]{wilson2008,klochkov2007}, it would be feasible to detect a
polarisation fraction of 20\% during the regular periastron passages,
while 8\% would be reachable if the source undergoes a giant outburst.
This would constitute a direct prove of the magnetic field intensity
even in absence and a clear spectral signature and would open a new
window to investigate why the scattering features are not an
ubiquitous phenomenon in magnetised pulsars.

The polarisation angle is strongly dependent on the $B$-field
orientation with respect to the line of sight and averaging over a
phase interval might lead to a lowering of the polarisation signal.
We argue, based on a rough comparison of theoretical computations
\citep{meszaros1988p}, that it could be sufficient to divide the pulse
profile in less than ten phase bins to measure a polarisation fraction
of a few 10\% in most bins and it would thus be sufficient to
invest a reasonable amount of observing time (100-200\,ks) on a bright
X-ray binary to achieve such a breakthrough result.

\subsection{Super-giant Fast X-ray Transients (SFXTs)}\label{sec:sfxt}

\subsubsection{Background and Previous Studies}

Together with the classical supergiant and Be/X-ray binaries, the
\textsl{INTEGRAL} satellite has provided evidence for a third class of
X-ray binary called the supergiant fast X-ray transients (SFXTs)
\citep{smith2004,sguera2005,negueruela2006}. These objects have an OB
supergiant donor star, but at odds with the persistent systems, they
show short hours-long periods of intense X-ray activity (flare,
$L_X\sim10^{36-37}$\,erg/s), often grouped in day-long outbursts, and
extended quiescent states ($\sim10^{32}$\,erg/s), with swings up to
$10^5$ in luminosity \citep[see,][for an extensive
summary]{romano2014}. An open debate exists on the nature of the
accreting object and the characteristics of the stellar wind from
which matter is funnelled: clumpy winds are thought to cause the very
variable accretion rate \citep{zand2005,walter2007,rampy2009}, but
some gating mechanism is probably in action to produce the very low
duty cycle of these objects compared to the classical systems
\citep{oskinova2007,grebenev2007,bozzo2008}. The spectral
characteristics of the objects and the detection of pulsation
\citep[from 20 to 1200\,s, references in][]{romano2014} in five out of
twelve confirmed systems strongly indicate that the compact object is
a neutron star.  It is therefore possible that the gating barrier is
provided by an ultra-strong magnetic field ($B \gg 10^{12}$\,G), which
causes matter to accumulate or be expelled at the magnetospheric
boundary. Evidence of flares linked to ingestion of clumps has been
reported in \textsl{XMM-Newton} and \textsl{Suzaku} observations
\citep{rampy2009,bodaghee2011,bozzo2011}, while the typical flare rise
time ($\sim 10$ min) is consistent with the free-fall time from the
Alf\'ven radius associated to magnetic field $B\sim10^{13}$\,G.

\subsubsection{\textsl{ASTRO-H} prospectives on SFXTs}

\begin{figure}[htb]
  \begin{center}
    \includegraphics[width=0.6\hsize]{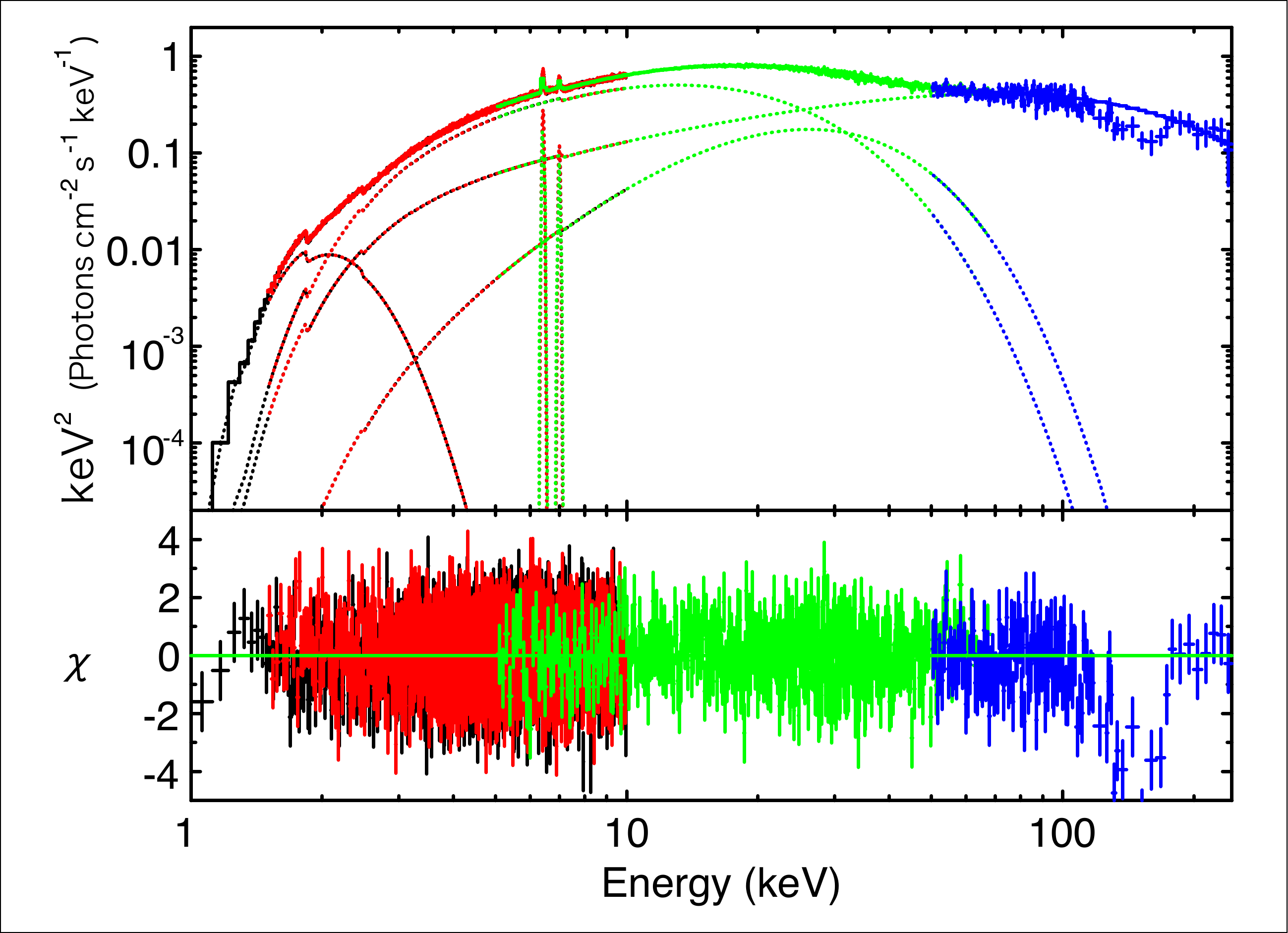}
  \end{center}
  \caption{Top: Simulation of 100 ksec \textsl{ASTRO-H} observation of long period pulsar, 4U 0114+65, with a CRSF at 150 keV. Bottom: Residuals of the simulation against a model with no CRSF. The residuals clearly deviate from the model at $\sim$ 150 keV and the CRSF will be sufficiently detectable with the SGD.}
  \label{fig:4u0114}
\end{figure}

\textsl{ASTRO-H} will provide unprecedented tools to study the very
nature of these objects. The HXI can provide good sensitivity to the
presence of a cyclotron scattering absorption feature in the hard
X-ray domain, with a possible extension using the SGD at higher energy
for very bright events. The SXS can provide a very useful diagnostic
on the presence of the emission Iron line, its ionisation status and
the proper motion of the clump impacting on the NS. The absorption can
be constrained by the SXI and SXS, similarly to the \textsl{Suzaku}
achievements \citep{bodaghee2011,rampy2009}. Such objects, if
confirmed, may be considered as aged magnetars in binary systems, and
will provide clues to the formation and evolution of magnetars
\citep{Enoto2010ApJ...722L.162E}, as well as to the origin of the NS magnetism
\citep{makishima1999}.

Other candidates to search for cyclotron scattering features are the
long period pulsars (e.g., 4U\,0114$+$65 and 4U\,2206$+$54).  Their
spectrum extends to 100 keV without appreciable cut-off, unlike those
of ordinary X-ray pulsars which show a steep cutoff around 20--40\,keV
and a CRSF at a higher energy.  We thus expect these LPPs to exhibit
CRSFs in energies above $\sim 100$\,keV \citep{makishima1999},
where the SGD will for the first time realize a sufficient
sensitivity.

\def\bi {\begin{itemize}}
\def\ei {\end{itemize}}
\def\ie {\textit{i.e. }}
\def\intgr{\textsl {INTEGRAL}}
\def\intgrsp{\textsl {INTEGRAL }}
\def\rxte{\textit {RXTE}}
\def\asca{\textit {ASCA}}
\def\ascasp{\textit {ASCA }}
\def\xmm{\textit {XMM-Newton}}
\def\xmmsp{\textit {XMM-Newton }}
\def\psrb{PSR~B1259--63} 
\def\psrbsp{PSR~B1259--63 } 
\def\ROSAT{\textit {ROSAT}}
\def\sax{\textit{Beppo}SAX}
\def\swift{\textit{Swift}}
\def\rxp {2RXP~J130159.6-635806}
\def\rxpsp {2RXP~J130159.6-635806 }
\def\deg {$^\circ$}
\def\gr{$\gamma$-ray}
\def\lsi {LSI~+61\deg~303}
\def\ls {LS~5039}
\def\hess {HESS~J0632+057}

\subsection{Gamma-ray Loud Binaries}\label{sec:gamma}
 
\subsubsection{Introduction} 

Gamma-ray-loud binary systems (GRLB) are X-ray binaries  which emit very-high
energy (VHE) \gr s. Four such systems \psrb, \ls, \lsi\ and HESS J0632+057, have 
been firmly detected as persistent or regularly variable TeV \gr\
emitters  \citep{aharon05,aharon06-ls,albert06-lsi, aharon06-innerGalaxy}. 

Observations of the {\it Fermi}/LAT telescope helped to reveal several more binaries emitting at very high energies. Among these sources are well-known microquasar Cyg X-3, symbiotic binary V 407 Cygni, colliding wind binary  $\eta$ Carina, newly discovered binary system 1FGL J1018.6-5856, but still the number of known GRLBs is very limited. 

The source of the high-energy activity of GRLBs is uncertain. It can be either accretion onto  or dissipation of rotation energy of the compact object. It is commonly assumed that the \gr\ emission is produced in result of interaction of the relativistic outflow from the compact object with the non-relativistic wind and radiation field of a companion massive star. Neither the nature of the compact object nor the geometry and physical properties of relativistic wind from this compact object  are known in most of the GRLBs. The only exception is \psrb\ system in which the compact object is known to be a young rotation powered pulsar which produces relativistic pulsar wind. 

In \cite{bednarek09} it was proposed that accreting magnetars in massive binaries can also generate TeV gamma-rays. In the inner magnetosphere of a magnetar the magnetic pressure can balance the gravitational pressure of the accreting matter, creating a very turbulent, magnetised transition region. This region provides good conditions for acceleration of electrons to relativistic energies. These accelerated relativistic electrons lose energy on the synchrotron process and the Inverse Compton scattering of the radiation from the nearby massive stellar companion, producing high energy radiation from X-rays up to TeV \gr s.  

Recently Burst Alert Telescope (BAT) on board \swift\ has detected a short burst from the direction of the TeV binary \lsi. The burst is visible in the 15 - 50 keV energy range, while no significant excess is observed above 50 keV. The total duration of the event is about 0.3 s. 
Previously such short flares have been also observed from the dirction of \lsi\ by \rxte\ \citep{smith09, li11}. In the paper of \cite{torres12} it was noticed that the properties of the burst observed by \swift/BAT (a very short duration and a thermal spectrum) are typical of magnetars. In their work the authors propose that due to the highly eccentric orbit the system is subject to flip-flop behaviour, from a rotationally powered regime in apastron to a propeller regime in the periastron. In this case in apastron an interwind shock leads to the normally observed \lsi\ behaviour, while during the periastron propeller is expected to efficiently accelerate particles only to sub-TeV energies, in agreement with the observations.  

More observations are needed to find out the true nature of these very interesting systems.
Below we discuss how \textsl{ASTRO-H} observations could help to solve some unresolved problems. We start with the case of \psrb, the only system in which we are sure about the nature of the compact object, and which we can use as a sample to test the other systems. Study of this system could provide a clue on the energy of the relativistic particles of the pulsar wind and give us a chance to study interactions of the winds in the highly variable environment.
After that we dicuss in more details \lsi\ and show the importance of measuring of its spectral variability in a broad energy range ( 1 - 100 keV) along the orbit.

\subsubsection{PSR B1259-63} 

In \psrb \ a 48 ms radio pulsar is in a highly eccentric 3.4 year orbit with a Be star LS 2883.  This system is known to be 
highly variable on an orbital time scale in radio (\citealt{johnston05} and references therein),
X-ray (\citealt{chernyakova09} and  references therein),  and TeV \citep{aharon05} energy ranges. 
The orbital multi-wavelength variability pattern is determined by the details of the interaction of relativistic pulsar wind with a strongly anisotropic wind of the companion Be star, composed of a fast, rarefied polar wind and a slow, dense equatorial  decretion disk. The disk of the Be star in the \psrb\ system
is believed to be  tilted with respect to the orbital plane. The line of intersection of the disk plane and
the orbital plane is oriented at  $\sim 90$\deg\ with respect to the major axis of the binary orbit  \citep{wang04} and the pulsar passes through the disk twice per orbit. 

\begin{figure}[h]
\includegraphics[angle=0,width=0.42\linewidth]{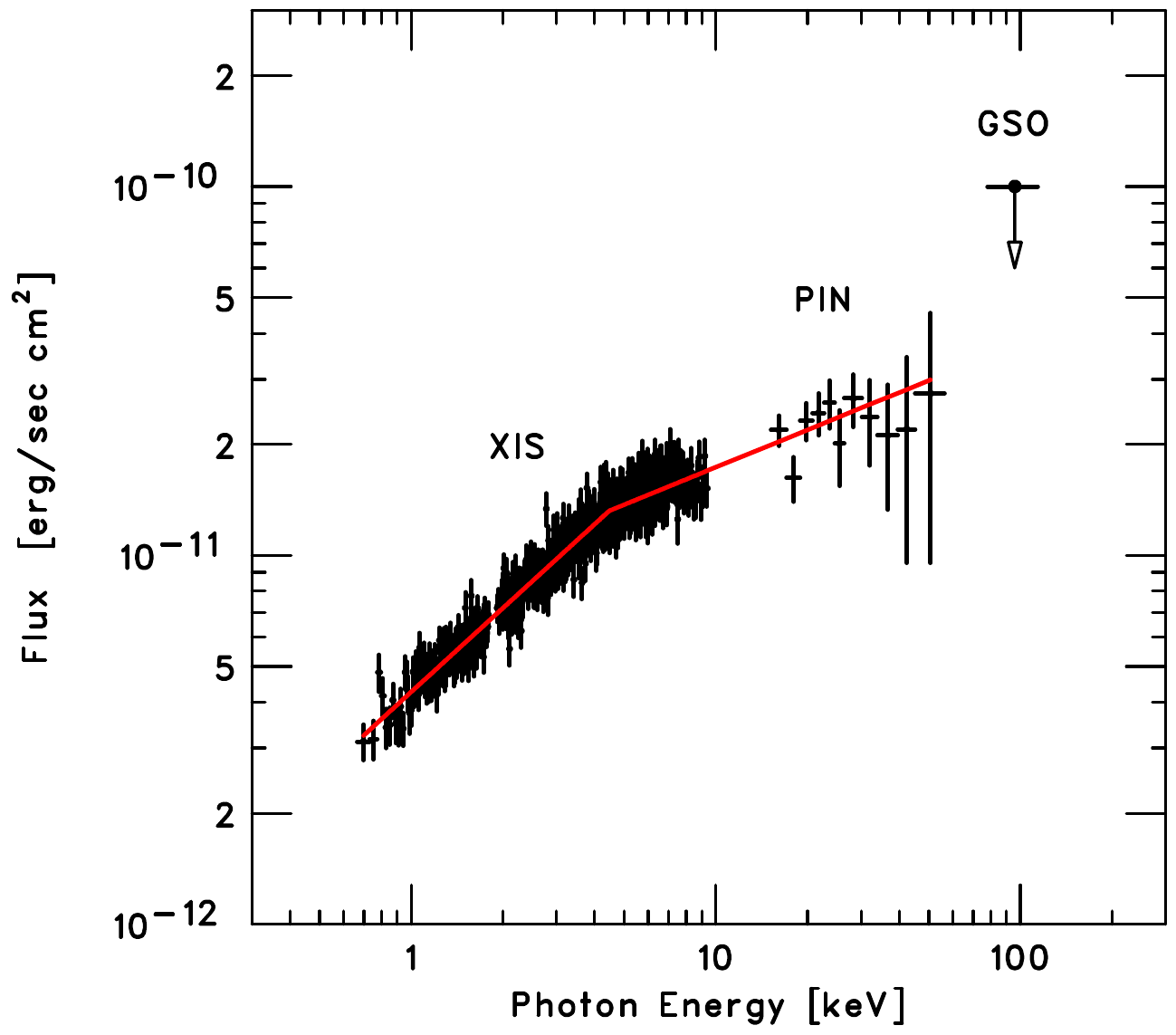}
\hspace{0.4cm}
\includegraphics[angle=90,width=0.55\linewidth]{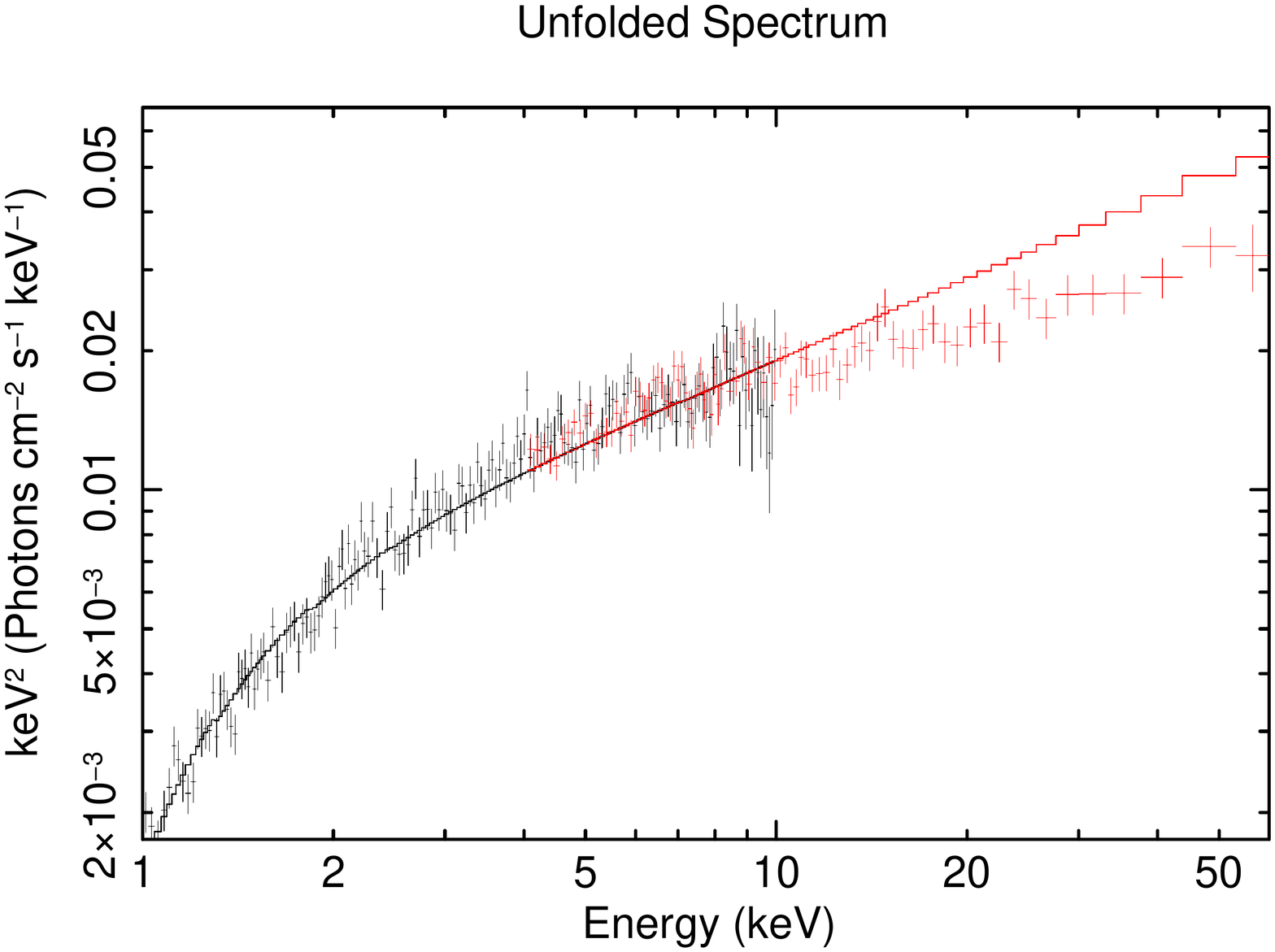}
\caption {\textit{left panel:}\textsl{Suzaku} observations of the spectral break in \psrb. Figure is taken from \cite{uchiyama09}. \textit{right panel:}Simulation of 10 ksec \textsl{ASTRO-H}/HXI observation of PSR B1259-63 as it crosses the disk with the parameters derived from the \textsl{Suzaku} observation. Fit with a single power law leads to the clear deviation of the data from the model.}
\label{psrb_uchiyama}
\end{figure}

Despite the intensive observational campaigns during the last three periastron passages (2004, 2007 and 2010) it was still not possible to conclude whether the observed X-ray emission has inverse Compton or synchrotron origin  
\citep{chernyakova09}. The answer to this question is very important for our understanding of the composition of the pulsar wind, as the Lorentz factor of the relativistic electrons varies from about 10 to 10$^6$ in these two models. Study of the variability of the broad band (1 -- 100 keV) spectrum of the source as the pulsar interacts with the disk before and after the periastron could give the missing clues to answer this question. \textsl{Suzaku} observations of the 2007 periastron passage show a break of the source spectrum as the pulsar was crossing the disk before the periastron \citep{uchiyama09}, see left panel of Figure \ref{psrb_uchiyama}. However presense of the nearby X-ray pulsar IGR J13020-6359, located only 10 minutes from the \psrb \ makes the results of \textsl{Suzaku}/HXD very model dependent, and independent measurements in this energy range with the imaging instrument would be a huge benefit. Right panel of Figure \ref{psrb_uchiyama} show simulation of 10 ks observation of \psrb with \textsl{ASTRO-H}. On this Figure we model the broken power law slope as observed with \textsl{Suzaku}. On the left Figure we show that if one would try to fit such a spectrum with a single power law the high energy data will clearly deviate from the model.

\subsubsection{LSI +61 303} 

The Be star binary \lsi\ is another GRLB system from which radio, X-ray  and very high-energy gamma-ray emission is observed. 
In \lsi\ the high-energy particle outflow is directly observed in the radio band, where angular resolution is sufficient to resolve the source and to detect variations of its morphology on the orbital period time scale. The observed morphological changes indicate that the outflow has
a variable morphology outflow filling a region the size $\sim 10^2-10^3$ times larger than the binary separation distance. The radio signal could not be used to trace the  outflow   down to the production site inside the binary orbit, because of the free-free absorption in the dense stellar wind environment \citep{zdz10}.  To understand the nature of the high-energy particles carrying  outflow one has to use  complementary high-energy data in X-ray and/or \gr\ bands.  

\begin{figure}[h]
	\begin{center}
	\includegraphics[angle=0,width=0.4\textwidth]{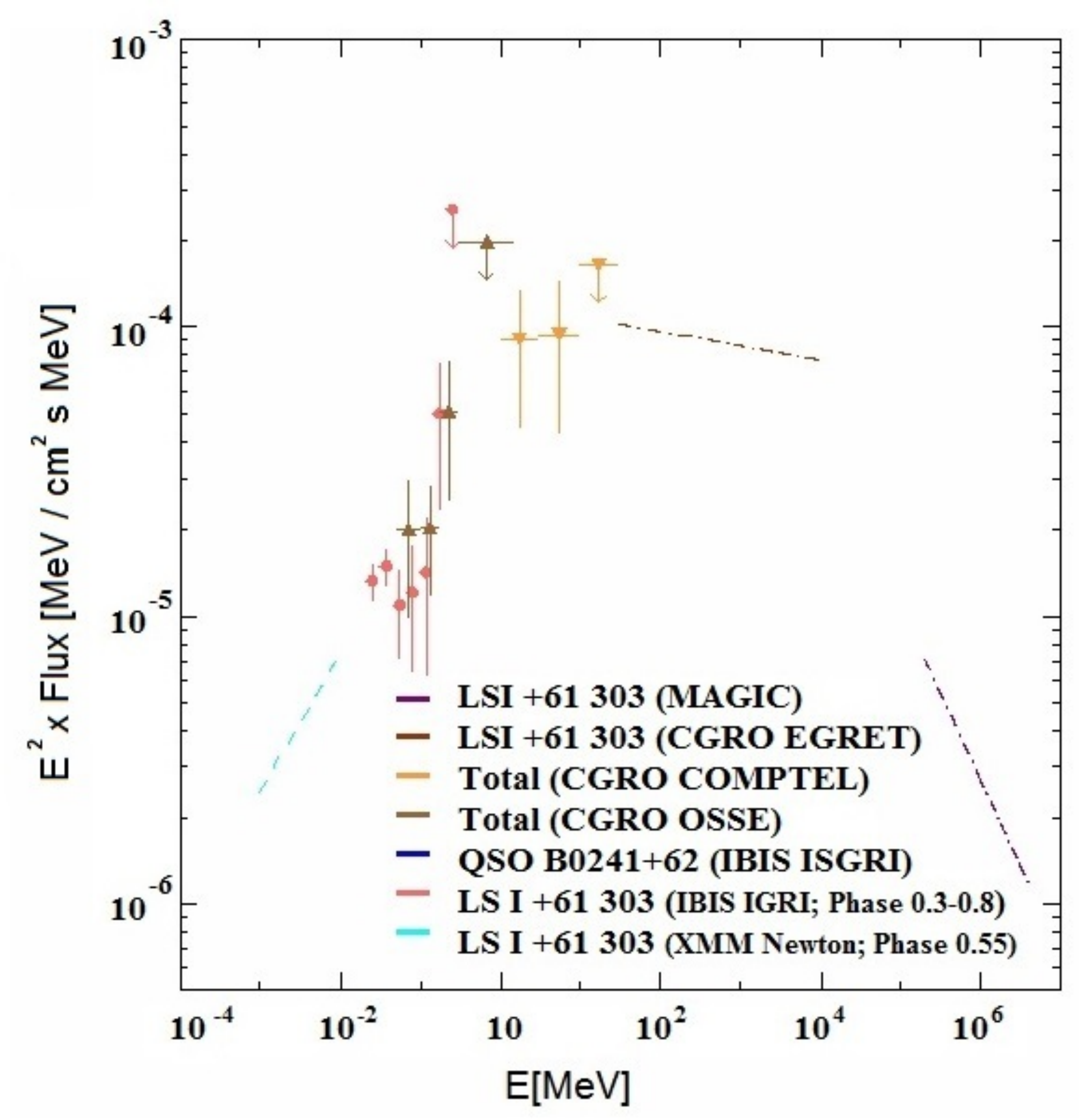}
	\includegraphics[angle=0,width=0.55\textwidth]{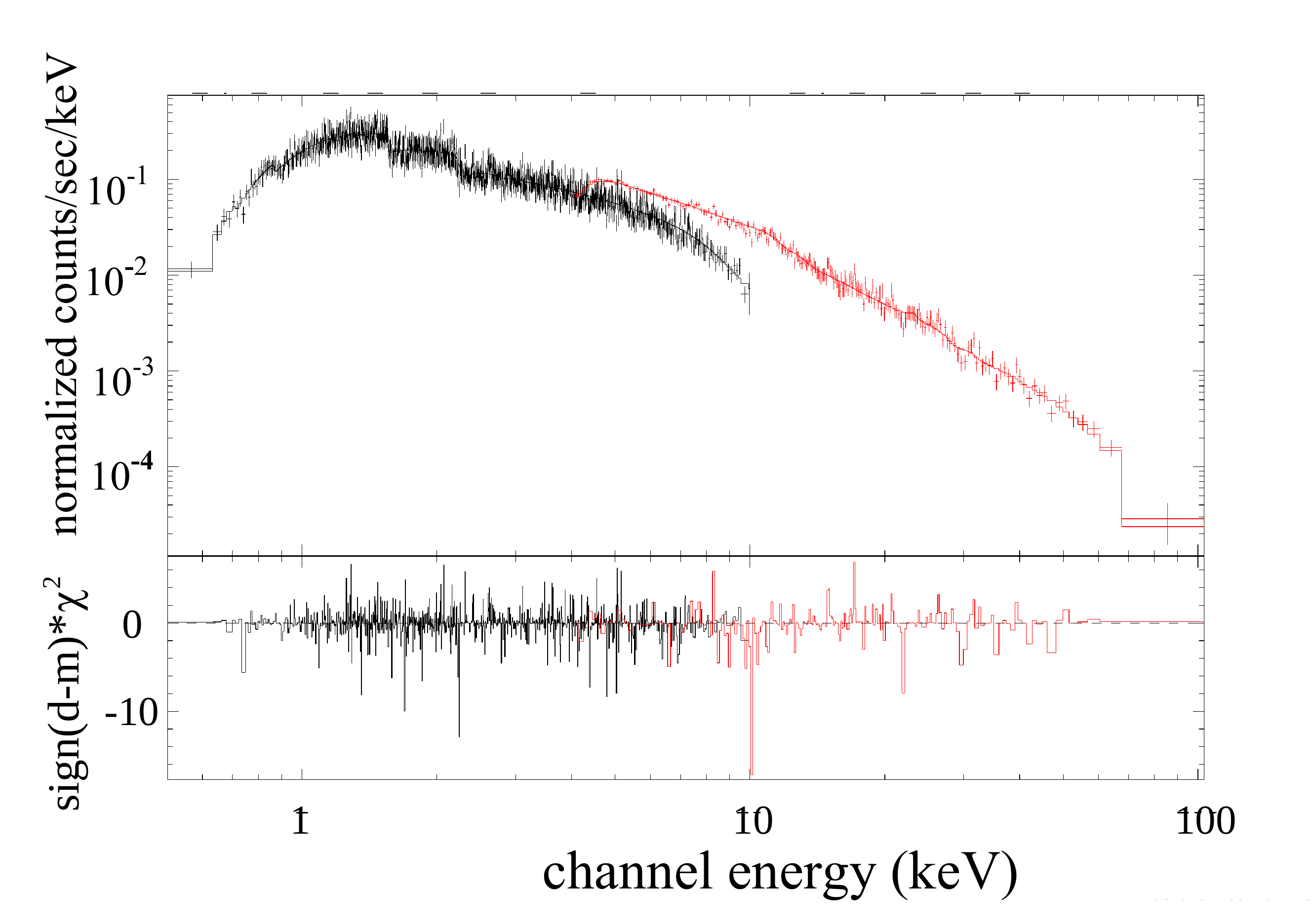}
	\caption {\textit{left:}Broad band spectrum of \lsi. \textsl{INTEGRAL} observations reveal a presence of a possible feature at ~50 keV. Figure is taken from W. Hermsen and L. Kuiper presentation at the first {\it Fermi} Symposium.
	\textit{right}:Simulation of 40 ks \textsl{ASTRO-H} observation of \lsi.}
	\label{fig:lsi}
	\end{center}
\end{figure}

Two major types of models of radio-to-X-ray activity of \lsi\ were proposed in
the literature. Models of the first type
assume that activity of the source is powered by accretion onto the compact
object.
Alternatively, the activity of the source could be explained by an interaction of a young rotation
powered pulsar with the wind from the companion Be star. 

If the system is an accreting neutron star or black hole, one expects to find a
cut-off powerlaw spectrum in the hard X-ray band.   The cut-off energy is
normally at $10-60$~keV for neutron stars and at $\sim 100$~keV for  black holes.   
If the jet and accretion
contributions to the X-ray spectrum are comparable, then emission from the
accretion disk should at least produce an observable  spectral feature 
(e.g. a bump, a break or turnover) in the  10-100~keV energy band. 

Currently the hard spectrum of \lsi\ was checked only with the \textsl{INTEGRAL}. Unfortunately the source is rather weak for the \textsl{INTEGRAL} and the only possibility to measure the spectrum was to sum up spectra collected through the years of observations. The resulted spectra didn't show a break, but have indicated a possible feature at around 50 keV, see left panel of Figure \ref{fig:lsi} presented by W. Hermsen and L. Kuiper at the first {\it Fermi} Symposium. Current \textsl{Suzaku} observations of the source are unable to constrain the spectrum of \lsi\ above 50 keV. Clearly more sensitive observations are needed to clarify the nature of the source, and \textsl{ASTRO-H} observations can help to solve the issue, see right panel of Figure \ref{fig:lsi}. On this panel we simulate a powerlaw spectrum with $N_H=0.5\times 10^{22} \mbox{cm}^{-2}$ , $\Gamma=1.5$ and $F_{2-10}=1.2 \times 10^{-11} \mbox{erg}\ \mbox{cm}^{-2}\ \mbox{s}^{-1}$. 

\section{Probes into Magnetars and their Environment}\label{sec:magnetars}

\subsection{What makes magnetars and how do they evolve? Probing their supernova progenitors, energetics and evolution}

\begin{figure}[h]
\centering
\includegraphics[scale=0.35]{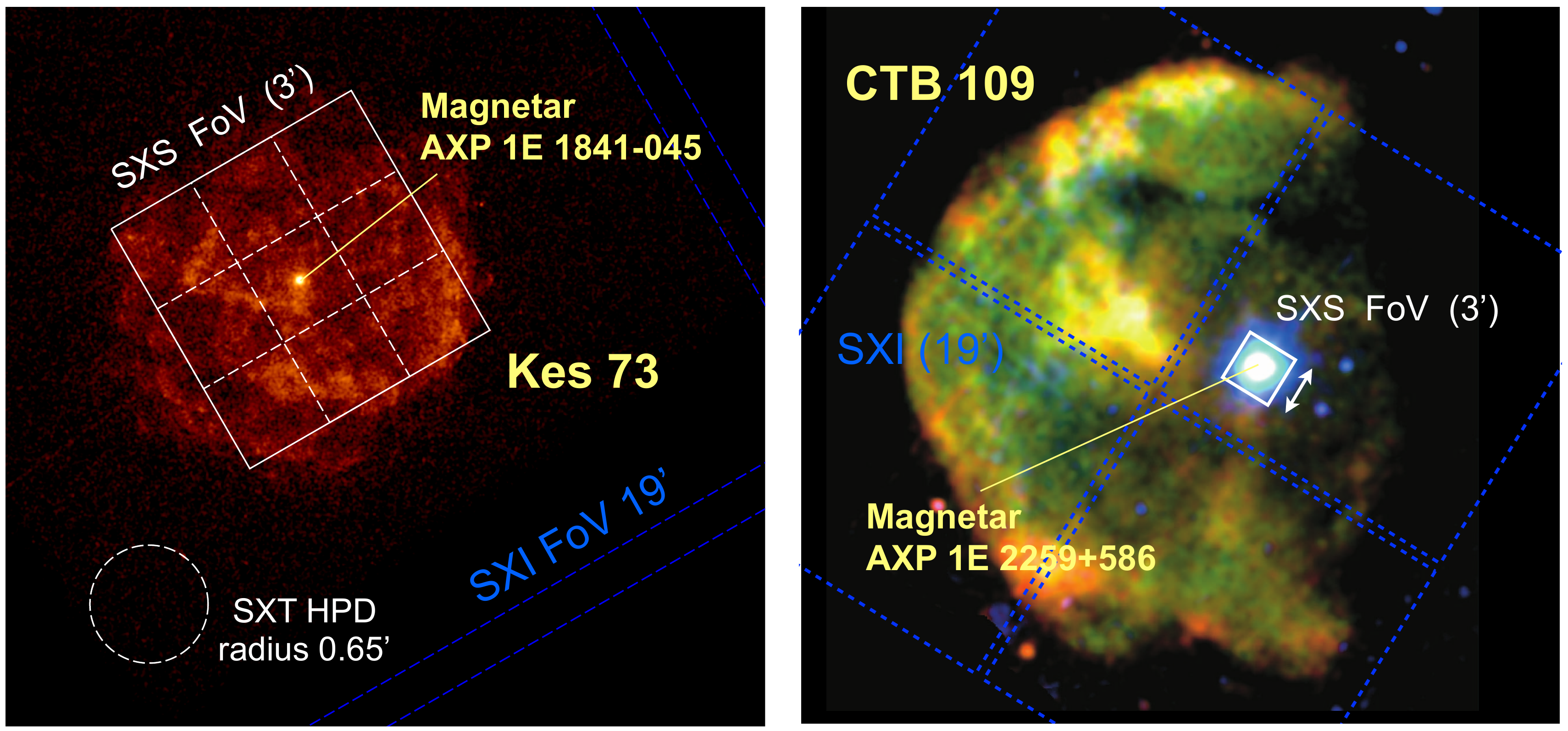}
\caption{Bright magnetar SNRs:
(left) The young SNR~Kes~73 observed with \textit{Chandra} with \textsl{ASTRO-H}'s SXS and SXI fields of view overlaid. 
The central source is the AXP 1E~1841--045.
(right) The evolved SNR~CTB~109 with \textit{XMM-Newton} \citep{2004ApJ...617..322S}
	with \textsl{ASTRO-H}'s SXS and SXI field of views.
The central source is the AXP 1E~2259+586.	
The dim western half part of this SNR is covered by a giant molecular cloud.
}
\label{snr_fov}
\end{figure}

The different manifestations of neutron star classes (see \S1 and Figure~1) can be potentially linked to the different
environments and progenitors of the supernova explosions creating these compact objects.
While there is currently no general consensus on the progenitors of highly magnetized neutron stars (including the high-magnetic field
radio pulsars and magnetars),
there is accumulating evidence, using multi-wavelength studies, for them to originate from very massive progenitors, 
i.e. with mass $\gtrsim$20 solar masses,
and expanding into a relatively low-density medium; see \cite{2013IAUS..291..480S}
for a summary.
X-ray spectroscopy of their associated supernova remnants (SNRs) is particularly a powerful tool in inferring their progenitor mass through the detection of
the X-ray emitting ejecta and comparison to nucleosynthesis models
 (see the \textit{Young Supernova Remnants} White Paper (WP\#7) for a more detailed discussion of this).
If indeed magnetars originate from very massive progenitors, as can be diagnosed with X-ray spectroscopic studies of their associated SNRs, then
this will lead to the much interesting implication that very massive stars do \textit{not} necessarily form black holes.

To date, only a handful SNRs are associated with magnetars, while the majority of the $\sim$330 known Galactic SNRs are associated with
rotation-powered neutron stars or other subclasses or neutron stars \citep{2013IAUS..291..251S, 2012AdSpR..49.1313F}. 
While this sparse SNR-magnetar association by itself holds some clues on the nature of these objects,
studying the energetics and properties of their associated SNR emission further addresses the magnetar nature of these sources.
In particular, one popular model for magnetars is that they are formed from proto-neutron stars with initial spin periods of only $\sim$0.6--3~ms.
The combination of convection and fast rotation helps build up the magnetic field to ultra-high values during the first tens of seconds following neutron star birth
\citep{1992ApJ...392L...9D, 1996AIPC..366..111D}. Such initial fast spin periods would imply larger-than-typical initial rotational energy of the neutron star, which in turn would lead
a more energetic SNR. Past and current X-ray spectroscopic studies, however, indicate that the SNRs
associated with highly magnetized neutron stars appear to have typical supernova kinetic energies in the 10$^{50}$--10$^{51}$~ergs \citep{2006MNRAS.370L..14V,2008AdSpR..41..503V, 2013IAUS..291..480S},
posing a challenge to the proto-neutron star model for magnetars. Furthermore,  more recent estimates for the explosion energies in two SNRs hosting an anomalous X-ray pulsar
(Kes~73/AXP 1E~1841--045) and a high magnetic field radio pulsar (G292.2--0.5/PSR J1119--6127) \citep{2014ApJ...781...41K,2012ApJ...754...96K}
confirm these ``typical" explosion energies. These studies also propose very massive progenitors ($\gtrsim$20 solar masses) based on the X-ray spectroscopic studies
of these SNRs with \textit{Chandra} and \textit{XMM-Newton}.

The past and current X-ray studies with \textit{Chandra}, \textit{XMM-Newton}, and \textsl{Suzaku}
have been however severely limited by poor statistics and the lack of spectral resolution and sensitivity needed for a more appropriate study and modelling of
the X-ray emitting supernova ejecta and surrounding medium.
Such an analysis can be best performed with the SXS on-board the \textsl{ASTRO-H} satellite.
Together with its coverage in the hard X-ray band thanks to the Hard X-ray Imager and the Soft Gamma-ray Dectector, \textsl{ASTRO-H} will provide a unique window to study \textit{simultaneously} the thermal plasma associated with the SNR,
and the associated compact objects with spectra characterized by both both a soft and a hard X-ray component and which occasionally emit outbursts that should impact their ``beambags", i.e. their associated SNRs.

\begin{figure}[t]
\centering
 \includegraphics[angle=0,width=1.0\textwidth]{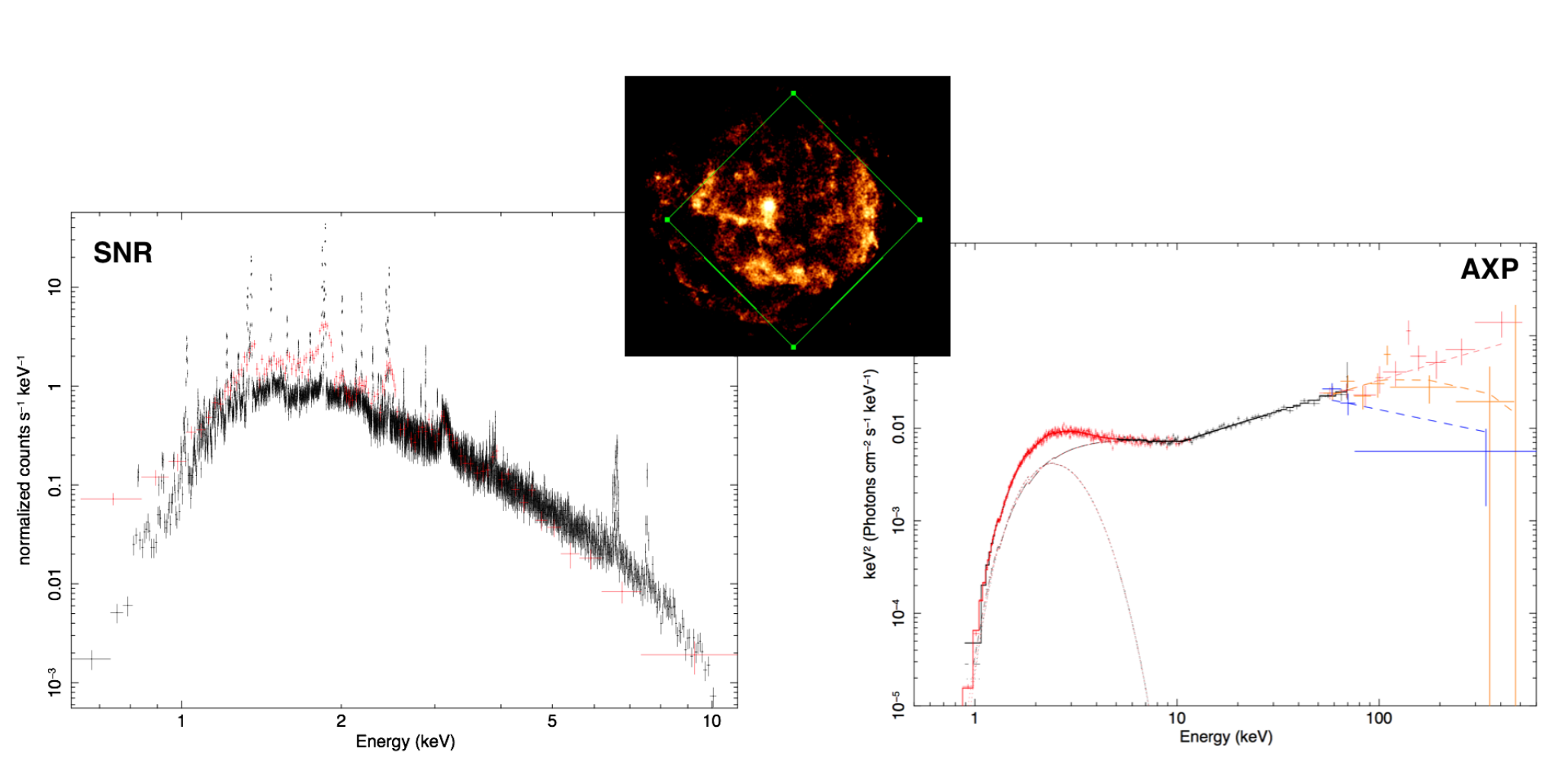}
\caption{Simulated spectra for the SNR Kes~73 and its associated AXP 1E1841--045 shown using a \textit{Chandra} observation (Kumar et al., 2014) with \textsl{ASTRO-H}'s SXS
field of view overlaid. 
(Left) A 100 ksec SXS simulation of the SNR Kes~73 (with the AXP's spectrum excluded) obtained using {\tt simx}, shown in comparison with the \textit{XMM-Newton} MOS1
spectrum.  The SXS spectrum should allow us to resolve the line emission (particularly from the Mg, Si and S complex) and detect for the first time strong line emission from Fe-K.
This will be needed to constrain the abundances and thus the mass of the progenitor star. 
(Right) Shown next for comparison the broadband SXS+HXI+SGD 100~ks simulated spectrum of the AXP 1E~1841--045
associated with the SNR Kes~73.
In the SGD band, we assumed three possibilities: no spectral break (red), a roll-over in the SGD band (orange), 
and a steep cutoff just above the HXI (or \textit{NuSTAR}) energy band. 
Studying both the SNR and the AXP with \textsl{ASTRO-H} will provide the first broadband view of this system. 
We note that the AXP's total flux is $\sim$10\% that of the SNR's flux in the SXS band.
The AXP's emission won't affect the line diagnosis of the SNR for abundance studies,
but will dominate the continuum in the hard X-ray band.}
\label{kes73axp}
\end{figure}

The SNR~Kes~73 hosting the AXP 1E1841-045 represents an ideal magnetar-SNR for \textsl{ASTRO-H} given its brightness, size, and previous detailed X-ray studies of both the SNR and AXP with CCD-type spectra.
In Figure~\ref{kes73axp}, we show the SXS field of view overlaid on the SNR, with the pointing aimed at covering the brightest and bulk of X-ray emission from the SNR, while also covering
the AXP.
Figure~\ref{kes73axp} shows a 100~ks SXS simulated spectrum of the diffuse emission from the SNR emission fitted with a two-component non-equilibrium ionization model
based on the \textit{Chandra} and \textit{XMM-Newton} study \citep{2014ApJ...781...41K}. The CCD spectrum shows evidence for a soft, ejecta-dominated component,
and a hot, low-ionization timescale component,  attributed to the shocked ISM or CSM blown by a massive progenitor. 
The simulated spectrum will provide new constraints on the ejecta abundances,
and thus on the mass of the progenitor star through a comparison to nucleosynthesis model yields such as those of \cite{1995ApJS..101..181W} and \cite{2006NuPhA.777..424N}.
As well, the Fe-K line complex will provide the first opportunity to diagnose the hot plasma conditions and confirm or refute the stellar wind blown bubble scenario.

The right panel of Figure~\ref{kes73axp}  shows the simulated broadband (SXS+HXI+SGD) spectrum of the central AXP~1E~1841--045, illustrating the additional magnetar science that
can be done with the same observation. While the line emission from the SNR dominates in the SXS band, the AXP's spectrum dominates in the hard band (HXI+SGD).
In particular, the spectrum will shed light on  a) the magnetic field of the AXP through a confirmation of the $\sim$30~keV
cyclotron (emission or absorption) feature detected recently with \textit{NuSTAR} \citep{An2013ApJ...779..163A}, and 
b) the nature of the hard X-ray emission through pinning down the spectral cutoff with the HXI+SGD combined spectrum.
If the 30~keV feature is attributed to a cyclotron line, the magnetic field is estimated to be 3$\times$10$^{12}$~G (electron cyclotron)
and 5$\times$10$^{15}$~G (proton cyclotron).
Furthermore, so far we have not yet distinguished †between the competing theoretical models for the origin of the magnetar's hard X-ray emission. 
For example, the e+/e-†outflow model \citep{2013ApJ...777..114B}
predicts the $\nu$--$F_{\nu}$ peak at†$\sim$7~MeV \citep{An2013ApJ...779..163A}
while the fall-back disk predicts it at $\sim$100--200 keV \citep{2014A&A...562A..62K}.
This will be an advantage over \textit{NuSTAR} due to the sensitivity of \textsl{ASTRO-H} above 70~keV. This science will be further discussed in Sections 3.2 and 3.3.

\begin{figure}[h]
\vspace{-5mm}
\centering
\includegraphics[width=80mm]{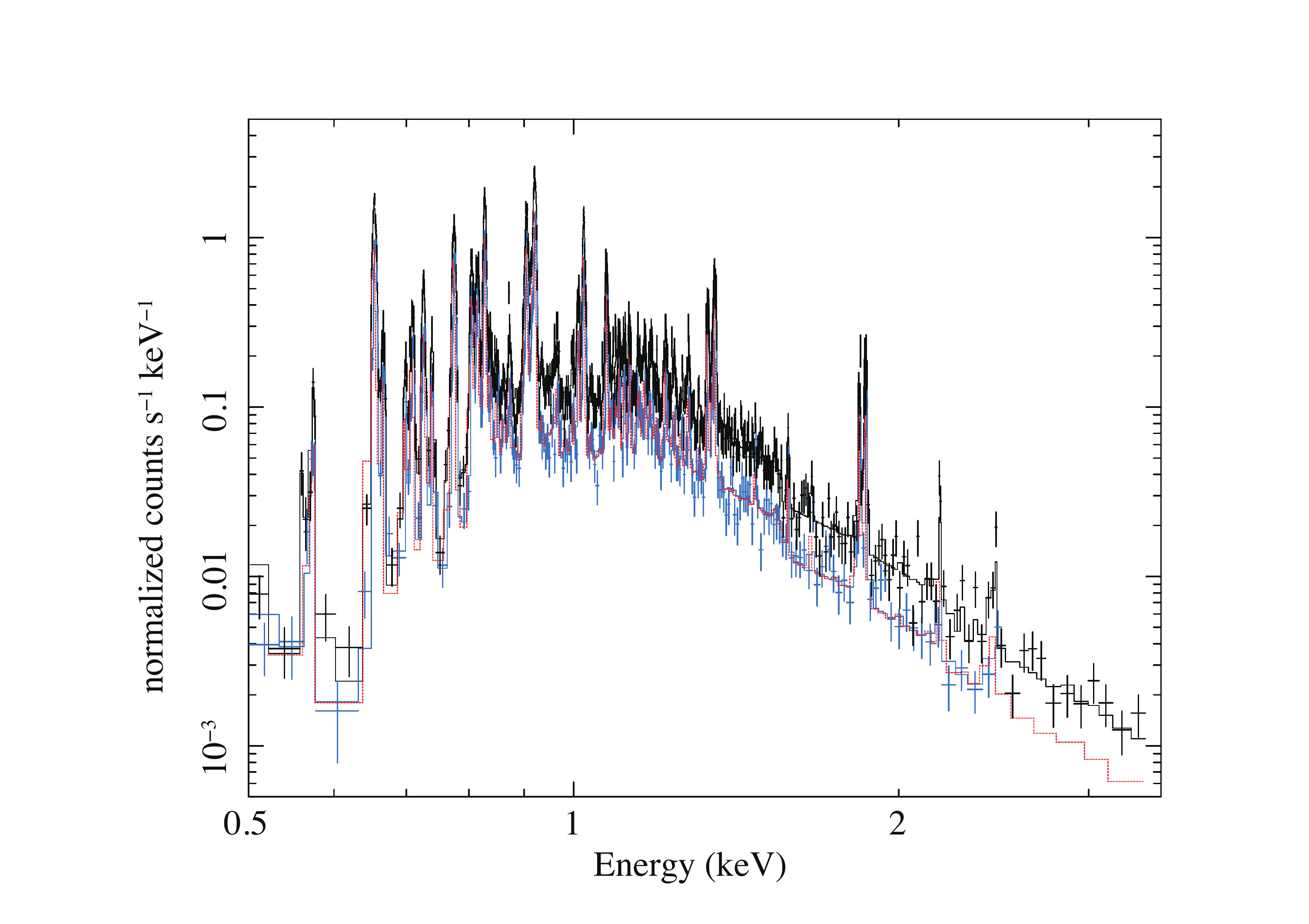}
\includegraphics[scale=0.27]{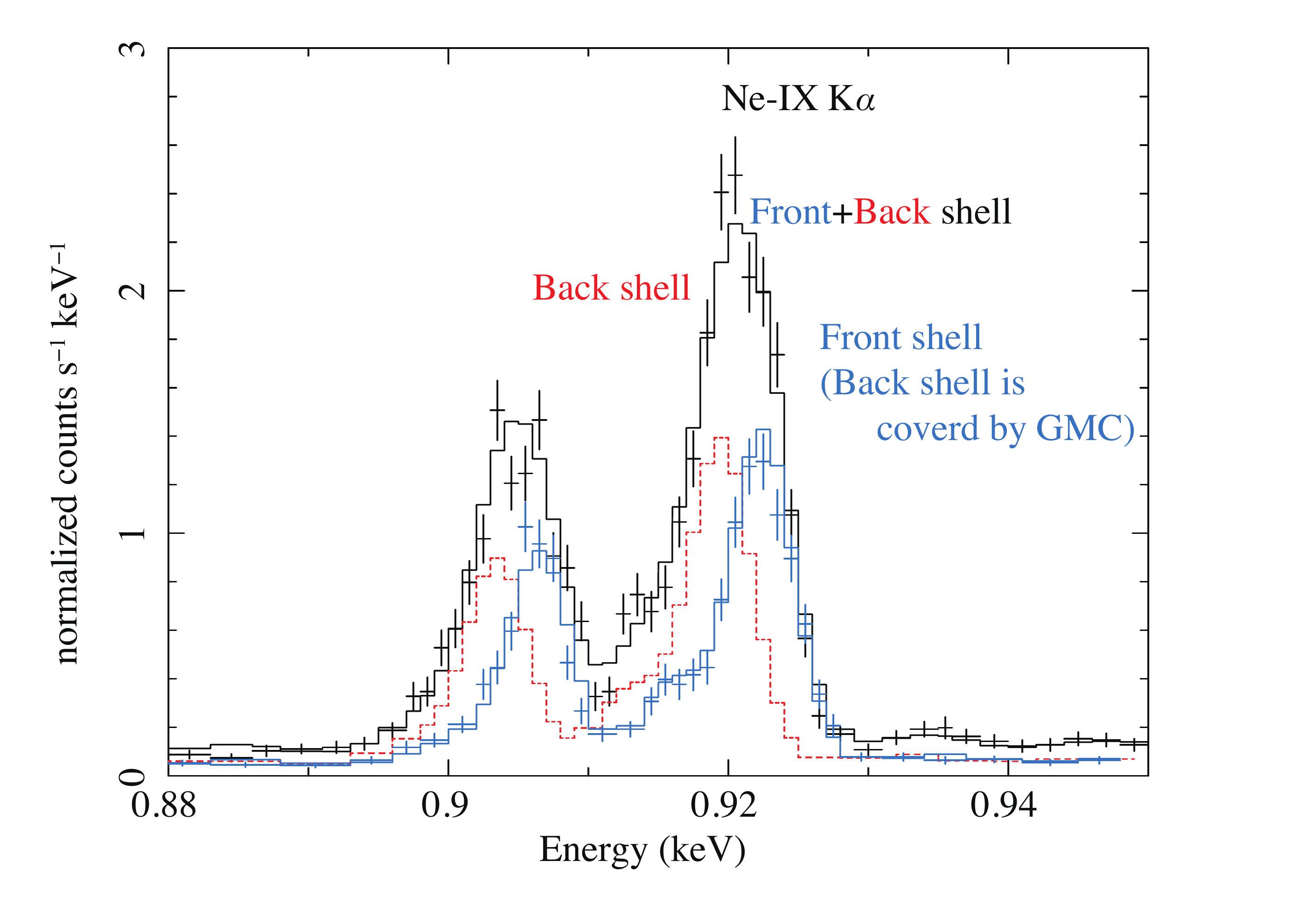}
\caption{
(left)
A 100~ks SXS simulated spectrum of the SNR CTB~109 using a two-component 
     absorbed non-equilibrium ionization model based on the \textsl{Suzaku} data.
(right)
SXS simulation of the Ne-IX lines based on the \textsl{Suzaku} spectra.
Red- and blue-shifted lines and their sum are shown in red, blue, and black colors, respectively.
Depending on the pointing position, we can observe both shell components (black) or only the front shell (blue),
	since the shell at far-western side is covered by the giant molecular cloud.
}
\label{ctb109_ne_line}
\end{figure}

CTB~109 is another relatively bright, but more evolved, SNR associated with a magnetar, AXP 1E2259+586 (see Figure~\ref{snr_fov}).  The remnant appears as a half-shell due to an obscuring molecular cloud on the western side.
A most recent study with \textit{Chandra} showed the presence of shock-heated ejecta with enhanced Si and Fe in and around the lobe adjacent to the AXP. The lobe is believed to
be created by the interaction of the SNR shock wave and the supernova ejecta with a dense and inhomogeneous medium in the SNR environment \citep{2013A&A...552A..45S}.
In Figure~\ref{ctb109_ne_line}, we show a simulated 100~ksec SXS spectrum based on a \textsl{Suzaku} observation, illustrating the wealth of lines that will be resolved with SXS.

One of the puzzling questions about magnetar SNRs is the discrepancy between the SNR's age (normally estimated from the shell's shock velocity) and the magnetar's age (estimated from
the characteristic age of the pulsar, $P/(2\dot{P})$). This discrepancy is particularly pronounced for the SNR CTB~109 (Sedov-estimated age$\sim$13--17~kyr) and its AXP 1E~2259+586 ($\sim$230~kyr), 
as recently confirmed with \textit{Chandra} \citep{2013A&A...552A..45S} \& \textsl{Suzaku} observations (Nakano et al., accepted). The answer to this puzzle lies
in understanding how magnetic fields evolve in magnetars and in acquiring accurate measurements of the associated SNR's shock velocity.

\begin{figure}[htb]
\begin{center}
 \includegraphics[width=70mm]{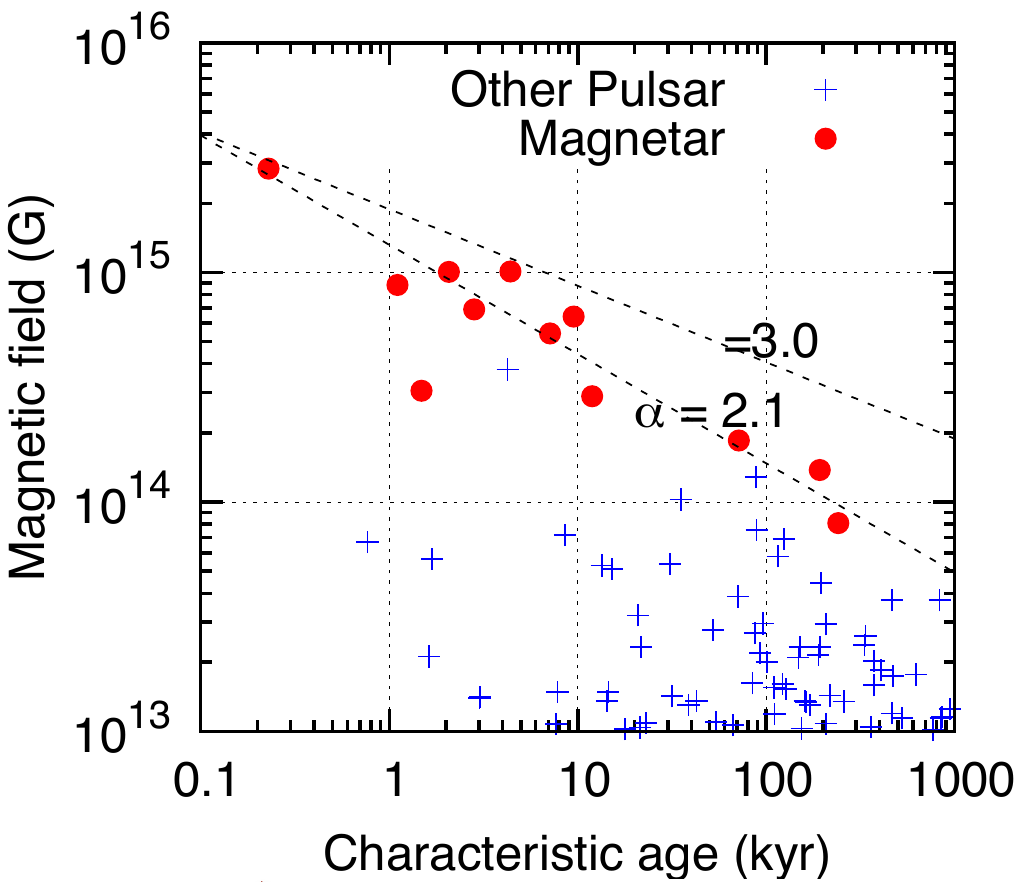}
\caption{
Magnetic field of magnetars and other (rotation-powered) pulsars calculated from the pulsar's period and its derivative 
(Nakano et al., submitted).
}
\label{magnetic_field_decay}
\end{center}
\end{figure}

Giant flares, short bursts, and bright persistent X-ray emission exceeding the spin-down power 
	are thought to originate from the magnetic energy dissipation. 
However, 
how the magnetic field of magnetars evolves with time, as the magnetar ages, is still being debated. 
The magnetic field measured from $P$ and $\dot{P}$ is known to decrease with
 the characteristic age, $P/(2\dot{P})$, as illustrated in Figure~\ref{magnetic_field_decay}.

While large characteristic ages can be potentially explained if one accounts for the magnetic field decay in magnetars 
\citep{2000ApJ...529L..29C}, 
the age measurement of associated SNRs provides an independent way of understanding the magnetic field evolution.
This can be done through an accurate measurement of the shock velocity, which will be also useful for
probing the supernova explosion energy shedding light on the origin of magnetars.
For example, in the Sedov phase of SNR evolution, the explosion energy and age
are given by $E \propto R^3 \upsilon^2$ and $\tau_\mathrm{snr} = 5R/2\upsilon$, respectively.
For the SNR CTB~109, the large molecular cloud on the western side is blocking half of the shell on the far-side
while the shell is fully visible on the eastern side.
As a result, depending on the pointing direction,
SXS is expected to measure blue- and red- shifted lines 
	originating from expanding shells towards and away from us. 
The 100~ks SXS simulation in Figure~\ref{ctb109_ne_line} assumes 
	a velocity of 500~km~s$^{-1}$, inferred from the
	\textit{Chandra} observation of CTB~109 \citep{2013A&A...552A..45S}.
The Doppler shift of the Ne-IX K lines can be detectable, 
	and simultaneous usage of multiple lines in the soft band increases the accuracy of the measurements.
For the compact object, we will be able to address its magnetic field strength and probe its hard X-ray emission, as detailed in Sections 3.2 and 3.3.

In summary, the young and more evolved SNRs Kes~73 and CTB~109, respectively hosting the magnetars 1E~1841--045 and 1E~2259+586 at their centers,
cover all the fundamental questions presented in this white paper (see the next sections for the compact objects' science).
The same method can be also applied to the other handful magnetar SNRs. This science is also relevant to WP\#7.

\subsection{Can we find direct evidence of the strong field?}

\begin{figure}[h]
\centering
\includegraphics[scale=0.35]{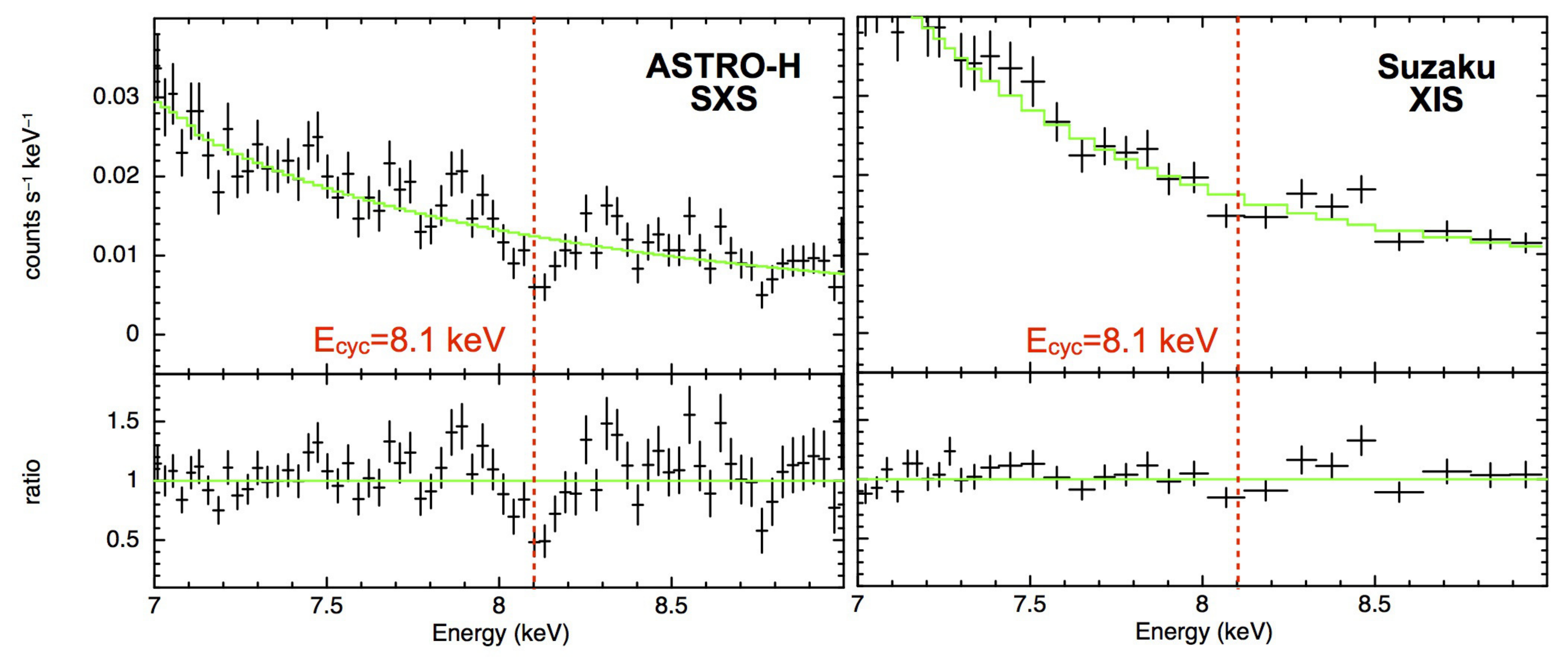}
\vspace{-0.3cm}
\caption{
Simulated possible proton CRSF at $E_{\rm cyc}=$8.1~keV 
	of \textsl{ASTRO-H}/SXS (left)
	and 
	of \textsl{Suzaku}/XIS (right).
Both Monte Carlo simulations employ the same model,	
	assuming a 100~ksec exposure 
	and a Gaussian feature on the continuum observed in 4U~0142+61,
	with a line width $\sigma_{\rm line}$=40 eV
	and an equivalent width of 60 eV.
}
\label{fig1:fig_4u0142_proton.eps}
\end{figure}

In the strong magnetic field ($B$$>$$B_{QED}$) of magnetars, 
	the electron cyclotron resonance scattering feature (CRSF) falls above the MeV range, 
	while the proton CRSF (Eq. \ref{eq:proton_cyclotron}) falls right in the soft X-ray band.
The dipole magnetic fields derived from $P$ and $\dot{P}$ reflects only from the poloidal component;
	this CRSF can provide a direct determination of the total surface $B$-field.
Early theoretical predictions suggested that 
	the proton CRSF exhibits 
	an equivalent width up to many hundred eV
	and 
	a relatively wider absorption width of $\triangle E/E_{pc}\sim 0.05 - 0.2$
	\citep{2001MNRAS.327.1081H, 2001ApJ...560..384Z}.	
Contrary to expectations for the detection of proton cyclotron lines as a direct measurement of the strong $B$-field,
	there are quite a few observational reports of the proton CRSF so far 
	from the quiescent magnetar spectra:
	$\sim$5, 10 keV from 1E~2259+586 \citep{Iwasawa1992PASJ...44....9I},
	$\sim$8.1 keV from 1RXS~J1708-4009 \citep{Rea2004NuPhS.132..554R},
	and 
	more recently an absorption feature from SGR~0418+5729 \citep{Tiengo2013Natur.500..312T}
	and a 25--35 keV (absorption or emission) weak feature from 1E~1841-045 in the SNR~Kes~73 \citep{An2013ApJ...779..163A}.
Some of those have not yet been confirmed by following observations  \citep{2007Ap&SS.308..505R,2011ysc..conf...43M}.
In the absence of a significant CRSF in most of the quiescent X-ray spectra of magnetars, 
	it was pointed out that 
	vacuum polarization effectively suppresses the strength of the proton CRSF,
	making the equivalent width nearly an order of magnitude lower than previously thought
	\citep{2002ApJ...566..373L, 2003ApJ...583..402O}.
In addition, the gradient of the $B$-field and effective temperatures would suppress these features as well. 
\textsl{ASTRO-H}'s SXS, combined with its broadband capability with HXI+SGD, will allow us to test for this 
and confirm previously reported features.

\begin{table}[t]
\caption{
Reported absorption-like features in magnetars, with reference to \citep{2011ysc..conf...43M}.
\vspace{2mm}
}
\label{table:proton_cyclotron_observations}
\vspace{-0.5cm}
\begin{center}
\footnotesize
\begin{tabular}{cp{2.5cm}cp{5.0cm}l}
\hline \hline
Object & $E_{\rm pc}$ (keV) & Detector & Note & Ref. \\
\hline
SGR~1806-20 & 
5.0, 7.5,  11.2, 17.5 & 
{\it RXTE}/PCA & in the harder part of a precursor &
(1) (2)
\\
SGR~1900+14 & 
6.4 & 
{\it RXTE}/PCA & during precursor to the main burst &
(3) 
\\
1RXS~J1708-4009 &
8.1 & 
{\it BeppoSAX} & the longest observation (200 ks) during rising phase & 
(4) (5) (6) 
\\
1E~1048.1-5937 &
14 & 
{\it RXTE}/PCA & emission, in a burst &
(7) 
\\
$\cdots$ & 
13 & 
{\it RXTE}/PCA & emission, at one part of a burst tail & 
(8)
\\
XTE~J1810-197 & 
12.6 &
{\it RXTE}/PCA & emission in a burst tail & 
(9) 
\\
4U~0142+61 & 
4, 8, 14 &
{\it RXTE}/PCA & 
emissions, in the most energetic among a sequence of bursts &
(10)
 \\
1E~2259+586 & 
5, 10 & 
{\it Ginga}/LAC & during flux increase &
(11)
\\
1E~1841--045 &
25--35 &
{\it NuSTAR} & phase-resolved spectrum &
(12) 
\\
\hline
\end{tabular}
\end{center}
\vspace{-2mm}
\begin{footnotesize}
References: 
(1) \citealt{Ibrahim2002ApJ...574L..51I}, 
(2) \citealt{Ibrahim2003ApJ...584L..17I},
(3) \citealt{Strohmayer2000ApJ...537L.111S}, 
(4) \citealt{2003ApJ...586L..65R}, 
(5) \citealt{Rea2004NuPhS.132..554R}, 
(6) \citealt{Oosterbroek2004ESASP.552..471O}, 
(7) \citealt{Gavriil2002Natur.419..142G}, 
(8) \citealt{Gavriil2006ApJ...641..418G}, 
(9) \citealt{Woods2005ApJ...629..985W}, 
(10) \citealt{Gavriil2008AIPC..983..234G}, 
(11) \citealt{Iwasawa1992PASJ...44....9I}, 
(12) \citealt{An2013ApJ...779..163A}
\end{footnotesize}
\end{table}%

\begin{figure}[htb]
\begin{center}
\includegraphics[width=0.6\hsize]{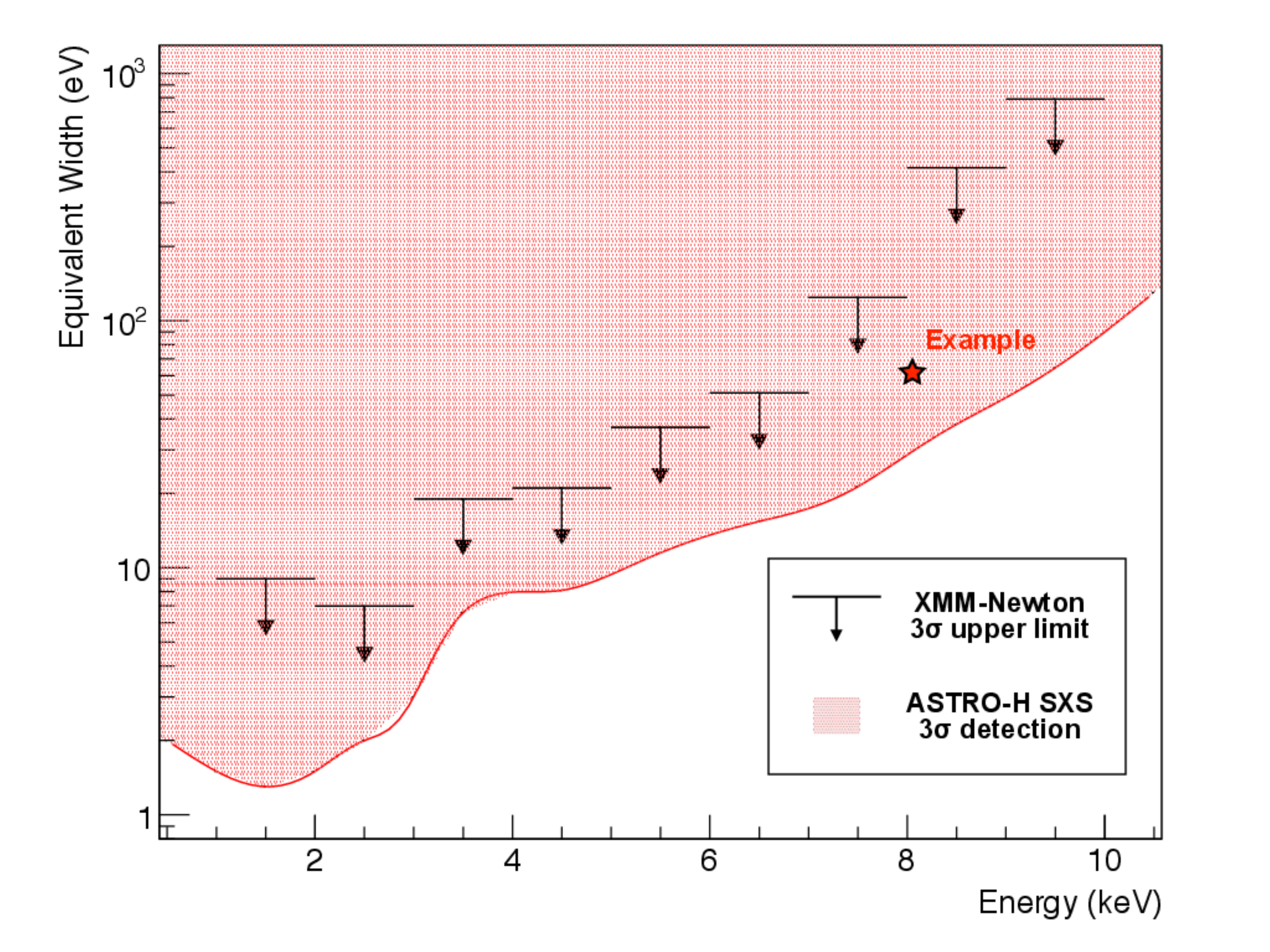}
\caption{
Detectable absorption features from the AXP 4U~0142+61. 
Feasibility to detect them ($>$3$\sigma$ detection region) in an \textsl{ASTRO-H} SXS 100 ksec observation
	are evaluated through the F-test probability with and without absorption features
	onto  the X-ray continuum of 4U~0142+61.
Gaussian absorption features are assumed at its width $\sigma_{\rm line}$=40~eV.	
The previously measured energy-dependent 3$\sigma$ upper limits are also shown as arrows 
	referring to the 46 ksec {\it XMM-Newton} observation in 2004 \citep{2007Ap&SS.308..505R},
	where assuming its possible absorption width at $\sigma_{\rm line}$=100~eV.	
The star mark corresponds to Figure~\ref{fig1:fig_4u0142_proton.eps} .	
}
\label{fig1:energy_vs_eqwidth.eps}
\end{center}
\end{figure}
		
More puzzling, 
	some proton CRSF have been reported from bursts spectra, 
	as shown in table \ref{table:proton_cyclotron_observations};
	e.g., at 5.0, 11.2, and 17.5 keV of SGR~1806$-$20 with {\it RXTE}  \citep{Ibrahim2002ApJ...574L..51I,Ibrahim2003ApJ...584L..17I}.
Even though the number of detections in bursts is larger than that of the persistent emission,
	these detections are still quite rare compared to numerous short bursts from magnetars.
In summary,
	the current observations have not yet achieved a consistent interpretation of the proton cyclotron features,
	requiring much deeper observations by high-resolution instruments.

The Soft X-ray Spectrometer (SXS) on board \textsl{ASTRO-H} can provide 
	a higher sensitivity search for the proton CRSF
	even with a shallower and/or narrower width than the previous studies.
Since the previous persistent and burst observations have not yet provided a consistent picture of 
	the proton CRSF (e.g., resonance energy, absorption width, equivalent width, and 
	spectral shapes),
	here we empirically assume the possible Gaussian absorption feature,
	and simulate the sensitivity to detect the lines.
Figure~\ref{fig1:fig_4u0142_proton.eps} shows Monte Carlo simulations of the absorption feature at 8.1 keV
	from the anomalous X-ray pulsar 4U~0142+61.
As shown in this plot,
	compared to the XIS on board \textsl{Suzaku},
	the high resolution of the SXS would 
	allow the detection of weaker features (small equivalent width).
Figure~\ref{fig1:energy_vs_eqwidth.eps}  represents an example of detectability of the proton CRSF 
	on the equivalent width versus energy plane.
Comparing to the previous {\it XMM-Newton} observations,
	the SXS potentially search for  features with a factor of 2--3 weaker equivalent width.

\subsubsection{On the link between magnetars and the other classes of neutron stars through a direct measurement of the magnetic field}

As mentioned in Section 1, the growing diversity of neutron stars includes an intriguing class of compact objects near the centres of core-collapse SNRs, referred to as CCOs (for Central Compact Objects).
These objects are typified by the CCO discovered with the first light \textit{Chandra} observation of the O-rich supernova remnant (SNR) Cas~A.
Originally, CCOs were thought to be ``relatives" of magnetars primarily based on the resemblance between their X-ray spectra in the 0.5--10 keV band, the relatively slow spin periods (in comparison to the typical rotation-powered pulsars), 
and the lack of radio emission and pulsar wind nebulae surrounding them.

\begin{figure}[htb]
\begin{center}
  \includegraphics[width=0.4\hsize]{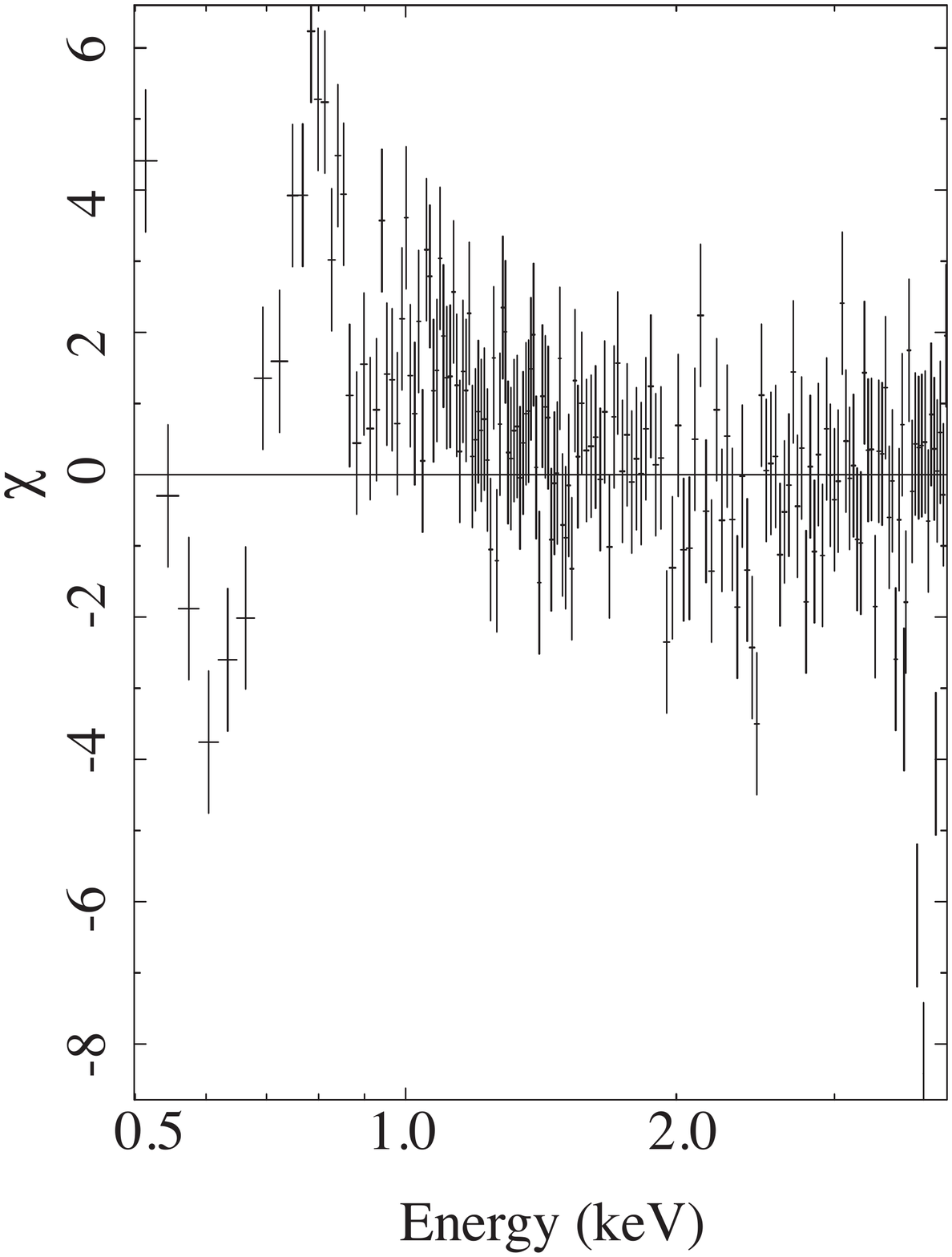}
\caption{
A 100~ksec SXS simulated spectrum of a CCO using a two-blackbody model plus an emission line, fitted with a two-blackbody model plus two cyclotron lines.
The residuals illustrate the difference between proposed models that can be addressed with the SXS.
We acknowledge the contribution of Adam Rogers (U. Manitoba) for this figure. 
}
\label{puppisa_emissionversuscyclotronabs}
\end{center}
\end{figure}

Recent dedicated timing observations of a few CCOs led to the suggestion that these are ``anti-magnetars", i.e. with a magnetic field much smaller than a typical magnetar's field, with inferred $B$$\sim$10$^{10}$--10$^{11}$~Gauss;
see \cite{2009ApJ...695L..35G} for the discovery of a 112~ms period in the CCO residing in the SNR Puppis~A. The measured $\dot{P}$ implies a surface dipole magnetic field $B$$<$9.8$\times$10$^{11}$~G, most recently refined to a value of
2.9$\times$10$^{11}$~Gauss \citep{2013ApJ...765...58G}.
Evidence is accumulating for other low-$B$ CCOs being anti-magnetars: 1E 1207.4--5209 in PKS~1209--51/52 and  
CXOU~J185238.6+004020 in the SNR Kes~79. In addition to timing measurements, X-ray spectroscopic studies of the CCO in Cas~A also supports the low-$B$ scenario \citep{2009Natur.462...71H}.
This study further indicates that the neutron star in Cas~A is covered with a non-magnetized atmosphere of Carbon, the product of nuclear burning of H and He.
Other, more exotic (quark star), models have been however proposed \citep{arXiv:1404.5063}, in light of the recent \textit{NuSTAR} study of the Cas~A SNR \citep{2014Natur.506..339G}.

As for magnetars, a direct measurement of the CCOs magnetic field comes from cyclotron resonance lines, and such low $B$ ($\sim$10$^{10}$--10$^{11}$~Gauss) are expected to show cyclotron features in the soft X-ray band, making
the SXS an ideal instrument to study these features.
Indeed cyclotron features have been discovered in a handful CCOs, including Puppis~A's CCO, RXJ~0822.0--4300, whose spectrum displays a phase dependent emission feature at 0.7--0.8 keV.
This feature has been modelled either as an emission line of energy $\sim$0.75 keV (hereafter {\tt emission}) or as a cyclotron absorption feature plus a harmonic with an energy of $E_0$$\sim$0.46 keV, hereafter {\tt cyclotron} 
\citep{2013ApJ...765...58G}.
It wasn't possible to distinguish between these models using \textit{Chandra} and \textit{XMM-Newton} data.
In Figure~\ref{puppisa_emissionversuscyclotronabs}, we illustrate the capability of the SXS in differentiating between the {\tt emission} model and the {\tt cyclotron} model using a 100~ksec exposure.
This particular CCO was selected based on its location near the centre of a large and not so bright SNR (in comparison for example to Cas A's CCO which will be difficult to resolve from the surrounding bright thermal emission from Cas~A).
We note however that this simulation does not take into account the presence of the SNR thermal plasma. 
This is meant as an illustrative example to show SXS's capability to differentiate  the different models proposed from fitting CCD-type spectra.
Other more ``isolated" targets (including other classes of neutron stars shown on Figure~1) with possible links to magnetars would be less contaminated by the X-ray emission from the SNR plasma.

In summary, SXS will open a new window to probe and understand the spectral features in an emerging and new class of (soft) X-ray emitting neutron stars, and confirm whether they are indeed anti-magnetars 
or descendants of magnetars.


\subsection{Unified understanding of the magnetar X-ray spectrum?}

The Hard X-ray component (HXC) was discovered from some magnetars above 10 keV 
	by \textsl{INTEGRAL} and {\it RXTE} \citep{2006ApJ...645..556K},
	and further confirmed by \textsl{Suzaku}.
The HXC is distinguished from the well known soft X-ray component (SXC)
	since the HXC is represented by a power-law (PL) with an extremely hard $\Gamma_{h} \sim 1$, 
	and extending $>$100 keV.
The {\it CGRO}/COMPTEL and {\it Fermi}/LAT upper-limits \citep{2010ApJ...725L..73A}
	suggest that the HXC must have a spectral break under $\sim$ 750 keV
	which has not yet been clearly detected.
The break energy is an important hint 
	to constrain the emission mechanism of the HXC; 
	e.g., thermal bremsstrahlung,
	synchrotron radiation 
	and resonant Compton up-scattering 
	\citep{2005ApJ...634..565T, 2007Ap&SS.308..109B}.
This is a unique science goal for the HXI and SGD.	

\begin{figure}[h]
\begin{center}
\includegraphics[width=72mm]{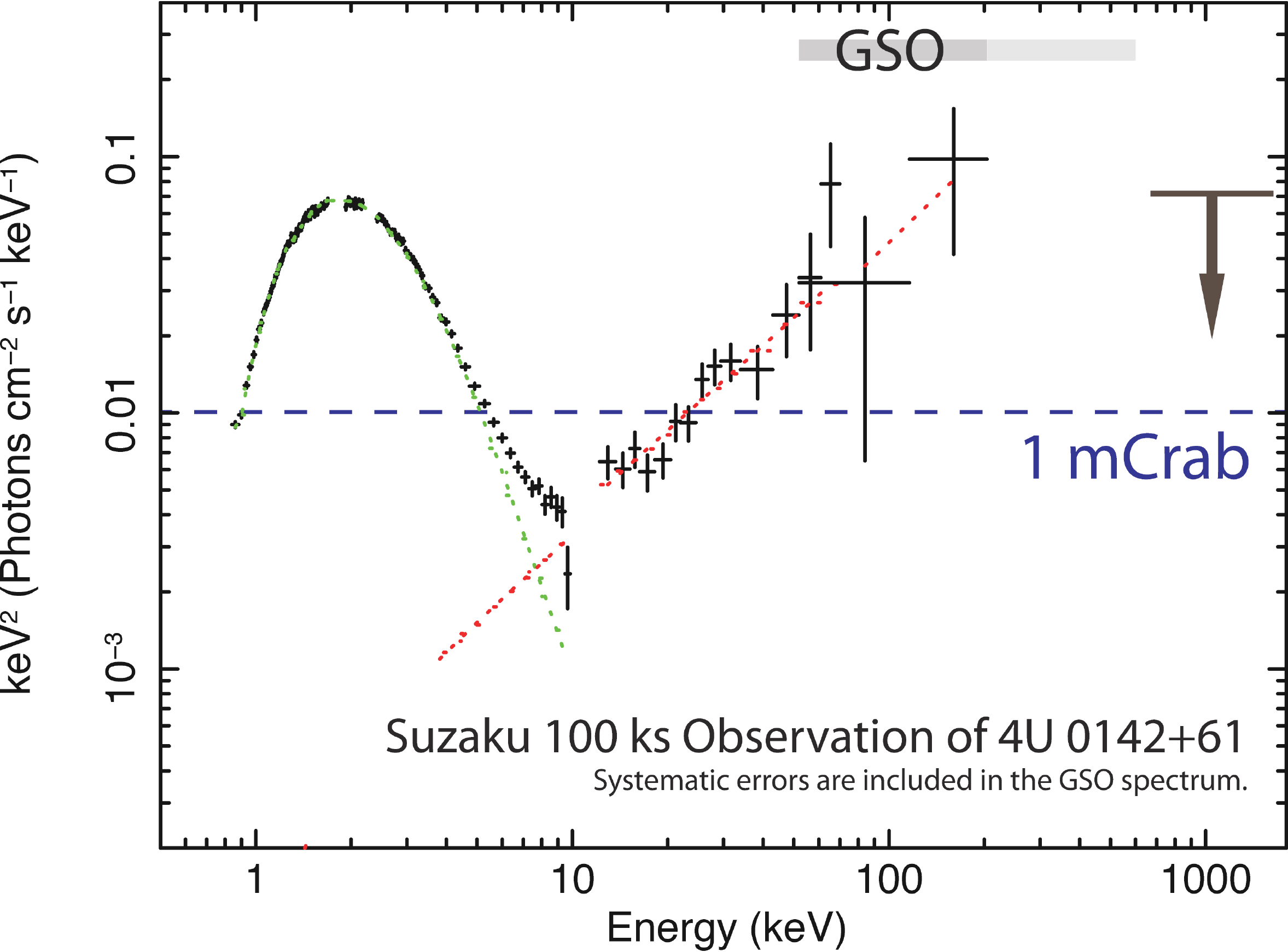}
\hspace{5mm}
\includegraphics[width=70mm]{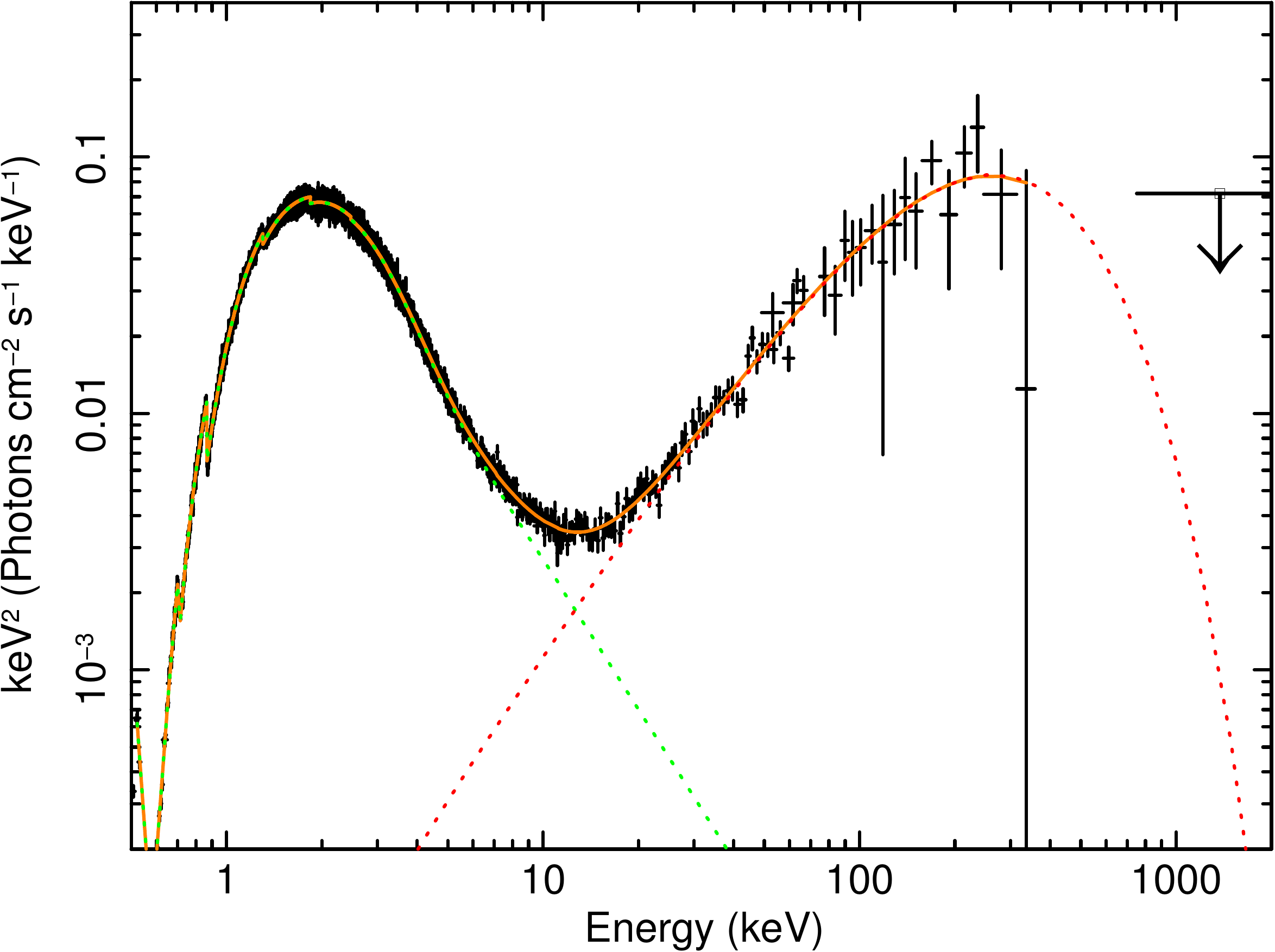}
\end{center}
\caption{
(Left) 
The 4U~0142+61 $\nu$F$\nu$ spectrum
	obtained from the \textsl{Suzaku} 100 ksec observation \citep{2011PASJ...63..387E}.
(Right) 	
Simulated 100~ks \textsl{ASTRO-H}  4U~0142+61 observation using the parameters of the \textsl{Suzaku} result.
The HXC is reproduced by a single PL of $\Gamma_{\rm h}=0.11$
	and the flux $F_{h}$ of 29.7 $\times$ $10^{-12}$ erg s$^{-1}$ cm$^{-2}$ in the 1-60 keV.
The exponential cutoff is further added at 150 keV to satisfy the COMPTEL upper limit.
Orange, dotted green, and red lines denote the 
	best-fit model, SXC model and HXC model, respectively.
The 2 $\sigma$ {\it CGRO}/COMPTEL upper-limits \citep{2006ApJ...645..556K} are also plotted.}
\label{fig:spec4u0142}
\end{figure}	

In the \textsl{Suzaku} era,
	the simultaneous spectroscopy of the SXC and HXC
	revealed a possible broadband spectrum evolution;
	the hardness ratio, defined as $F_{\rm h}/F_{\rm s}$,	
	is positively (or negatively) correlated with their magnetic field (or pulsar characteristic age).
The $\Gamma_{\rm h}$ of the HXC is 
anti-correlated with the characteristic age.
Although the interpretation of these correlations has not yet been established,
	one possible 
	explanation is a down-cascade of the sub-MeV photons 	through photon splittings.
Sub-MeV photons can be generated via electron-positron annihilation 
	or resonant Compton up-scattering near the stellar surface,
	and are repeatedly splitting into hard X-rays in the QED field.
In addition,
	 \cite{Nakagawa2011PASJ...63S.813N} suggested a similarity of the 
	 hard X-ray spectrum 
	 between the persistent emission and an accumulated weaker short bursts
	 from the activated magnetar, SGR~0501+4516 (\S3.4).

Furthermore, 
	the broad-band coverage both of SXC and HXC can potentially provide 
	a clue for the hidden toroidal magnetic field inside magnetars.
In the recent \textsl{Suzaku} observation \cite{Makishima2014PhRvL.112q1102M},
	an evidence for a free precession
	was detected from a prototypical magnetar 4U~0142+61.
In the 15--40 keV hard X-rays, its 8.69 sec pulsations 
	suffered slow phase modulations by ±0.7 sec, with a period of $\sim$15 h.
When this is interpreted as free precession of the magnetar,
	the object is inferred to deviate from spherical symmetry 
	by $\sim 1.6\times 10^{-4}$ in its moments of inertia. 
This deformation, 
	suggests a strong toroidal magnetic field, $\sim 10^{16}$ G in the stellar interior
	when ascribed to magnetic pressure.

\begin{figure}[htb]
\begin{center}
  \includegraphics[width=0.45\hsize]{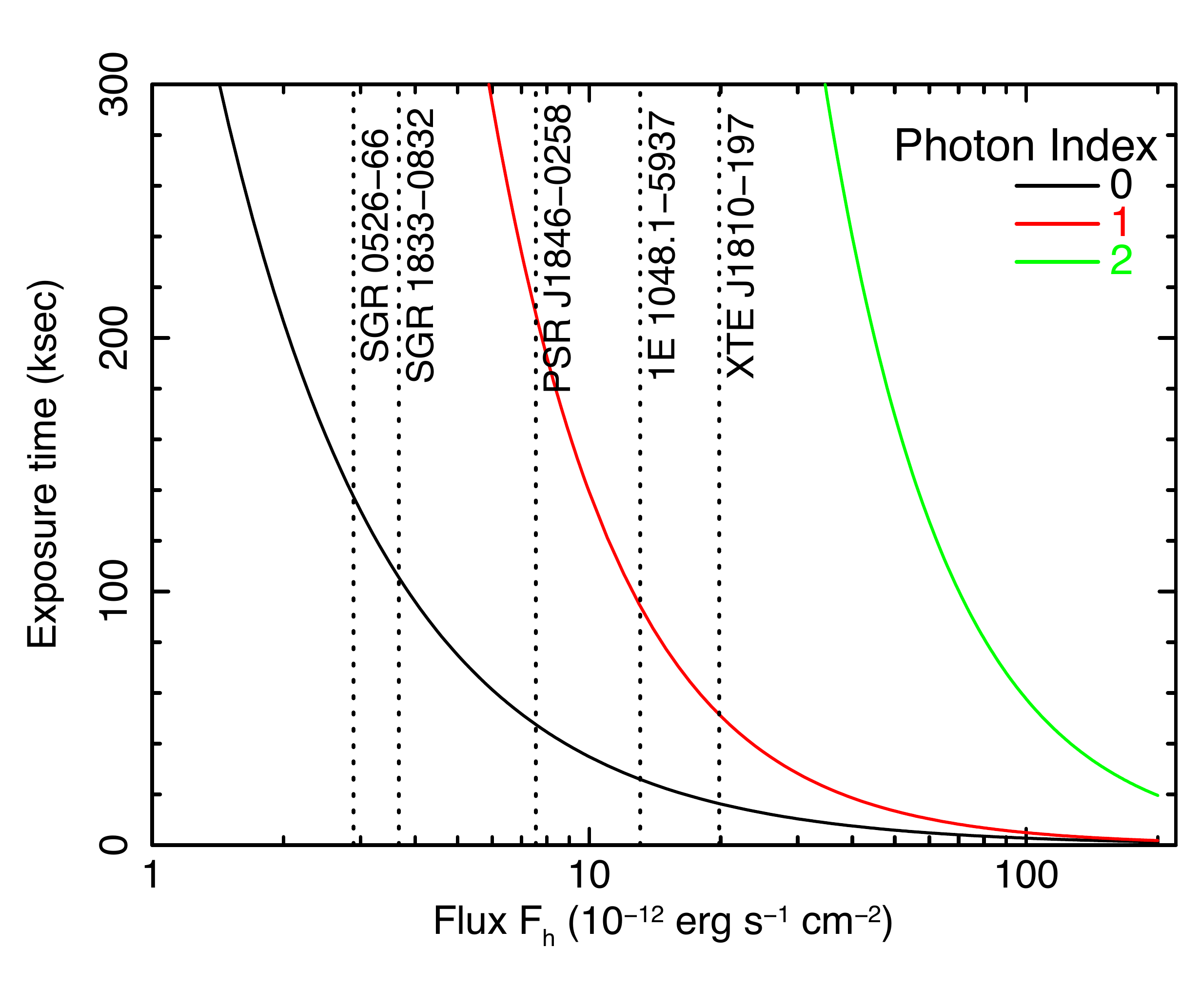}  
  \caption{The HXC detectability with the HXI and SGD. Required
    exposure for HXC 3$\sigma$ detections is plotted as a function of
    the 10--100 keV flux $F_{h}$, assuming a single PL with
    $\Gamma_{\rm h}=2$ (green), 1 (red), and 0 (black) and 3\%
    systematic uncertainty of the non X-ray background. Dotted lines
    indicate assumed $F_{h}$ of 5 magnetars assuming the correlation
    suggested by \textsl{Suzaku} \citep{Enoto2010ApJ...722L.162E}.}
\label{fig:feasibility}
\end{center}
\end{figure}

\begin{figure}[h]
\begin{center}
\includegraphics[width=72mm]{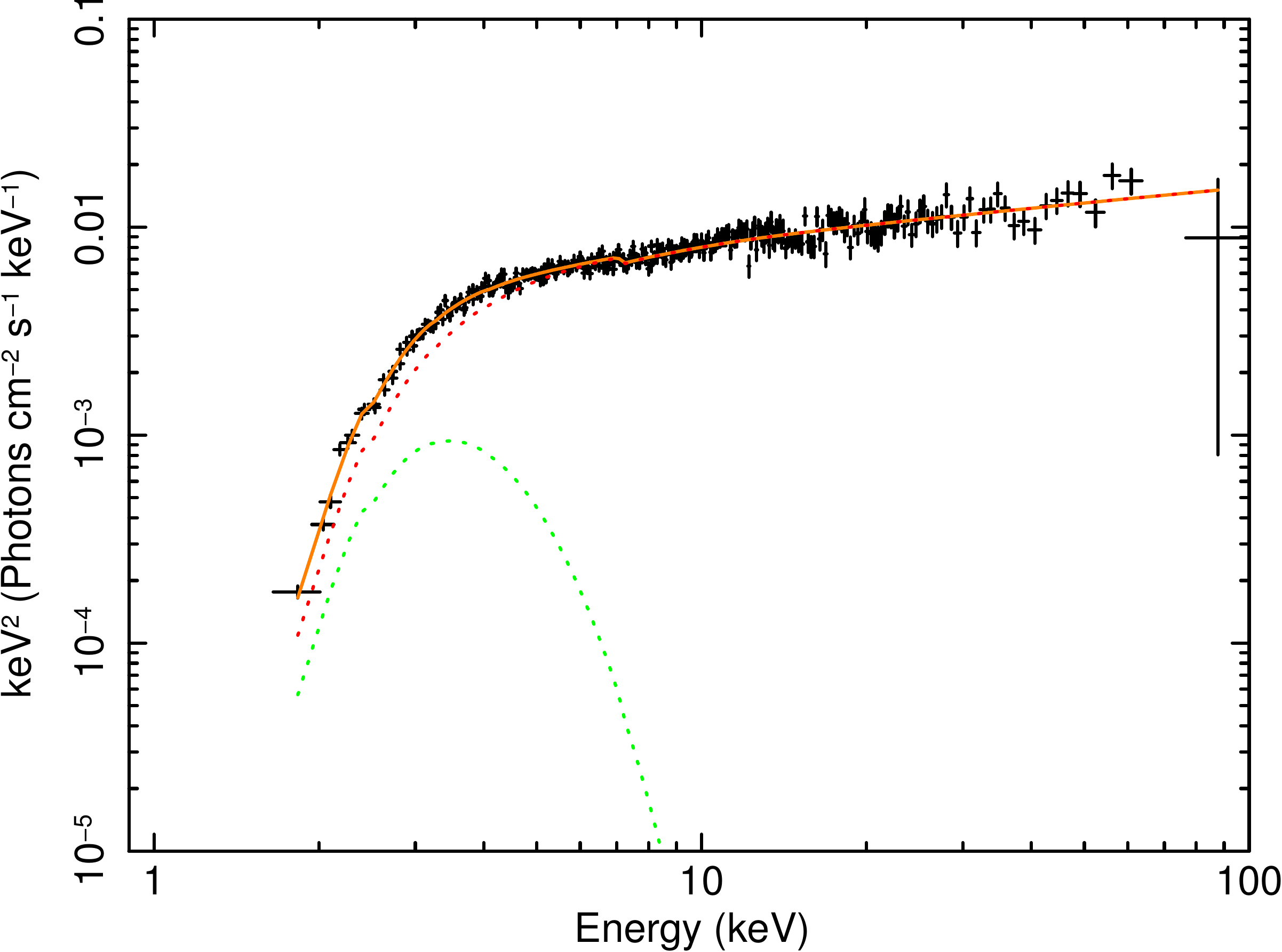}
\hspace{5mm}
\includegraphics[width=70mm]{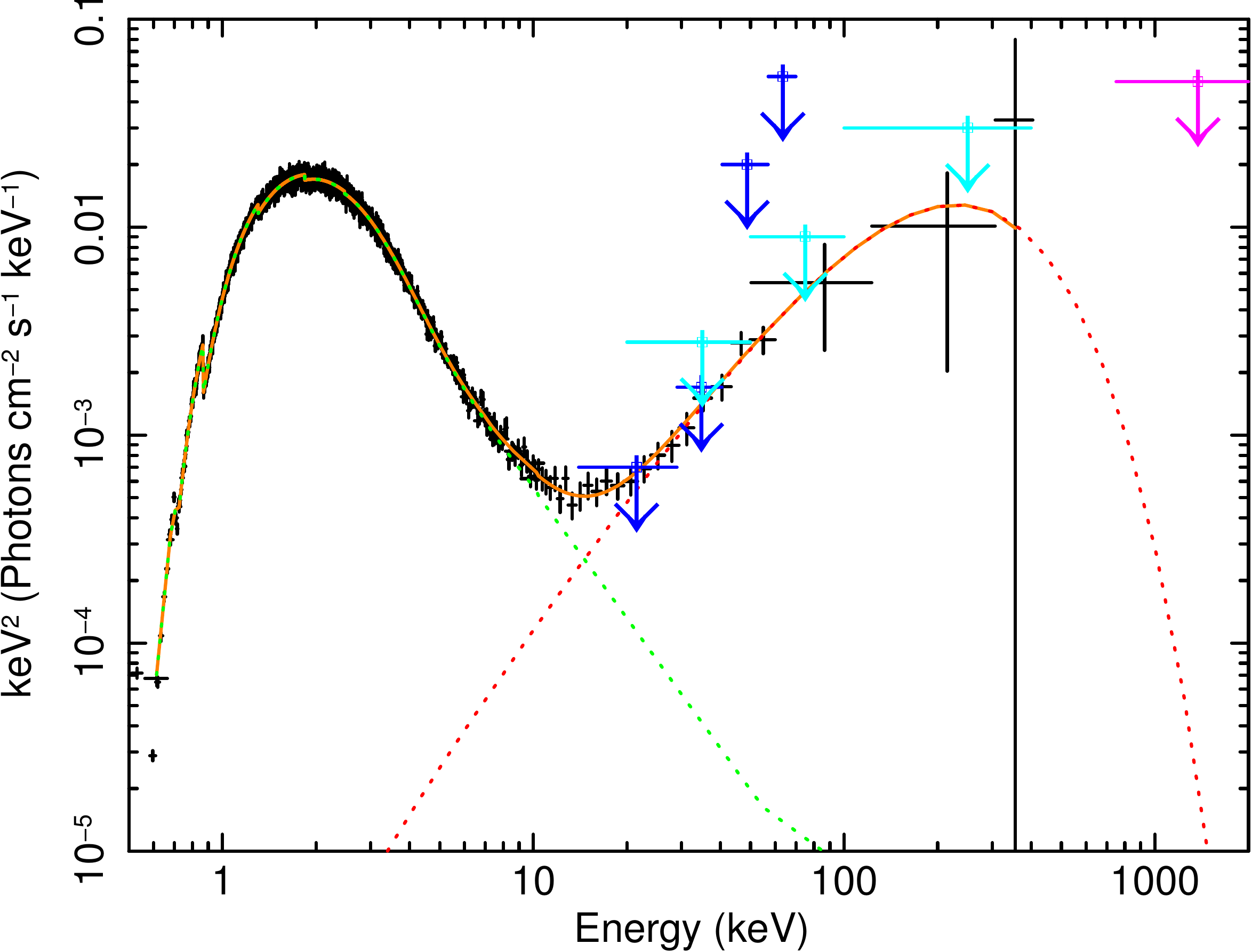}
\end{center}
\caption{
(Left) Expected $\nu$F$_{\nu}$ form of SGR 1806$-$20 observed by \textsl{ASTRO-H}. 
An absorbed blackbody plus a PL model is employed as a best-fit model 
	based on the \textsl{Suzaku} observation.
The orange, green dotted and red dotted curves
	are the total best fit model, SXC and HXC, respectively.
An absorbed blackbody plus a power-law model was employed.
(Right)
Same as left panel but for 1E 2259$+$586. 
The upper-limits are shown from \textsl{Suzaku} \citep[blue,][]{Enoto2010ApJ...722L.162E},
	 \textsl{INTEGRAL} \citep[light blue,][]{2006ApJ...645..556K} 
	 and {\it CGRO} (magenta). 
	 [After compiling this white paper, the NuSTAR reported the hard X-rays \citep{2014ApJ...789...75V}]
}
\label{fig:specsgr1806}
\end{figure}

As a first demonstration of the power of the \textsl{ASTRO-H} observation,
	we compared the $\nu F_{\nu}$ \textsl{Suzaku} and  \textsl{ASTRO-H} spectra of a bright anomalous X-ray pulsar 4U~0142+61 
	in Figure~\ref{fig:spec4u0142}.
Although the HXC was detected by \textsl{Suzaku} \citep{2011PASJ...63..387E},
	its spectral information is still poor above 80 keV (see Figure~\ref{fig:spec4u0142} left).
On the other hand, 	
	as shown in Figure~\ref{fig:spec4u0142} right,
	\textsl{ASTRO-H} will detect the HXC up to 400 keV with 3 $\sigma$ level,
	when assuming 100 ksec exposure.
In order to further estimate detectability of the cutoff,
	we compare the models with and without an exponential cut-off,
	indicating that \textsl{ASTRO-H} achieves the detection of the cut-off feature. 
Another bright magnetar, SGR~1806$-$20, 
	was also simulated in Figure~\ref{fig:specsgr1806} (left).
To obtain the same statistical significance,	
	the required 50 ksec observation by \textsl{Suzaku}
	is reduced only to 30 ksec by \textsl{ASTRO-H}.	
Although the \textsl{Suzaku}/HXD and \textsl{INTEGRAL} has already detected the HXC up to $\sim$20 keV
	\citep{2007A&A...476..321E, 2009PASJ...61S.387N, Enoto2010ApJ...722L.162E}, 
	\textsl{ASTRO-H} can extend it up to 100 keV,
	presumably providing a precise measurement 
	of the HXC and confirming the proposed correlation by \textsl{Suzaku}.
	In Section 3.1, we also show the HXI+SGD simulation of the bright AXP 1E~1841--045, located in the SNR~Kes~73, to illustrate the ability of \textsl{ASTRO-H} to pin down the nature
	of its hard X-ray emission (e.g in the light of magnetar versus fossil disk accretion models; see Figure~14, right).
	
	\textsl{ASTRO-H} has the advantage over \textit{NuSTAR} in that it will allow a \textit{simultaneous} broadband coverage needed to study the SXC and HXC together (see also the potential for studying these sources with the Soft Gamma-ray Detector on board \textsl{ASTRO-H}, \S3.4)

Contrary to the above three examples,
	the HXC from a famous anomalous X-ray pulsar,  1E2259$+$586 located in the SNR~CTB~109 (see Section 3.1), 
	has not yet been detected
	despite intensive observations with \textsl{INTEGRAL}, \textsl{RXTE} and \textsl{Suzaku}.
	\citep{2006ApJ...645..556K,2011PASJ...63..387E}.
If the correlation proposed by \textsl{Suzaku} is true,
	 1E 2259$+$586 may emit the HXC 
	 at $F_{h}=3.9\times 10^{-12}$ erg s$^{-1}$ cm$^{-2}$ 
	 based on the confirmed SXC 
	 $F_{s}=7.3\times 10^{-11}$ erg s$^{-1}$ cm$^{-2}$ with its characteristic age, 230 kyr.
Here we employed an absorbed blackbody plus a power-law for the SXC
	and another hard power-law for the HXC.
If there is no spectral break in the HXC and its photon index is between -1 and 1,
	\textsl{ASTRO-H}/HXI will clearly detect the HXC at least up to 50 keV with $\sim$120 ks exposure time.
On the other hand,
	without the spectral break,
	the HXC exceeded the the upper limits from COMPTEL (together with \textsl{Suzaku} and \textsl{INTEGRAL}),
	 as seen in 4U~0141+61.
Figure \ref{fig:specsgr1806} (right) shows $\nu$F$\nu$ spectrum 
	when adding an exponential cut-off feature at 150 keV,
	together with upper-limits obtained from 
	previous observations \citep{Enoto2010ApJ...722L.162E, 2006ApJ...645..556K}.
Even in this case,
	\textsl{ASTRO-H} will detect the HXC with a certain significance level,
	if the photon index is roughly 0, which is consistent with the correlation by \textsl{Suzaku}.	
Since 1E 2259$+$586 has the oldest characteristic age among magnetars observed with \textsl{Suzaku}, 
	revealing the HXC spectrum of the object means to test 
	whether the relation is applicable for all magnetars or not.

We finally evaluated detectability of the HXC from dim  sources in 10-100 keV range.
Figure \ref{fig:feasibility} shows required exposure time to detect the HXC with a $3\sigma$ significance.
Five magnetars indicated in the figure 
	were observed with several X-ray observatories, 
	but their weak HXC has not yet been reported so far.
Assuming hard X-ray fluxes calculated by the broadband spectral correlation suggested by \textsl{Suzaku},
	\textsl{ASTRO-H} will catch the HXC with a realistic exposure time (i.e., $\sim$100 ksec).
It should be also noted that 
	the imaging capability of the HXI enables us 
	to extract the hard X-ray spectrum from objects previously contaminated from nearby sources, 
	e.g., CXO J164710.2$-$455216.
Such a systematic study by \textsl{ASTRO-H} will help us reach a unified hard X-ray nature for this class.

\subsection{How the burst activity is related with the nature of magnetars?}

One of the prominent magnetar features are sporadic X-ray outbursts (flares) 
	accompanied by short bursts
	which both are thought to be directly related to the magnetic energy release.
However, 	
	the persistent HXC and short bursts during X-ray outbursts
	are poorly understood.

\subsubsection{Unresolved spectral change during the X-ray outburst}
\begin{figure}[h]
\centering
\includegraphics[scale=0.50]{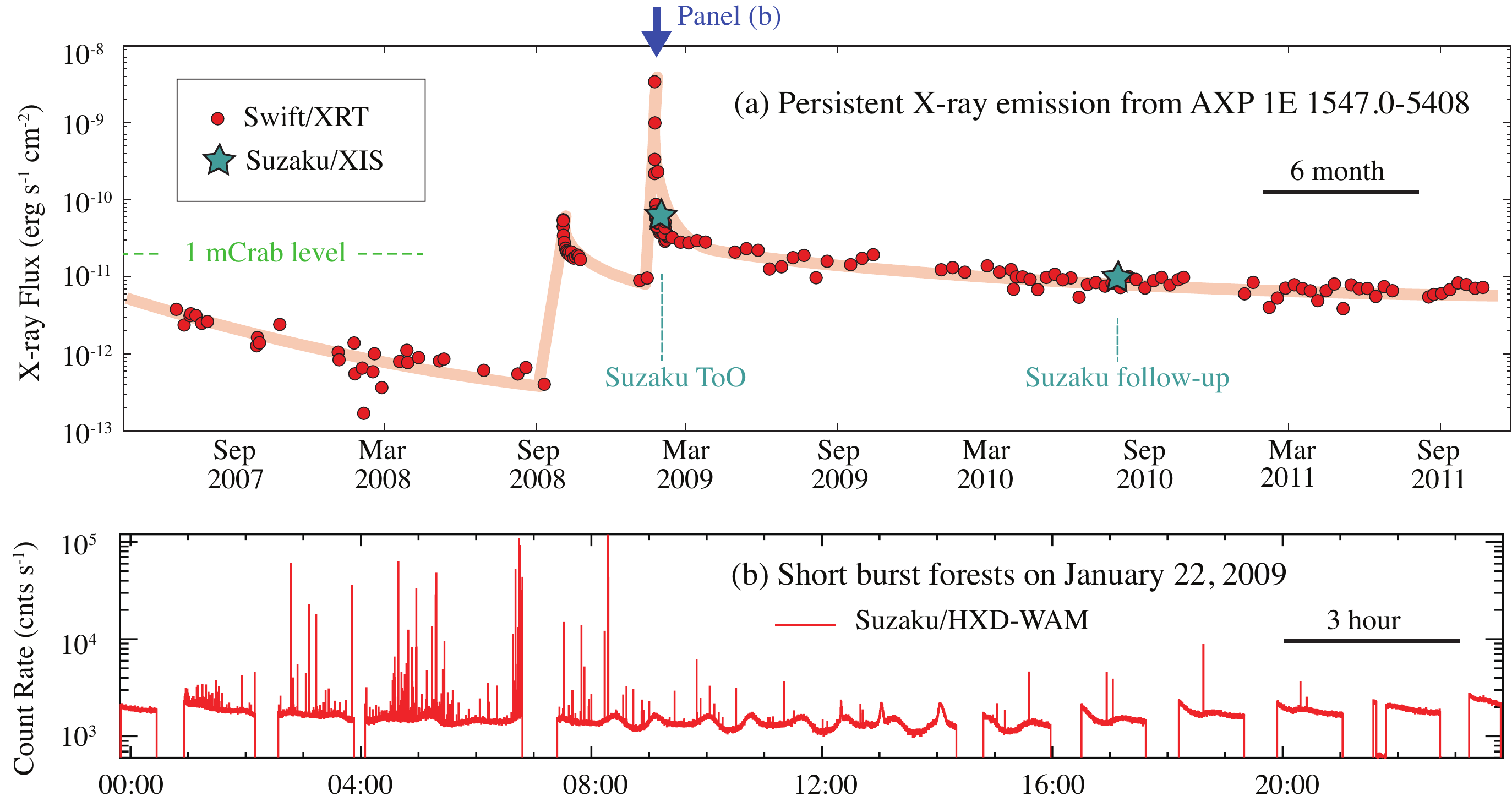}
\caption{
(a) A long-term SXC monitoring of AXP 1E~1547.0$-$5408 with {\it Swift}/XRT, with the absorbed 2-10 keV X-ray flux shown. Two \textsl{Suzaku} observations are also shown in green stars \citep{2010PASJ...62..475E}. (b) Short burst forests from this source recorded by \textsl{Suzaku}/HXD-WAM on January 22, 2009, indicated by a blue arrow in panel (a).
}
\label{fig1:1e1547_longhistory.eps}
\end{figure}

\begin{figure}[t]
\centering
\includegraphics[scale=0.25]{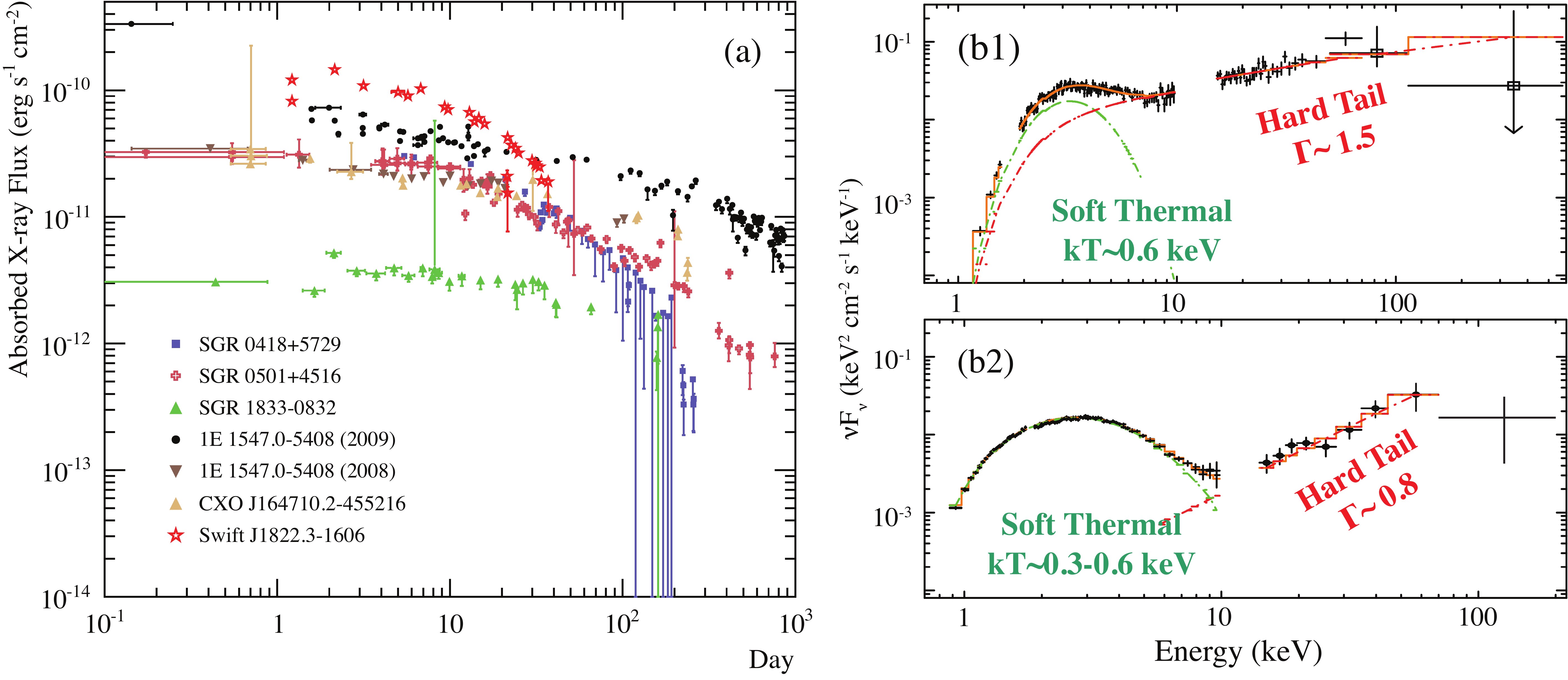}
\caption{
(a) Absorbed 2-10 keV X-ray fluxes of  transient magnetars monitored with Swift/XRT, {\it RXTE}/PCA, and \textsl{Suzaku}/XIS. The time onset corresponds to the first detection of magnetar short bursts from these sources, mainly detected with {\it Swift}/BAT. (b) \textsl{Suzaku} spectra during recent two transient activities of 1E~1547.0$-$5408 (b1) and SGR~0501+4516 (b2) \citep{2010PASJ...62..475E, 2010ApJ...715..665E}.
}
\label{fig1:magnetar_flux_decay.eps}
\end{figure}

The persistent SXC of magnetars sometimes increases 
	during the outburst by one to two orders of magnitude with unpredictable timing.
Such a transient enhancement lasts typically for a few months, with gradual decay.
Figure \ref{fig1:1e1547_longhistory.eps}a shows
	two recent sudden X-ray activations of the AXP 1E~1547.0-5408.
As shown in Figure~\ref{fig1:1e1547_longhistory.eps}b,
	it is often accompanied, in its early phase, by short burst activities.
These burst-active states have been already observed from some magnetars;
	XTE~J1810$-$197 \citep{2004ApJ...605..368G},
	CXOU~J164710.2$-$455216 \citep{2007ApJ...664..448I},
	SGR~0501+4516 \citep{2009ApJ...693L.122E},
	and 1E~1547.0-5408 \citep{2009ApJ...696L..74M}.
More recently, 
	there has been an increasing number of reports 
	from magnetars with much weaker dipole fields 
	($\lesssim 4.4\times 10^{13}$~G):
	SGR~0418+5729 \citep{2010Sci...330..944R},
	SGR~1833-0832 \citep{2010ApJ...718..331G},
	Swift~J1822.3-1606 \citep{2012ApJ...754...27R},
	and 
	Swift~J1834.9$-$0846 \citep{2012ApJ...748...26K}.
Figure~\ref{fig1:magnetar_flux_decay.eps}a presents 
	recent examples of the SXC flux decay of magnetars.

During the SXC outburst,
	an enhanced HXC above 10 keV has also been reported
	\citep{2009MNRAS.396.2419R, 2010ApJ...715..665E, 2010PASJ...62..475E, 2012ApJ...748..133K}.
While the SXC has been monitored in detail by {\it RXTE} and {\it Swift}  during the decay phase,	
	the HXC detections are still quite rare:
	only a few observations by \textsl{INTEGRAL} and \textsl{Suzaku} from SGR~0501+4516 in 2008 and 1E~1547.0-5408 in 2009.
Actual spectra recorded by \textsl{Suzaku} are shown in Figure~\ref{fig1:magnetar_flux_decay.eps}b and c,
	where the HXC was successfully detected with a flux of
	$2.7\times 10^{-11}$ erg s$^{-1}$ cm$^{-2}$ and
	$1.1\times 10^{-10}$ erg s$^{-1}$ cm$^{-2}$ in the 15--50 keV.
However,
	they are just snapshots during the decay phase,
	and it is not yet clear how the HXC evolves during the outburst state
	and how the HXC is physically related with the SXC.
Thus, the broadband spectral coverage for the SXC and HXC 
	is expected to help resolve the emission mechanism 
	and the postulated dissipation process of the magnetic energy.

\begin{figure}[htb]
\begin{center}
  \includegraphics[width=0.6\hsize]{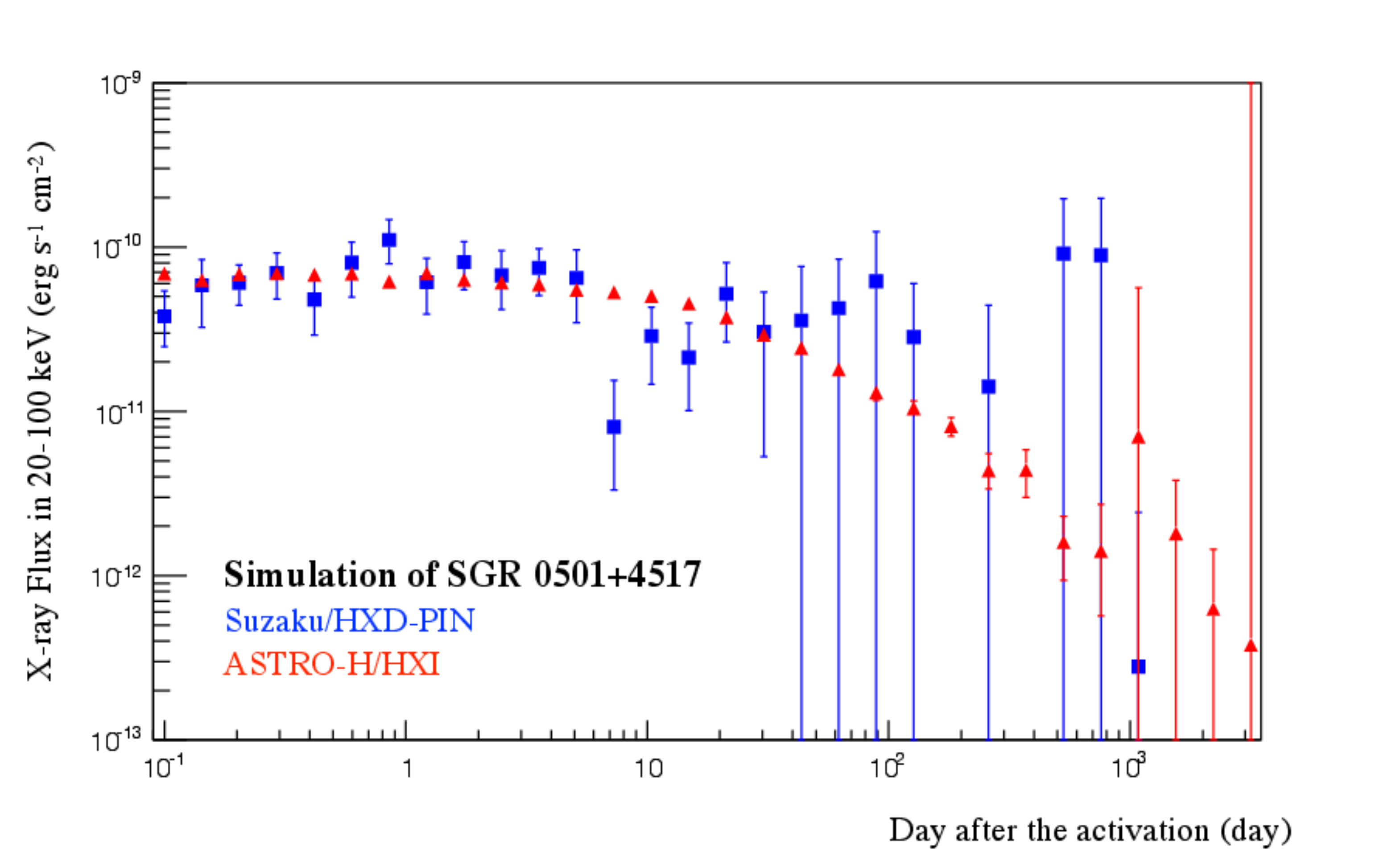}
\caption{
Simulations of the HXC monitoring toward an activated SGR~0501+4516 in 2008.
The spectral shape is assumed to be the same as that observed by \textsl{Suzaku} in 2008,
	and normalized to have the same decaying speed as the SXC.
Blue and Red data points are simulated X-ray fluxes in the 20--100 keV
	measured by \textsl{Suzaku}/HXD-PIN and \textsl{ASTRO-H}/HXI, respectively. 
Both exposures are set to be 40 ksec (90\% confidence level errors).
}
\label{fig1:fig_sim_sgr0501_decay.eps}
\end{center}
\end{figure}

Previous \textsl{Suzaku} ToOs (Figure~\ref{fig1:magnetar_flux_decay.eps})
        of 1E~1547.0$-$5408 and SGR~0501+4516 
          were performed $\sim$4 and $\sim$7 days after the onset of outburst, respectively.
The HXC quickly decays to $\sim$0.1 mCrab or less
	below the \textsl{Suzaku} detection limit within a few weeks.	
To compare \textsl{Suzaku} and \textsl{ASTRO-H}, 
	we simulated, in Figure~\ref{fig1:fig_sim_sgr0501_decay.eps}, 
	the HXC monitoring of the X-ray outburst. While \textsl{Suzaku} HXD-PIN can not detect the HXC 
	when $\sim$3 weeks have passed after the onset of the outburst, 
	\textsl{ASTRO-H} HXI still provides sufficient counts
	to detect the HXC
	within nearly a few ``years" after the onset.
These provide us, for the first time, with a detailed measurement
	of the HXC spectral evolution during the decay phase of magnetars.
	
So far,
	2--3 magnetar outbursts have been detected per year.
Recent accumulated discoveries of weak-field magnetars
	further suggest that 
	there is a large hidden population of this kind of sources.
The above continuous monitoring into the HXC 
	provides us a way to investigate the connection between the quiescent and 
	activated magnetars,
	which is vital to understand the evolutionary path of the magnetar class.	
The prompt one or two ToO observations can provide us the physical conditions 
	at an activated 	state,
	while the follow-up observations at once or twice per year 
	during subsequent two years
	can further give us the decay trend of the hard X-rays (e.g., photon index, decay speed),
	e.g., Swift J1822.3-1606 \citep{2012ApJ...754...27R}, 
	Swift J1834.9−0846 \citep{2012ApJ...748...26K}, 
	and 3XMM J185246.6+003317 (Zhou et al., 2014; Rea et al., 2014).

\subsubsection{Magnetar signatures in the short bursts and giant flares}
\begin{figure}[htb]
\begin{center}
  \includegraphics[width=60mm]{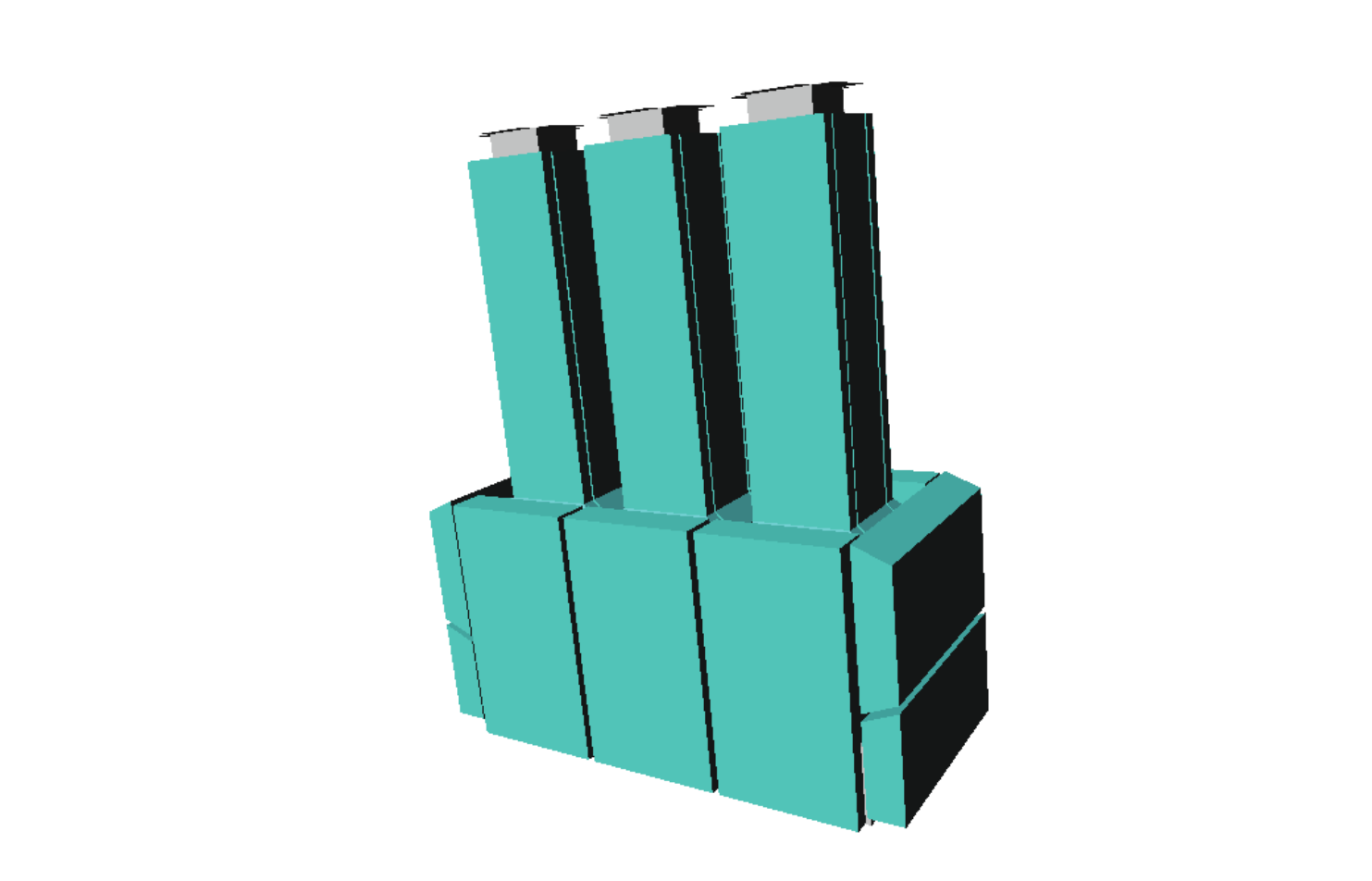}
\caption{
A schematic picture of Soft Gamma-ray Detector (SGD). The main detector of SGD is surrounded by
    large and thick 25 BGO crystals.
}
\label{SGDfig}
\end{center}
\end{figure}

One characteristic form of magnetar X-ray radiation is 
	sporadic emission of bursts
	with a typical duration from $\sim$0.1 second to a few hundred seconds.
The burst mechanisms is thought to be related to the
	rearrangement of the $B$-field 
	due to reconnections
	or motions of the NS crust (e.g., star quake).
These bursts are phenomenologically classified into three types: 
	``giant flares" ($L_{\rm x} > 10^{45}$ erg s$^{-1}$, lasting about a few hundred seconds),
	``intermediate flares" ($L_{\rm x} \sim 10^{42}-10^{43}$ erg s$^{-1}$, lasting a few seconds),
	and frequently occurring ``short bursts"
	($L_{\rm x} \sim 10^{38}-10^{41}$ erg  s$^{-1}$, $\sim$0.1-sec durations).
These explosive events often show luminosities exceeding the Eddington limit for 
	a NS of $1.4M_\odot$,
	$L_{\rm Edd} \sim 1.8\times 10^{38}$ erg s$^{-1}$,
	presumably due to suppression of  the electron scattering cross sections in the strong $B$-field.
These short bursts are thus attractive targets during the ToO observations.

\begin{figure}[t]  
     \begin{minipage}{0.48\hsize}   
     \begin{center}    
     \includegraphics[width=70mm]{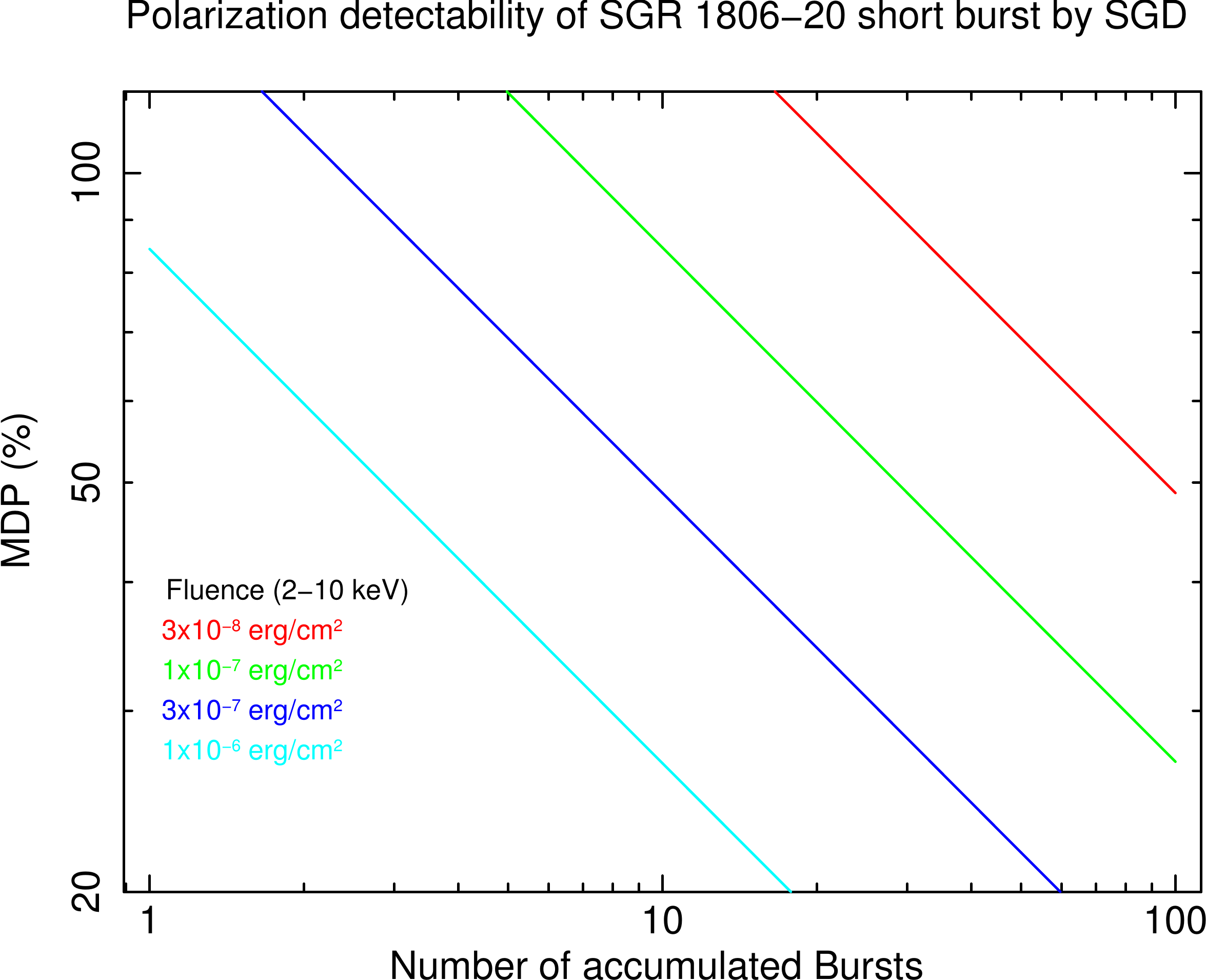}   
     \end{center}   
     \caption{Simulated polarization detectability from accumulated short burst events by SGD. Spectrum from 1806-20 bursts were assuming two black body model: $\frac{R_{\rm HT}}{R_{\rm LT}}^{2}$ = 0.01, $kT_{\rm LT}$ = 4 keV, $kT_{\rm HT}$ = 11 keV (Nakagawa et al., 2007) in varied burst intensities.}   
     \label{fig:nobvsmdp}  
     \end{minipage} 
     \hspace{5mm}
      \begin{minipage}{0.48\hsize}   
     \begin{center}    
     \includegraphics[width=70mm]{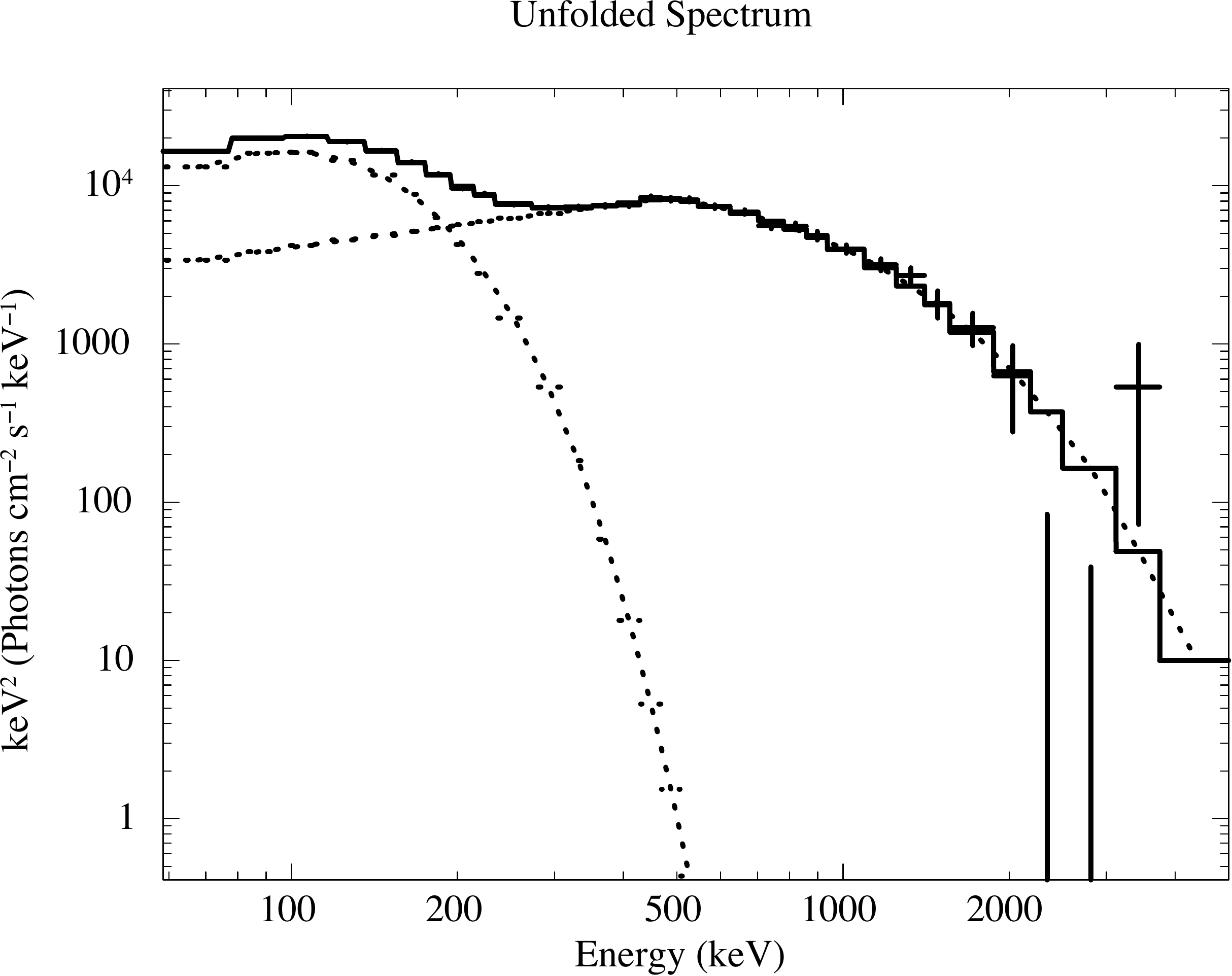} 
     \end{center}   
     \caption{Spectral simulation of one of the bright bursts of AXP 1E1547-5408 with the SGD shield detectors.
      Assumed spectral model is blackbody plus power-law model as reported in Yasuda et al. (in prep). Dashed-lines shows each spectral model component and the solid line represents the total model.}   
     \label{SGDWAM_1E1547simspec}  
     \end{minipage}
\end{figure}

{\bf Polarization of bursts:} High polarization degree 
	is expected from magnetars due to the high $B$-field. 
We simulated the polarization detectability of short bursts with SGD 
	using a simulator provided by the SGD hardware team. 
We assumed short bursts from SGR 1806$-$20 ($N_{\rm H}$ = 6$\times10^{22}$cm$^{-2}$, distance = 15 kpc) 
with spectral parameters of two-blackbody model 
(\citep{Nakagawa2007PASJ...59..653N}: $R_{\rm HT}^2/R_{\rm LT}^{2}$ = 0.01, $kT_{\rm LT}$ = 4 keV, $kT_{\rm HT}$ = 11 keV ). Figure \ref{fig:nobvsmdp} shows the expected polarization in different intensities 
when burst events accumulated. 
The burst rate of SGR~1806$-$20 in the active state was $\sim$2 bursts/day 
	observed by HETE-2 \citep{Nakagawa2007PASJ...59..653N}. 
In this simulation,
	we can detect the polarization of bursts from SGR/AXP when we observe bright bursts in the active state.
	
{\bf Proton CRSF:} Proton CRSF is a potential interesting target of the short burst using the \textsl{ASTRO-H} SXS as already discussed in \S3.1.

{\bf Is the persistent emission composed of unresolved short bursts?} The enhanced persistent and burst emissions have been 
	simultaneously observed in many activated magnetars.
However, 
	it is not yet clear how these two emission forms are physically related to each other.
One interesting possibility is that 
	the persistent emission is 
	composed of a large number of small bursts that are not individually detectable.
Such a possibility has been examined 
	using a cumulative number-intensity distribution of short bursts.
The observational information has so far remained insufficient to evaluate this possibility,
	since the studied short bursts 
	are so bright (with fluence $> 10^{-7}$ ergs cm$^{-2}$) 
	and infrequent 
	that their time-averaged flux is much lower than that of the persistent emission.
It is interesting 
	to examine, from observations of activated magnetars, 
	whether weaker short bursts become similar in spectral shape to 
	the persistent X-ray emission,
	as recently found in SGR~0501$+$4516 \citep{Nakagawa2011PASJ...63S.813N}.
The high sensitivity of \textsl{ASTRO-H} can provide further studies 
	connecting the persistent to burst emissions.
	
{\bf Advantages of the SGD shield detector:}
So far there is no clear detection of MeV photons from magnetar bursts except for during giant flares.
MeV photons of magnetar bursts are an important key 
	to investigate the radiation mechanism and physics in the strong $B$-field
	because photon splitting effect may suppress the high energy photons. 
In order to explore the MeV photons from short bursts, 
	more photon statistics with fine time resolution is essentially needed. 
The \textsl{Suzaku} Wide-band All-sky Monitor (WAM)
	has reported a hint of MeV photons from one strong short burst from AXP 1E1547$-$5408 (Yasuda et al., in prep),
	but the lack of time resolution prevented a detailed investigation.
The Soft Gamma-ray Detector (SGD) onboard \textsl{ASTRO-H}
	is surrounded by large Bi$_4$Ge$_3$O$_{12}$ crystals 
	to reduce the cosmic and gamma-ray backgrounds (Figure~ \ref{SGDfig}).
This active shield with a wide field-of-view 
	is  available as an all-sky monitor covering $\sim$200 keV to 5 MeV.
The most notable features are 
	a large effective area  ($\sim$800 cm$^2$ even at 1 MeV) roughly twice to the \textsl{Suzaku} WAM,
	fine time resolution (16 ms),
	and fine spectral resolution (32 channel than 4 channel of the \textsl{Suzaku} WAM).
The \textsl{Suzaku} WAM has to disable triggers 
	until the onboard triggered data is transferred to the spacecraft memory at the next SAA passage,
	while the SGD active shield can minimize this latency and the triggered data is immediately transferred.
These features become a powerful tool to study magnetar short burst 
	especially with large photon statistics and fine time resolution.	

Figure \ref{SGDWAM_1E1547simspec} shows a simulated spectrum of 
	the intense short burst from AXP 1E1547$-$5408, which
	has a signature of MeV photons\footnote{Due to readout deadtime and counter rollover effect, the data from the SGD
shield detectors should be subject to some corrections such as pile-up or carry over of observed counts. The estimated maximum
brightness of bursts which free from such corrections is about 100 Crab, and 1000 Crab would be also observable
with some corrections. From spectral simulations, a possible detection limit is found to be about 1 Crab. }.
The SGD monitor can clearly detect up to 2 MeV,
	and distinguish different spectral models, such as 2BB and BB+PL.
Figure \ref{fig:nobvsmdp}
shows the observable flux range of SGD shield detectors. 
We can expect that bright short bursts and intermediate flares  are good targets for SGD shield detectors.


\section*{Acknowledgements}
We thank Matthias K\"uhnel (Remeis Observatory \& FAU, Germany) for substantial contributions to preparing Figures 3 and 8 as well as for valuable comments on an earlier version of the manuscript.

\section{Appendix}
\subsection{Acronym}

\begin{description}
 \item[AXP]\mbox{}\\   Anomalous X-ray Pulsar.  
 \item[BAT]\mbox{}\\  Burst Alart Telescope onboard {\it Swift}.                          
 \item[CCD]\mbox{}\\  Charge Coupled Devise.      
 \item[CCO]\mbox{}\\  Central Compact Object.  
 \item[CRSF]\mbox{}\\ Cyclotron Resonance Scattering Feature.  
 \item[CSM]\mbox{}\\  Circumstellar Medium.  
 \item[GRLB]\mbox{}\\  Gamma-ray Loud  Binary.      
 \item[HMXB]\mbox{}\\   High Mass X-ray Binary.      
 \item[HXC]\mbox{}\\  Hard X-ray Component of magnetar X-ray spectra ($>$10 keV).      
 \item[HXD]\mbox{}\\  Hard X-ray Detector onboard {\it Suzaku}.      
 \item[HXI]\mbox{}\\  Hard X-ray Imager onboard {\it ASTRO-H}.      
 \item[ISM]\mbox{}\\  Interstellar Medium.                          .                              
 \item[LMXB]\mbox{}\\  Low Mass X-ray Binary.      
 \item[LPP]\mbox{}\\  Long Period Pulsar.                          
 \item[NS]\mbox{}\\   Neutron Star.  
 \item[PCA]\mbox{}\\  Proportional Counter Array onboard {\it RXTE}.  
 \item[QED]\mbox{}\\  Quantum Electrodynamics.
 \item[RPP]\mbox{}\\   Rotation Powered Pulsar.   
 \item[RRAT]\mbox{}\\  Rotating Radio Transient.  
 \item[{\it RXTE}]\mbox{}\\  Rossi X-ray Timing Explorer.      
 \item[SCF-effect]\mbox{}\\  Self Charge Filling effect.      
 \item[SFXT]\mbox{}\\   Supergiant Fast X-ray Transient.      
 \item[SGD]\mbox{}\\  Soft Gamma-ray Detector onboard {\it ASTRO-H}.                          
 \item[SGR]\mbox{}\\   Soft Gamma Repeater.  
 \item[SNR]\mbox{}\\  Supernova Remnant.
 \item[SXC]\mbox{}\\   Soft X-ray Component of magnetar X-ray spectra ($<$10 keV).      
 \item[SXI]\mbox{}\\  Soft X-ray Imager onboard {\it ASTRO-H}.      
 \item[SXS]\mbox{}\\  Soft X-ray Spectrometer onboard {\it ASTRO-H}.      
 \item[ToO]\mbox{}\\  Target of Opportunity.
 \item[VHE $\gamma$-ray]\mbox{}\\  Vert High Energy $\gamma$-ray.                          
 \item[WAM]\mbox{}\\  Wide-band All-sky Monitor onboard {\it Suzaku}.
 \item[XDINS]\mbox{}\\  X-ray Dim Isolated Neutron Star.  
 \item[XIS]\mbox{}\\  X-ray Imaging Spectrometer onboard {\it Suzaku}.      
 \item[XRBP]\mbox{}\\   X-ray Binary Pulsar.      

\end{description}

\clearpage
\begin{multicols}{2}
{\footnotesize

}
\end{multicols}

\end{document}